\documentclass{jfm}
\usepackage{graphicx}
\usepackage{float}
\usepackage{epstopdf, epsfig}
\usepackage{amsfonts}
\usepackage{amsmath}
\usepackage{tikz,bm}
\usetikzlibrary{decorations.markings}
\usetikzlibrary{arrows.meta,shapes.misc,decorations.pathmorphing,calc}
\usepackage{physics}
\usepackage{xcolor}
\usepackage{breqn}
\usepackage{subcaption}
\usepackage{ulem}
\usepackage{comment}
\usepackage{soul}
\usepackage{cleveref}
\usepackage[export]{adjustbox}

\newcommand{\lv}[1]{\hat{\bm{#1}}}

\newcommand{\rev}[1]{{\color{black}#1}}

\shorttitle{Inertial particle trapping in rotating vortex pair}
\shortauthor{S. Kapoor, D. Jaganathan, R. Govindarajan}

\title{Trapping and extreme clustering of \rev{finitely-dense} inertial particles near a rotating vortex pair}

\author{Saumav Kapoor,
  Divya Jaganathan
  \and Rama Govindarajan
  \corresp{\email{rama@icts.res.in}}}

\affiliation{International Centre for Theoretical Sciences, Tata Institute of Fundamental Research, Bengaluru 560089, India}

\begin{document}

\maketitle

\begin{abstract} 
Small heavy particles cannot get attracted into a region of closed streamlines in a non-accelerating frame \citep{haller2010}. In a rotating system, however, particles can get trapped \rev{\citep{angilella2010physica}} in the vicinity of vortices. We perform numerical simulations to examine trapping of inertial particles in a prototypical rotating flow described by a rotating pair of Lamb-Oseen vortices of identical strength, \rev{in the absence of gravity}. Our parameter space includes the particle Stokes number $St$, which is a measure of the particle's inertia, and a density parameter $R$, which measures the particle's density relative to the fluid. In particular, we study the regime $0<R<1$ and $0<St<1$, which corresponds to an inertial particle that is \textit{finitely} denser than the fluid. We show that in this regime, a significant fraction of particles can be trapped indefinitely close to the vortices, and display extreme clustering into objects of smaller dimension: attracting fixed points and limit cycles of different periods including chaotic attractors. As $St$ increases for a given $R$, we may have an incomplete or complete period-doubling route to chaos, as well as an unusual period-halving route back to a fixed point attractor. The fraction of trapped particles can be a non-monotonic function of $St$, and we may even have windows in $St$ for which no particle trapping occurs. At $St$ larger than a critical value, beyond which trapping ceases to exist, significant fractions of particles can spend long but finite times in the vortex vicinity. The inclusion of the Basset-Boussinesq history (BBH) force is imperative in our study due to the finite density of the particle. We observe that the BBH force significantly increases the basin of attraction over which trapping occurs, and also widens the range of $St$ for which trapping can be realized. Extreme clustering can be of significance in a host of physical applications, including planetesimal formation by aggregation of dust in protoplanetary disks, and aggregation of phytoplankton in the ocean. Our findings in the prototypical model provide impetus to conduct experiments and further numerical investigations to understand clustering of inertial particles.
\end{abstract}

\section{Introduction}\label{introduction}
Turbulent flows with suspended inertial (\rev{finitely small}) particles of varying relative densities are ubiquitous in both natural and industrial systems. Examples of such particulate suspensions include water droplets in clouds, near-neutrally buoyant microplastics and phytoplankton in the ocean, dust in protoplanetary disks, and air-drying systems of powdered food, fertilizers and pesticides. These particles, typically denser than the suspending fluid, can exhibit clustering. Clustering results in enhanced possibility for inter-particle collisions, which are critical for various natural phenomena including raindrop formation \rev{\citep{Falkovich2002,Wilkinson2006}}, reproduction among small organisms \rev{\citep{guastoReview12}} and planet formation \rev{\citep{tanga96, Bracco1999}}.  \rev{There have been several studies on particle dispersion in direct numerical simulations of turbulent flows \citep{squiresEaton91, Marshall2005, bec07}}. \rev{An alternative approach to model the turbulence has also been widely taken.} Intense vortices, generated by vortex stretching, are the building blocks of turbulent flows \citep{moffatt1994}. These prevalent turbulent flow structures are sampled by the suspended inertial particles, and this can influence their clustering. Therefore, the ultimate goal of understanding inertial particle dynamics in turbulent flows is \rev{equally} well-served by studying motion of particles in model vortical flows \rev{\citep{calvo1993,kiong94,MMN1995,rajuMeiburg97,Marshall1998,alekseyReview22}}, as the underlying physics can be well-elucidated. 

Our purpose is to understand the effect of systemic rotation on particle clustering. We present an argument for why the physics in a rotating system is worth studying. Conventional wisdom suggests that inertial particles denser than the suspending fluid centrifuge out of vortical regions and cluster in regions of high strain \citep{squiresEaton91, wangMaxey93, readeCollins00, aliseda02}. An explanation for this is provided in \cite{haller2008,haller2010} for the case of particles of small inertia, characterized by small Stokes number $St$ (ratio of particle relaxation timescale to a characteristic flow timescale). In the field description of particle velocity $\lv{v}$, \rev{an approximation that is allowed when $St\ll 1$ \citep{fieldapprox_Maxey87,fieldapprox_Druzhinin95,fieldapprox_Ferry01}}, its divergence at any spatial point $\lv{x}$ is given by
\begin{equation}
     \lv{\nabla} \cdot \lv{v} = - St\left\{|\lv{S}(\lv{x},t)|^2 -|\lv{\omega}(\lv{x},t)|^2\right\}.
 \end{equation}
Here, $\lv{S}$ and $\lv{\omega}$ are the strain-rate tensor and the \rev{rotation-rate} tensor respectively of the underlying incompressible flow field $\lv{u}(\lv{x},t)$, $t$ is time, $|.|$ is the Euclidean matrix norm, \rev{and $\hat{(\cdot)}$ refers to quantities in the laboratory-fixed frame.} A positive divergence implies the evacuation of a neighbourhood, while
a negative divergence implies clustering. They further argued that the net divergence from any region encompassed by a closed streamline is positive, i.e., there can thus be no clustering in the neighbourhood of an elliptic fixed point in the laboratory frame. In particular, we may conclude that particles \rev{of $St \ll 1$} will evacuate the vicinity of an isolated vortex and constantly move further from it. \rev{In the presence of background rotation, the criterion gets modified.} \cite{croorVortexPair2014} showed that
 \begin{equation}{\label{clusteringCriterionRot}}
     \nabla \cdot \bm{v} = -St\left\{|\bm{S}(\bm{x},t)|^2 -|\bm{\omega}(\bm{x},t)|^2 + 2\Omega^2\right\} \; \equiv St Q_{rot}~,
 \end{equation}
where $\Omega$ is the constant angular speed of the frame of reference, and \rev{quantities without the over-hat are written in the rotating frame. Thus} particles of small $St$ can cluster into regions within closed streamlines. This opens up the possibility that a significant loading of particles can be trapped in the vicinity of vortices in rotating systems for long times. Therefore, the physics  of particle collisions, coalescence and growth in rotating systems can be significantly different. The above equation also defines the Okubo-Weiss parameter $Q_{rot}$ in the rotating frame of reference. 

A pair of co-rotating point vortices, executing motion on a circle at a constant angular velocity, is a prototypical flow description which permits clustering in atypical locations due to the condition \ref{clusteringCriterionRot}. The dynamics of infinitely heavy point-like particles (finite $St$, $\rho_p/\rho_f \rightarrow \infty$) suspended in such \rev{vortical flows with systemic rotation} has been the subject of several studies. \rev{\cite{angilella2010physica} and} \cite{croorVortexPair2014} considered point vortex-pair of equal strengths, while \cite{angilella2010} considered unequal strengths. They showed that such inertial particles can get trapped at various attracting fixed points in the rotating frame, whose exact locations vary with the Stokes number. Further, an attracting fixed point may cease to exist beyond a critical Stokes number or give way to multiple attracting points. \cite{angilella2014} showed that particles can undergo transient clustering (and chaos) near co-rotating vortices in the presence of a wall. \rev{\cite{nath2024clustering}} find similar behaviour for infinitely dense inertial particles near a single non-axisymmetric (elliptical) vortex in the presence of shear. \rev{\cite{tanga96} proposed the trapping of heavy dust particles in the vortices present in rotating solar nebulae as a mechanism for planetesimal formation.} Along similar lines, \cite{bec2023} have reported enhanced clustering of heavy inertial particles in Keplerian turbulence with rotation and shear, which model gaseous systems in protoplanetary disks.  

Our background flow consists of a pair of co-rotating Lamb-Oseen vortices of identical \rev{circulation, $\Gamma$}. The fact that Lamb-Oseen vortex closely emulates a typical vortical structure seen in 2D turbulence \citep{gallayWayne02, ramadugu2022} legitimises our choice. The width of the vortices is taken to be sufficiently small compared to their separation. \rev{Two identical vortices which are initially far apart undergo merger in four stages \citep{cerretelliWilliamson2003}. In the first diffusive stage they maintain their individual Gaussian structure and mutual separation while executing constant angular velocity motion on a circle. In two dimensions, as the flow Reynolds number, $Re \equiv \Gamma/\nu$, where $\nu$ is the kinematic viscosity of the fluid, is made arbitrarily large, the first stage of merger can last for an arbitrarily long time. Once the vortices diffuse to a radius of about $0.3$ times their separation, the second stage of merger begins, and the large-scale motion is no longer periodic.} For simplicity, we assume a high enough flow Reynolds number such that the vortices \rev{execute circular motion with constant angular velocity} during our simulation time.  We distinguish our work from earlier studies in our consideration of inertial particles that are finitely-dense $(\rho_p/\rho_f<\infty)$. Consequently, we have an additional dimensionless parameter in our problem besides the Stokes number, namely the density factor $R$, which is a measure of particle to fluid density ratio. We model the dynamics of inertial particles in our study using the Maxey-Riley equation (hereafter referred to as the `MRE') \rev{which includes the Basset-Boussinesq history (BBH) force}. A majority of inertial particle studies employ the reduced MRE, i.e.,
omit the BBH force. For finite $\rho_p/\rho_f$, however, the effects of the BBH force could become significant, and are expected to be pronounced for near-neutrally buoyant particles ($\rho_p \sim \rho_f$). In order to understand the dynamics as well as to isolate the effects of the history force, we study both the reduced MRE without the BBH force and the MRE with the BBH force. \rev{The reduced MRE represents a dynamical system in the position-velocity state space, i.e., given the present state of the particle, the future state is uniquely determined.  But, upon the inclusion of the BBH force, the resulting integro-differential equation enforces non-local dynamics in time. Indeed, the entire past trajectory is required to determine a future state of the particle.} Our recent studies \citep{prasath2019,jaganathan2023} enable us to interpret the full MRE as a dynamical system embedded in an extended space. This reinterpretation offers computational advantages, as we shall briefly discuss in our numerical methods \rev{\cref{sec1p1}}. We note here that \citet{chong2013} and \citet{daitche2014} have previously conducted studies on inertial particle clustering with the BBH force included in their models in different flows. 

Our study finds a host of new clustering features that occur in rotating flows of finitely-dense particle suspensions. Particles of finite density have a higher propensity to be trapped forever in the system than infinitely dense particles. A significant fraction of particles in the system participate in extreme and permanent clustering \rev{on to attractors}, up to Stokes numbers of order one. The final clusters, \rev{or attractors}, rotate with the system. They can be point-like, in the form of attracting fixed points, or \rev{annulus}-like, in the form of limit cycles of \rev{varying periodicities, or chaotic attractors. Depending on the Stokes number and the density ratio, there are a variety of transitions from one type of attractor to another. Beyond a critical Stokes number, no particles are trapped forever, but there can be} long-lasting transients. \rev{Particle trapping is} significantly enhanced\rev{, and the particle attractors often qualitatively altered,} by the inclusion of the BBH force. 

Since we refer to trapping and clustering repeatedly, it is useful to distinguish them. Trapping refers to the condition of particles to be constrained to a particular predefined region. Clustering, on the other hand, refers to a collection of particles progressively occupying smaller volumes with time. A clustering set of particles need not remain in a fixed region, whereas trapped particles need not cluster. 

The rest of the paper is organized as follows. In \cref{sec:2}, we describe the physical model of inertial particles in co-rotating vortex pair and the associated governing equations. We also outline the numerical methods and analysis tools used in the study. In \cref{sec:3,sec:leak}, we discuss the trapping dynamics observed in the model with and without the BBH force for particles of different inertia and densities. We conclude in \cref{sec:summary} with a discussion on our observations and the limitations of the model.

\section{Governing equations for the flow and particles}\label{sec:2}
\begin{figure}
    \centering
    \begin{minipage}[b]{.31\textwidth}   
    \includegraphics[width=\textwidth]{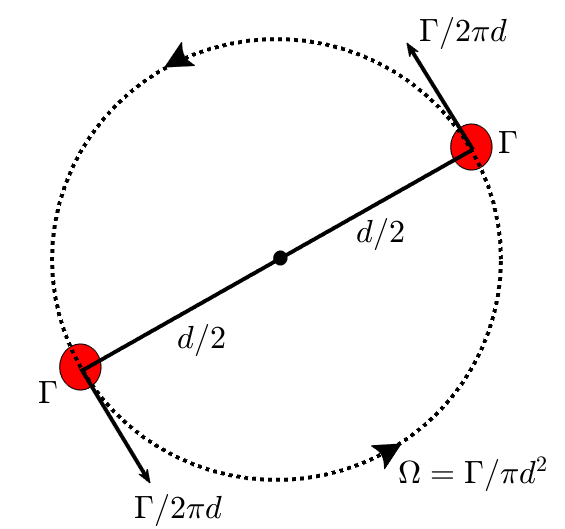}
    \subcaption{}
    \end{minipage}
    \:
    \begin{minipage}[b]{.275\textwidth}   
    \includegraphics[width=\textwidth]{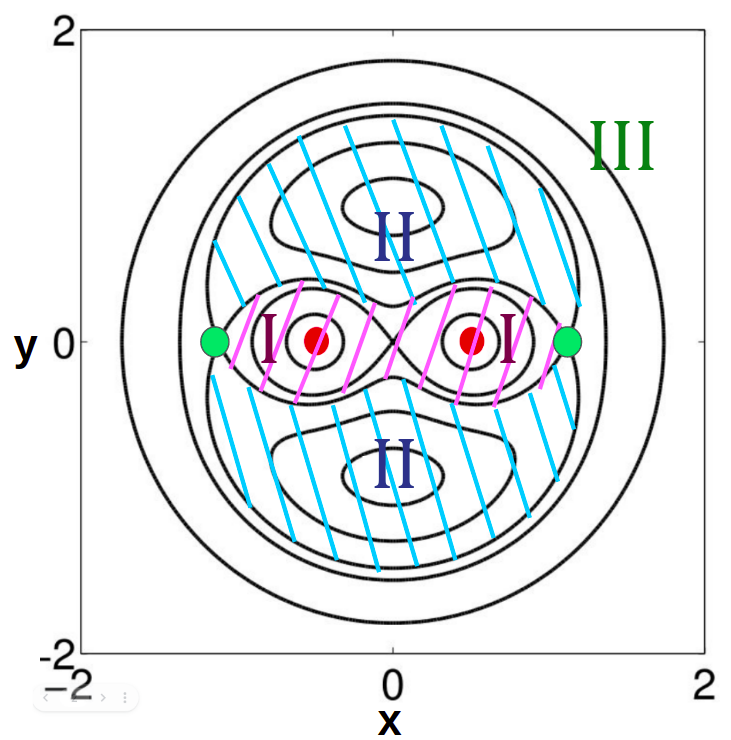}
    \subcaption{}
    \end{minipage} 
    \:
    \begin{minipage}[b]{.31\textwidth}   
    \includegraphics[width=\textwidth]{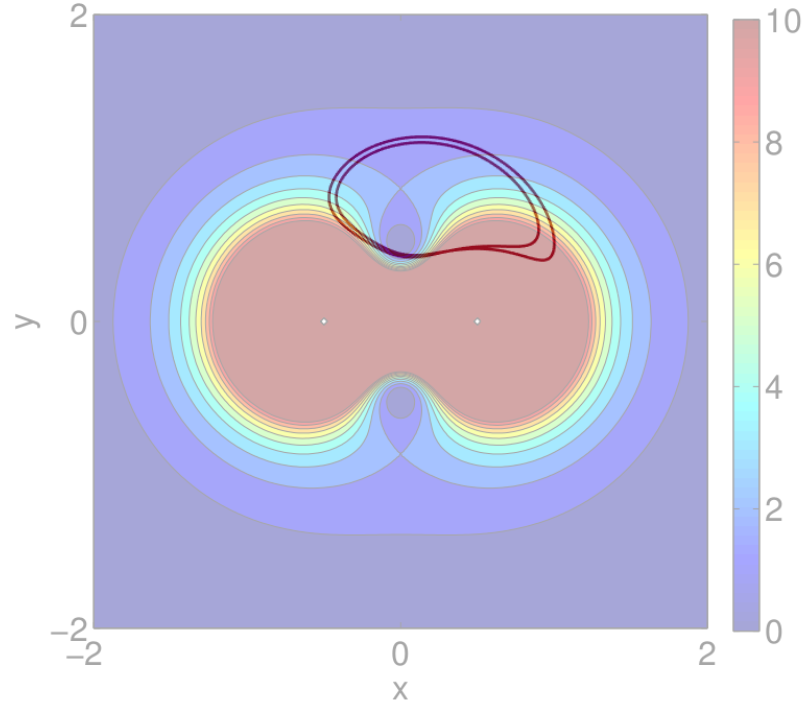}
    \subcaption{}
    \end{minipage}  
    \caption{(a) Schematic showing two identical vortices executing circular motion at a constant rate. The coordinate system rotates with them. (b) Vortex locations (red dots) and representative tracer-particle trajectories (\rev{closed orbits shown in} black lines) are shown in the rotating frame of reference. Region II is the primary host for the attracting orbits of inertial particles. The green points are hyperbolic fixed points from which heteroclinic orbits emanate, which separate regions I, II and III. Region III contains simple closed orbits encircling both vortices. (c) Negative of the Okubo-Weiss paramater $Q_{rot}$ overlaid by a representative limit cycle (attractor) of inertial particle trajectories.}
    \label{fig:schematic}
\end{figure}

The Lamb-Oseen vortices in the pair are of identical strength $\Gamma$ and core-width $b$, with their centres separated by a distance $d$, chosen such that $b\ll d$, as shown in fig.~\ref{fig:schematic}(a). In accordance to the Biot-Savart law, these vortices revolve around each other on a circle of diameter $d$, with an angular speed $\Omega=\Gamma/\pi d^2$, while maintaining a constant mutual angular separation of $\pi$. The corresponding time period of rotation is $T=2\pi/\Omega$. 

The separation length $d$, the time period of rotation $T=2\pi/\Omega$, and their ratio $U=d/T$ provide natural length, time, and velocity scales to non-dimensionalise the system. In the non-dimensional form, the background flow field is given by
\begin{equation}{\label{vortexpairfield}}
    \lv{u}(\lv{x},t) = \pi \bm{e}_z \times \Bigg[(1-e^{-|\lv{x}-\lv{X}|^2/b^2})\frac{\lv{x}-\lv{X}}{|\lv{x}-\lv{X}|^2} + (1-e^{-|\lv{x}+\lv{X}|^2/b^2})\frac{\lv{x}+\lv{X}}{|\lv{x}+\lv{X}|^2} \Bigg],
\end{equation}
where the instantaneous vortex centres are at $ \lv{X} = (\cos(2\pi t)/2, \;\sin(2\pi t)/2)$, and $\bm{e}_z$ is the unit vector perpendicular to the plane of the vortices. The non-dimensional vortex width is set to $b=0.1$ throughout our analysis, without loss of generality. 

We are interested in the dynamics of inertial particles in the above unsteady background flow \rev{in the absence of gravity}.  We model the particles as rigid and spherical with radius $a$, and of negligible particle slip Reynolds number, i.e, $Re_p=a |\bm{v}_d(t)-\bm{u}_d(\bm{r}_d,t)|/\nu \ll 1$. Here, \rev{ $\bm{u}(.,t)$ is the background fluid velocity,} and $\bm{v}(t)$ and $\bm{r}(t)$ are the instantaneous particle velocity and location respectively\rev{, which are dimensional when indicated with subscript `$d$', non-dimensional otherwise}. We also assume negligible shear Reynolds number, $Re_s=a^2 s/\nu$, where $s=|\nabla \bf{u}_d|$ is a measure of velocity gradients in the flow field. These are fair assumptions for \rev{sufficiently} small particles. \rev{Further, we assume} that the particles are in dilute suspension, allowing us to neglect their mutual interaction as well as their effect on the flow \rev{(one-way coupling)}. The dynamics of an inertial particle in such a suspension, under the above assumptions, is governed by the Maxey-Riley equations \citep{maxey1983, gatignol1983} given in non-dimensional form as,
\begin{subequations}
\begin{align}
\frac{d\lv{r}}{dt} &= \lv{v}~,\\
\nonumber \frac{d \lv{v}}{dt} &= -\frac{(\lv{v}-\lv{u}(\lv{r}))}{St}+R \frac{D\lv{u}}{Dt}(\lv{r})\\
& -\sqrt{\frac{3R}{\pi St}} \Bigg[\frac{(\lv{v}_0-\lv{u}_0)}{\sqrt{t}}+ \int_0^{t} \:ds
\frac{1}{\sqrt{t-s}} \left\{\frac{d}{ds}(\lv{v}(s)-\lv{u}(\lv{r}(s)))\right\} \Bigg] ~,\label{nondimensionalMR}
\end{align}
\end{subequations}
where the subscript $0$ stands for a quantity at the initial time. The three forcing terms on the right-hand side of \cref{nondimensionalMR} are the viscous Stokes drag, the force due to local fluid acceleration (which includes the added mass and pressure drag effects), and the BBH force respectively. We ignore the Fax\'en corrections, which account for the differential flow curvature effects across the diameter of the particle, assuming that the particle is sufficiently small. The two non-dimensional numbers that feature in the equation are the density factor $R$, and Stokes number $St$, defined as:
\begin{equation}
 R \equiv 3/(2\beta+1), \quad St \equiv \tau_p/T
 \label{Reqn}
\end{equation} 
where $\beta=\rho_p/\rho_f$ is the ratio of particle and fluid densities, $\tau_p=a^2/(3\nu R)$ is the relaxation time of the particle, and $T$ is the time period of vortex rotation. Note that $R \rightarrow 0$ for an infinitely dense particle, while $R=1$ for a neutrally buoyant particle and $R=3$ for a light particle such as a bubble. The limit $St \rightarrow 0$ corresponds to a tracer/non-inertial particle which faithfully follows the fluid streamlines. We shall work in the regime $0 < R < 1$ and $0<St<1$, which corresponds to an inertial particle that is \textit{finitely} denser than the fluid.

The dynamics is better brought to light by our choice of frame of reference. We choose a reference-frame rotating with the non-dimensional angular velocity of the vortex-pair. In this co-rotating frame, the background flow is steady, and the stationary vortices are centred at $(\pm 1/2, 0)$. The representative streamlines of the stationary flow are shown in fig. \ref{fig:schematic}(b), which also defines the $x$ and $y$ coordinates. We may define three water-tight regions based on the behaviour of tracer particles. Region I includes the close vicinity of the vortices; tracer particles seeded in this region execute closed trajectories encompassing the vortex closest to them, and are influenced primarily by that vortex. In region II, which is of primary interest to us, the tracer particles move on \rev{closed orbits} passing through their initial positions. They are, on average, equally influenced by the two vortices. Region III is the far-field, where tracers execute closed orbits encircling both vortices. As we go further from the origin and into Region III, tracer particles increasingly perceive the system as a single vortex of twice the strength. In fig.~\ref{fig:schematic}(c), we plot \rev{the modified Okubo-Weiss parameter, $\Omega_{rot}$, which is most negative in the red region. According to \cref{clusteringCriterionRot}, heavy particles of $St \to 0$ will have higher propensity to cluster in the red region}. Overlaid on this plot is a typical limit cycle for finitely-dense inertial particles, \rev{where particles reach asymptotically in time}. This suggests that finitely-dense particles of finite Stokes number can cluster in regions well outside that predicted by \cref{clusteringCriterionRot}, which is valid only for very small Stokes number. Upon comparing the locations of the closed streamlines in fig.~\ref{fig:schematic}(b) to the limit cycle in fig.~\ref{fig:schematic}(c), we demonstrate that particles can cluster within closed streamlines enclosing elliptic fixed points in a rotating frame.

In the rotating frame, the non-dimensional background flow field is given by the transformation
\begin{align}{\label{vel2rot}}
    \lv{u}(\lv{x}) - 2\pi \bm{e}_z\times \lv{x} \rightarrow \bm{u}(\bm{x})
\end{align}
whereas the equation of motion \cref{nondimensionalMR} for the particle, upon defining a slip velocity $\bm{v}_{rel} \equiv \bm{v}-\bm{u(\bm{r})}$, reads as:
\begin{subequations}{\label{nondimensional_rotMR}}
\begin{align}
\frac{d \bm{r}}{dt} &=\bm{v},\\
\frac{d \bm{v}}{dt}&=-\frac{\bm{v}_{rel}}{St}+R\bigg\{\frac{D\bm{u}}{Dt}(\bm{r})+4\pi\bm{e}_z\times\bm{u}(\bm{r})-4\pi^2\bm{r}\bigg\}\nonumber\\
&\quad\ - \sqrt{\frac{3R}{\pi St}}\qty[\frac{\bm{v}_{rel,0}}{\sqrt{t}}+\int_{0}^{t}ds\ \frac{d\bm{v}_{rel}(s)/ds+(2\pi\bm{e}_z)\times\bm{v}_{rel}(s)}{\sqrt{t-s}}]\nonumber\\
&\quad - 4\pi \bm{e}_z\times\bm{v}+4\pi^2\bm{r}.
\end{align}
\end{subequations}
Note that the variables in \cref{nondimensional_rotMR} are now measured in the co-rotating frame.
\subsection{Numerical methods} \label{sec1p1}
\begin{figure}
    \centering
    \subfloat[]{\includegraphics[width=0.4\textwidth]{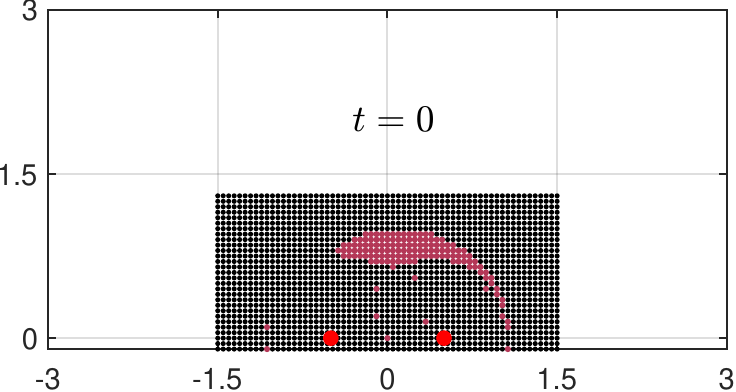}}
    \quad
     \subfloat[]{\includegraphics[width=0.4\textwidth]{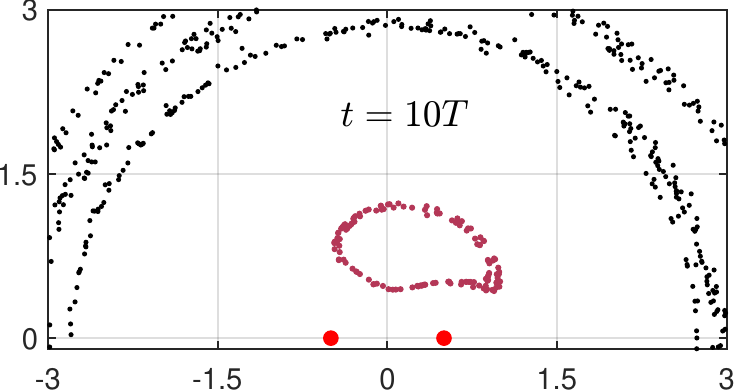}}
    \caption{Typical evolution of an ensemble of inertial particles $(\rho_p/\rho_f>1)$ in the position space, in the rotating frame of an identical vortex pair. The particles are uniformly seeded near the vortex pair as shown in (a). \rev{Particles after 10 time periods of rotation, evolved under reduced MRE, are shown in (b).} A fraction of particles (coloured maroon) get trapped to an attractor such as a fixed point or a  limit cycle. However, a majority of particles (coloured black) are centrifuged out in \rev{spiralling} orbits. The former set of particles forms our primary focus.}
    \label{fig:typicalstateevolution}
\end{figure}
We perform numerical simulations of inertial particles for a range of Stokes number and density ratios. Without the BBH force, \cref{nondimensional_rotMR} \rev{reduces to a nonlinear ODE}. Therefore, for the part of the analysis where we exclude the BBH force, we use \rev{the standard} fourth-order Runge-Kutta scheme  to integrate particle trajectories in accordance with \cref{nondimensional_rotMR}. The time-step is chosen between $\Delta t = 10^{-3}$ and $10^{-4}$. However, with the inclusion of the BBH force\rev{, the equation of motion is an integro-differential equation. This precludes the direct use of standard time-integrators, such as the Runge-Kutta schemes, to integrate the history-dependent particle trajectories without incurring quadratically growing computational cost and a linearly increasing memory storage cost.} We therefore use the explicit time-integrator for the MRE prescribed in \citet{jaganathan2023}. This explicit integrator possesses the benefits of standard integrators, in particular a nominal linear growth rate of computational costs with simulation time and a time-independent memory storage requirement. The algorithm involves rewriting \cref{nondimensionalMR} as a local-in-time system of equations following a Markovian embedding procedure. The embedding introduces an auxiliary variable which encodes the history of particle trajectory exactly and evolves according to an ordinary differential equation. Consequently, the resultant set of equations represents a dynamical system in an abstract extended space. We solve \cref{nondimensionalMR} using the second-order Runge-Kutta time-differencing method (with $10^{-4} < \Delta t < 10^{-3}$) in \citet{jaganathan2023} in the laboratory frame, and then transform the variables to their counterparts in the rotating frame. 

For detecting an attractor, we initially place sufficient number of particles \rev{$(> 2500)$} on a uniform grid over a chosen spatial region in \rev{$[-1.5,1.5] \times [0,1.5]$}. In the results presented, the initial particle velocity is set to zero in the rotating frame. We evolve their trajectories over a long enough time to achieve motion on an attractor. We use the last $5\%$ of the trajectory to calculate the properties of the attractor. Fixed points are easy to detect in our simulations, since the velocity of a particle in the rotating frame goes to zero as it approaches a fixed point. To detect limit cycles, we note that at its extremities in the $x$-direction, we must have the $x$ component of the particle velocity $v_x=0$ in the rotating frame. We count the number of distinct $x$ locations at which $v_x=0$ and divide by two to get the period of the limit cycle. When every such location is distinct, we have a chaotic attractor. We point out that in the event of a basin of attraction (BoA) being very small, there is a chance that we may have missed the attractor entirely. Therefore, we may not have found the exhaustive set of all attractors, but that was not the purpose of our study. Without the BBH force, a few tens of non-dimensional time are typically sufficient for particles to converge to an attractor, whereas with the inclusion of the BBH force the system takes longer to converge to the final attractor. \rev{The time taken for this depends on the resolution we require. By $\sim 10T$ the attractors are clearly delineated, but we run the simulations sometimes for $\sim 500 T$ for near-perfect convergence.} Given reasonable access to compute power, this study would have been prohibitive by the brute force method of solving for the BBH force for a large ensemble of particles and long integration-times, and speaks to the efficacy of our numerical method. 

In fig.~\ref{fig:typicalstateevolution}, we show a typical evolution of an ensemble of particles in the position space. In fig.~\ref{fig:typicalstateevolution}(a), we have a uniformly seeded particle ensemble with $R=0.84, St=0.22$, each particle coloured either in maroon or black. \Cref{fig:typicalstateevolution}(b) shows their respective positions after 10 time periods of rotation: the maroon patch of particles has converged to an attractor (a limit cycle here) whereas the black patch of particles has centrifuged out \rev{spirally}. Since we are interested in clustering and trapping behaviour of particles, centrifuging particles (coloured black in the figures) are excluded from our study and the results therein. 

\rev{In the upcoming sections, we restrict our discussion to a co-rotating vortex pair of identical strengths, and to the case of particle velocity initialized to zero in the rotating frame. Unequal vortex strengths and different initial conditions are discussed in \cref{app:A,app:B} respectively. We see that while the behaviour is qualitatively similar, significant quantitative variations can exist. Thus our model flow is to be treated as one bringing out general physical features of trapping and clustering, and not as a predictive tool.} 

\section{Particle trapping dynamics}\label{sec:3}
\begin{figure}
    \centering
    \begin{minipage}[b]{.3\textwidth}   
    \includegraphics[width=\textwidth]{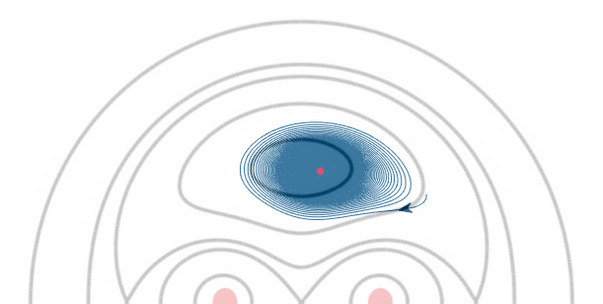}
    \end{minipage}
    \:
     \begin{minipage}[b]{.3\textwidth} 
     \includegraphics[width=\textwidth]{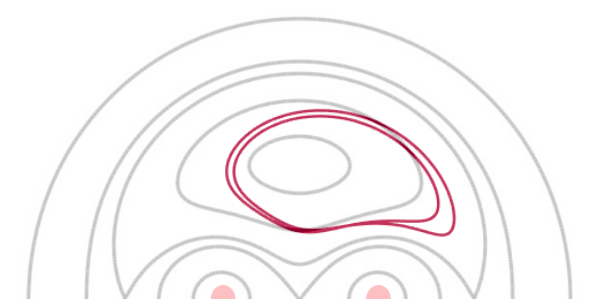}  
    \end{minipage}
    \:
    \begin{minipage}[b]{.3\textwidth}   
    \includegraphics[width=\textwidth]{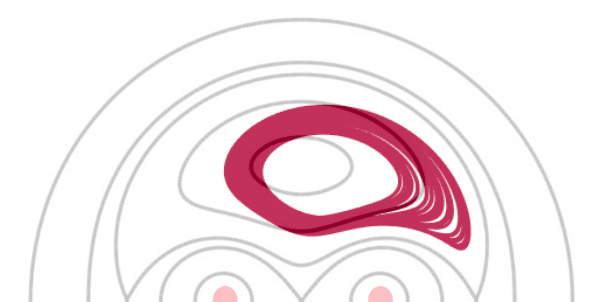}
    \end{minipage}
    
    \vspace{0.1in}
    
    \begin{minipage}[b]{.31\textwidth}   
    \includegraphics[width=\textwidth]{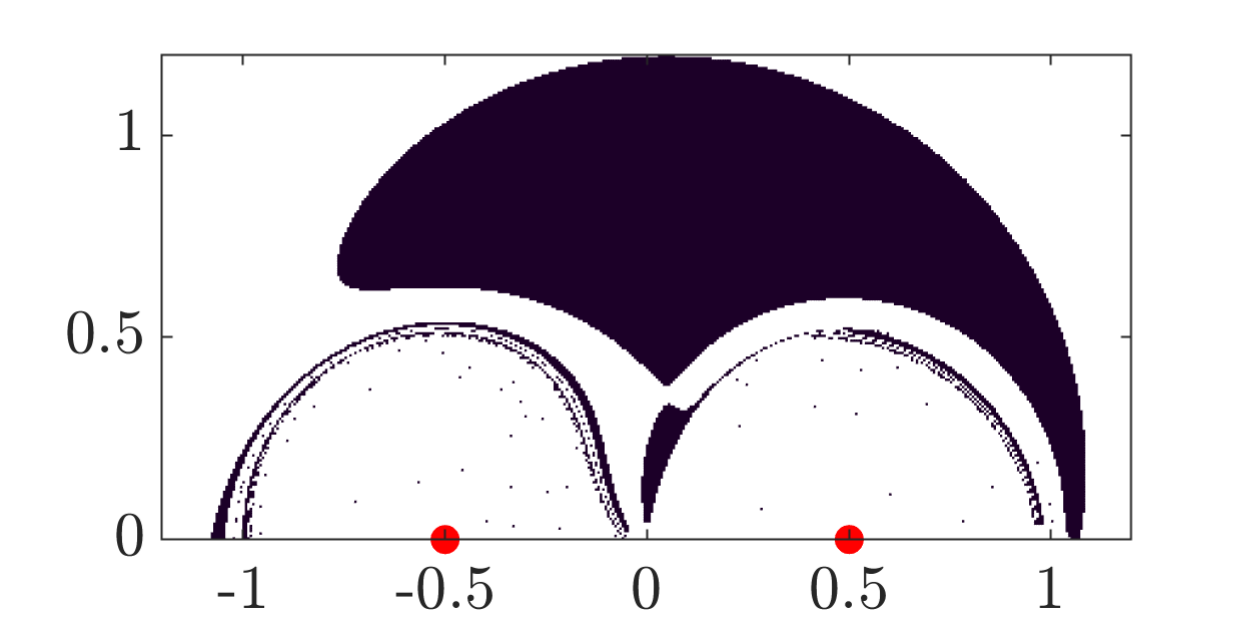}
    \subcaption{$St=0.09$}
    \end{minipage}
    \!
    \begin{minipage}[b]{.31\textwidth} 
    \includegraphics[width=\textwidth]{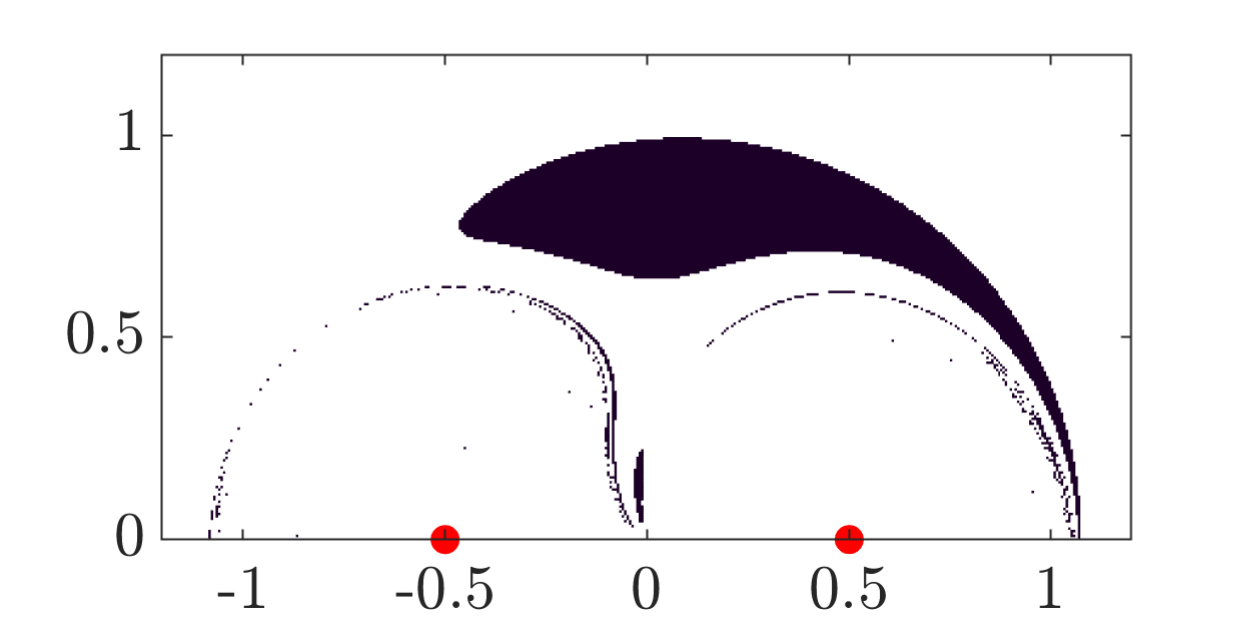}
    \subcaption{$St=0.22$}
    \end{minipage}
    \!
    \begin{minipage}[b]{.31\textwidth}   
    \includegraphics[width=\textwidth]{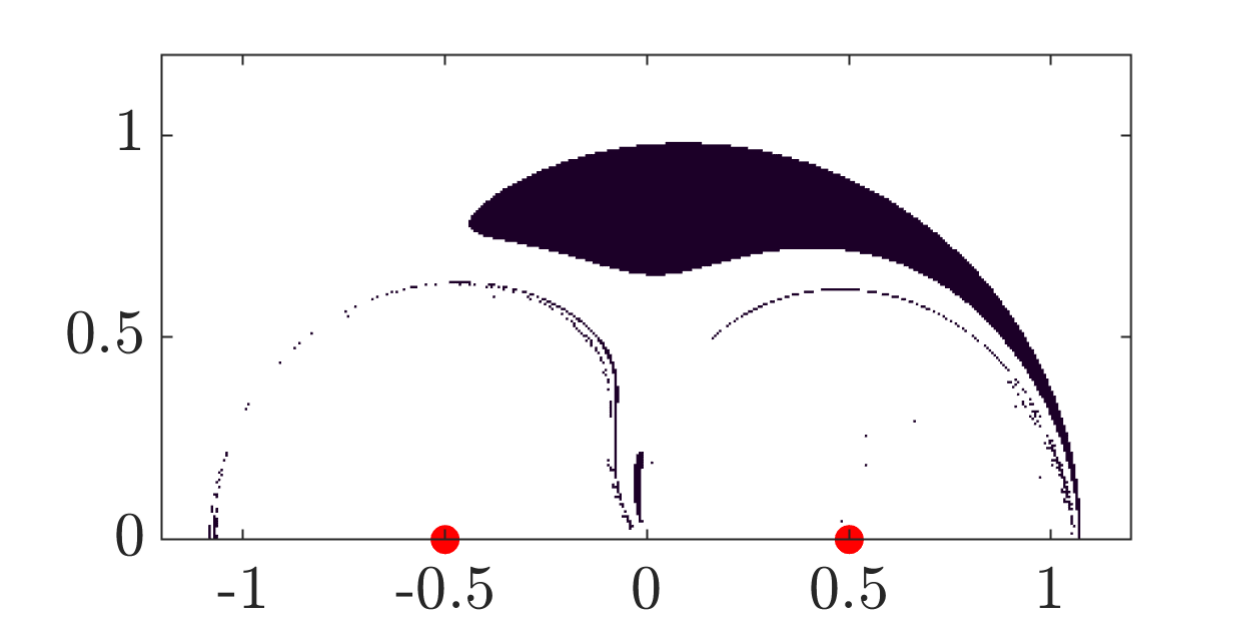}
    \subcaption{$St=0.24$}
    \end{minipage}
    \caption{Typical asymptotic states shown in maroon (top) and their corresponding basins of attraction (bottom) for a finitely-dense inertial particle with density factor $R=0.84 \: \rev{(\rho_p/\rho_f \approx 1.3)}$ and varying Stokes number $St$, without the BBH force. Red dots indicate the vortex centres in the rotating frame. In (a) the particle spirals (shown in blue) into a fixed point attractor, whereas in (b) and (c) the particle is trapped into a limit cycle of period 2 and a strange attractor respectively. The orbits are overlaid on the separatrices of the background flow for clarity of their scale and location. Mirror-symmetric patterns exist in the lower half-plane. }
    \label{fig:trajectories}
\end{figure}
Region II in fig.~\ref{fig:schematic}(b) is of special interest in the context of particle trapping. Attracting orbits of various descriptions are contained within this region, allowing trapping of particles for long times. Moreover, a high level of clustering happens in this region, which is of significance in different contexts. We discuss region I no further, except to mention that inertial particles which begin within them are expected to display the standard centrifuging behaviour to leave the vicinity after a brief transient \citep{croorSingleVortex2015}.

The Stokes number $St$ and the density parameter $R$ are the pertinent non-dimensional numbers in our context. At the initial time, particles of a fixed $St$ and $R$ are placed in a dense uniform grid across a region of interest, and their asymptotic behaviour is categorised. Broadly, higher Stokes number particles quickly exit the region while those at lower Stokes number can either be trapped in the vicinity forever, or spend varying amounts of time in the vicinity before leaking out. 

We begin by examining the dynamics in the absence of the BBH force. Under this approximation, we have a finite-dimensional nonlinear dynamical system, and standard principles for the behaviour of such systems apply. The case where $R=0.84$, i.e., each particle is $1.285$ times denser than the fluid, is discussed first since it displays what we term as canonical behaviour in this context, namely that the attractor undergoes successive period-doubling bifurcations to chaos. Typical attractors for particles of increasing Stokes number are shown in \rev{the top panels of} fig~\ref{fig:trajectories}: a fixed point, a period-2 limit cycle and a chaotic (strange) attractor. We remark that these attractors as shown are in a rotating frame. What appears as a fixed point in fig.~\ref{fig:trajectories}(a) is actually a point which undergoes periodic motion along a circle in the laboratory-fixed frame. Thus, particles which collect here are in continuous motion. Similarly, \rev{what appears as a limit cycle in the rotating frame fills an annular region in the laboratory frame}. 
All particles starting within the corresponding basins of attraction (BoA) shown in \rev{the lower panels of} fig.~\ref{fig:trajectories} asymptotically reach \rev{their respective} attractors and never leave the vicinity. 

\begin{figure}
\centering
\begin{minipage}[b]{.55\textwidth}
   \includegraphics[width=\textwidth]{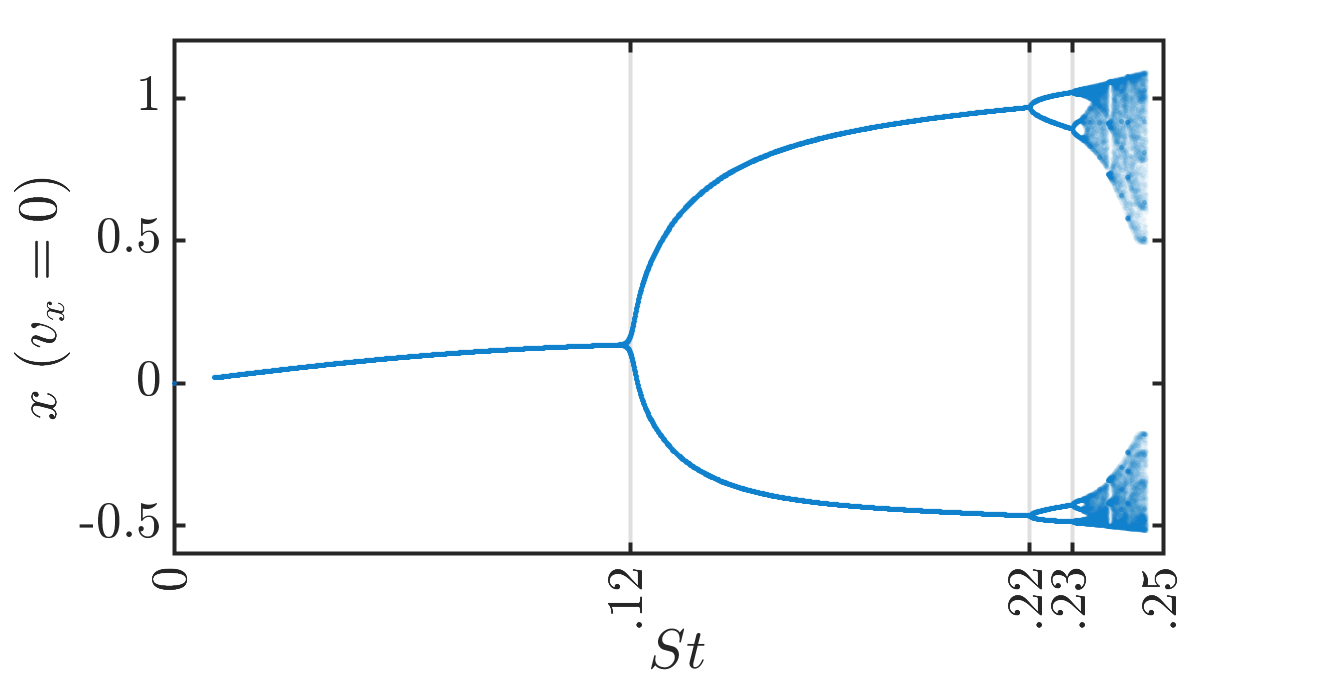}
\end{minipage}    
\caption{Bifurcation diagram for $R=0.84 \:\rev{(\rho_p/\rho_f \approx 1.3)}$ without the BBH force. An attracting fixed point exists below $St=0.12$, while for $0.12 < St < 0.22$ we have a period-1 limit cycle, followed by ever more complex limit cycles as the Stokes number increases. There are no asymptotic attractors beyond $St_{crit}=0.24675$.}
\label{bif0.84}
\end{figure}

\begin{figure}
\centering
\includegraphics[width=0.45\textwidth]{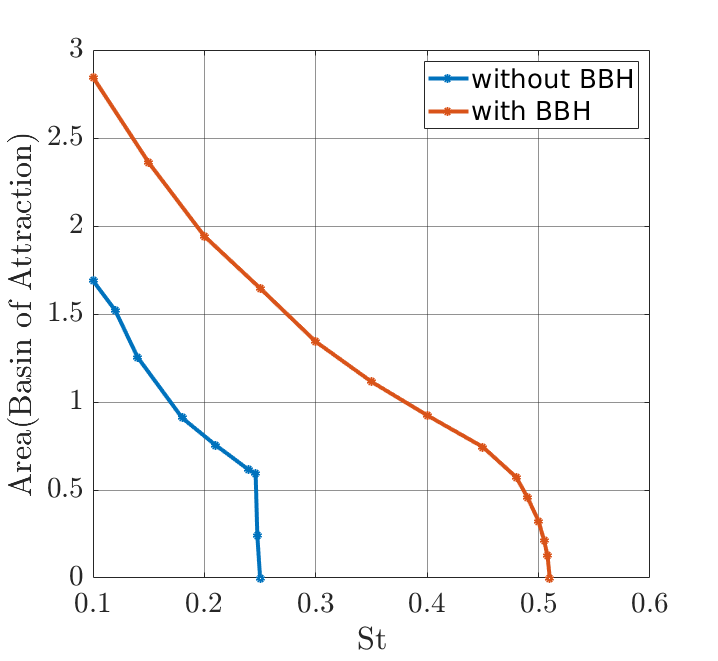}
    \caption{Variation of the area of the basin of attraction with the Stokes number for $R=0.84 \:\rev{(\rho_p/\rho_f \approx 1.3)}$, with and without the BBH force.}
 \label{fig BoA_area_0.84} 
\end{figure}

Next, we construct a bifurcation diagram, shown in fig.~\ref{bif0.84} for $R=0.84 \:\rev{(\rho_p/\rho_f \approx 1.3)}$. Below $St\approx 0.12$, the attractor is a fixed point, and beyond we have limit cycles of increasing complexity. The extrema on the horizontal axis of the limit cycles are plotted on the ordinate of the figure. A textbook period-doubling route to chaos ensues. We have checked that the Stokes number gap between successive bifurcations goes down asymptotically as the Feigenbaum number, with chaos setting in at $St=0.232$. It is seen that the BoA for the higher Stokes number is smaller (see \cref{fig:trajectories,fig BoA_area_0.84}). 

A general observation which is relevant for all the attractors we find is as follows. The attractors are manifolds of dimension lower than two. This means all particles which initially occupy a two-dimensional BoA not only remain in the vicinity indefinitely, but actually converge on to objects of smaller dimension. This focusing of particles is a signature of caustics formation, and is indicative of extreme clustering. The clustering thus achieved can enormously enhance opportunities for collision and coalescence. Whether for carbonaceous material in the ocean participating in carbon fixing, swimmers who benefit from clustering in their quest for reproduction, or dust in protoplanetary disks agglomerating into planetesimals, such attractors are thus of consequence. 

\begin{figure}
\centering
\includegraphics[width=0.6\textwidth]{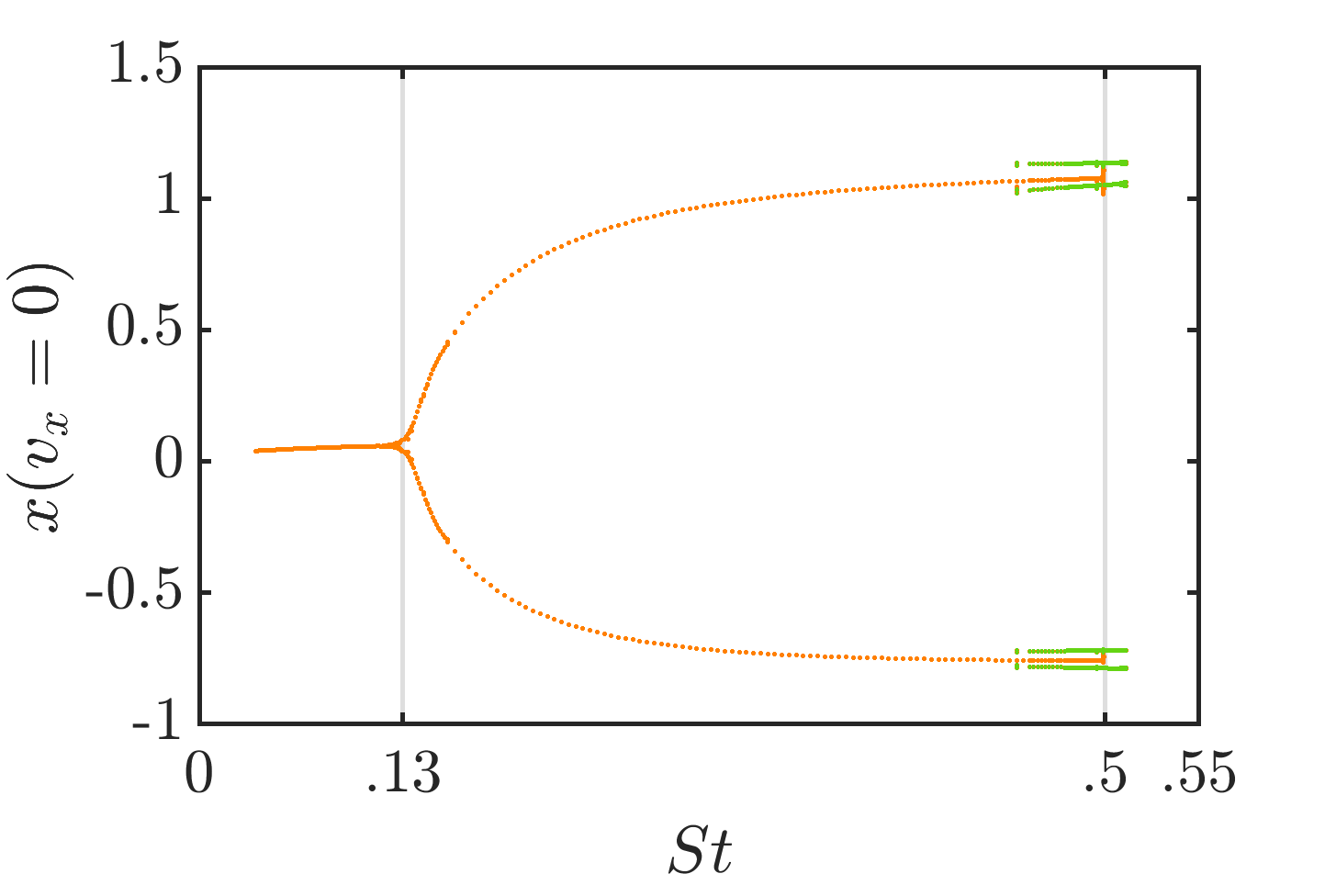}  \caption{\rev{Bifurcation diagram for $R=0.84 \:\rev{(\rho_p/\rho_f \approx 1.3)}$ with the inclusion of the BBH force. Trapping prevails for a wider range of Stokes number then without the BBH force (compare with Fig.~\ref{bif0.84}). A period-2 limit cycle (shown in green), with a very small basin of attraction, co-exists with the period-1 limit cycle (in orange) near $St_{crit}\approx 0.5$.}}  \label{bif_bbh_0.84}
\end{figure}

The stage is now set to discuss the physics we miss when the BBH force is neglected, as well as to bring to light the non-monotonic and counter-intuitive response of the system to both $St$ and $R$. Fig. \ref{bif_bbh_0.84} shows the bifurcation diagram for $R=0.84 \:\rev{(\rho_p/\rho_f \approx 1.3)}$ with the inclusion of the BBH force. Though the dynamics now is not a standard dynamical system in the position-velocity state space, we obtain fixed points and limit cycles. The contrast with fig.~\ref{bif0.84} is self-evident. Interestingly, the bifurcation from a fixed point to a limit cycle occurs at a similar Stokes number with and without the BBH force. However, the period-1 limit cycle persists with the BBH force up to a rather large Stokes number of $St_{crit}\approx 0.5$ whereas without the BBH force, there was no attractor beyond $St \approx 0.25$. The fact that trapping of particles of relatively large inertia takes place in this simple vortical system is remarkable, and underlines the need for including the BBH force in our studies. As the Stokes number approaches the critical value, we find the BoA splitting into two with a very small BoA corresponding to a period-2 limit cycle \rev{(shown in green in fig.\ref{bif_bbh_0.84})}, while the vast majority of particles are attracted to the period-1 cycle. As previously seen in \rev{the lower panels of} \cref{fig:trajectories}, the BoA is a sensitive function of the Stokes number. 

The area of the BoA is a direct measure of the fraction of particles which get trapped in an attractor. The area of such BoA is obtained for a range of Stokes numbers, and shown in fig.~\ref{fig BoA_area_0.84}, with and without the BBH force, for $R=0.84 \:\rev{(\rho_p/\rho_f \approx 1.3)}$. The measurement involves storing the initial locations of all particles which get trapped in the attractor, and calculating the area of the region over which the initial locations are spread.  Whether with or without the BBH force, as the Stokes number becomes higher, i.e., particles become more inertial, their propensity to leave the vicinity monotonically increases. So the BoAs shrink steadily. At this value of $R$, this feature is as would be intuitively expected, but we shall soon see different behaviour \rev{for particles that are denser}. There is a sharp cut-off at a Stokes number, which we refer to as $St_{crit}$, beyond which no particles are trapped. 
The $St_{crit}\approx0.25$ for the case without the BBH force, and is significantly greater at $St_{crit}\approx0.5$ when the BBH force is included. Close to $St_{crit}$, the BoA shows a sensitive dependence on Stokes number, i.e., a rapid shrinking of the area of the BoA to zero. The behaviour past \rev{the critical Stokes number} is ``leaky", \rev{i.e., particles slowly escape from the region of interest.} Notably, besides missing the significant trapping of particles of larger inertia, the fraction of particles trapped is seen to be grossly underestimated at all Stokes numbers by neglecting the BBH force. In the case with the BBH force the period-doubling route is left incomplete. 
\begin{figure}
    \centering
       \begin{minipage}[b]{.47\textwidth}
    \includegraphics[width=\textwidth]{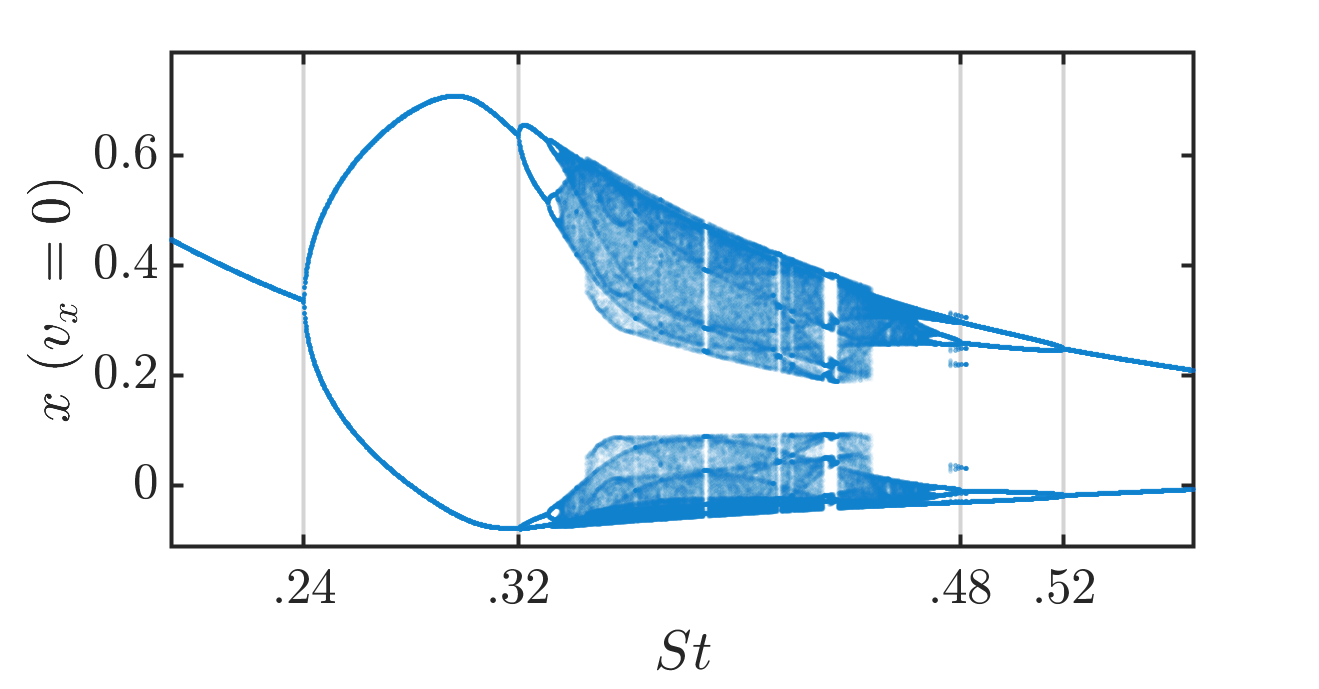}
    \subcaption{$R=0.5 \:\rev{(\rho_p/\rho_f = 2.5)}$ without BBH}
    \end{minipage}
    \begin{minipage}[b]{.47\textwidth}
    \includegraphics[width=\textwidth]{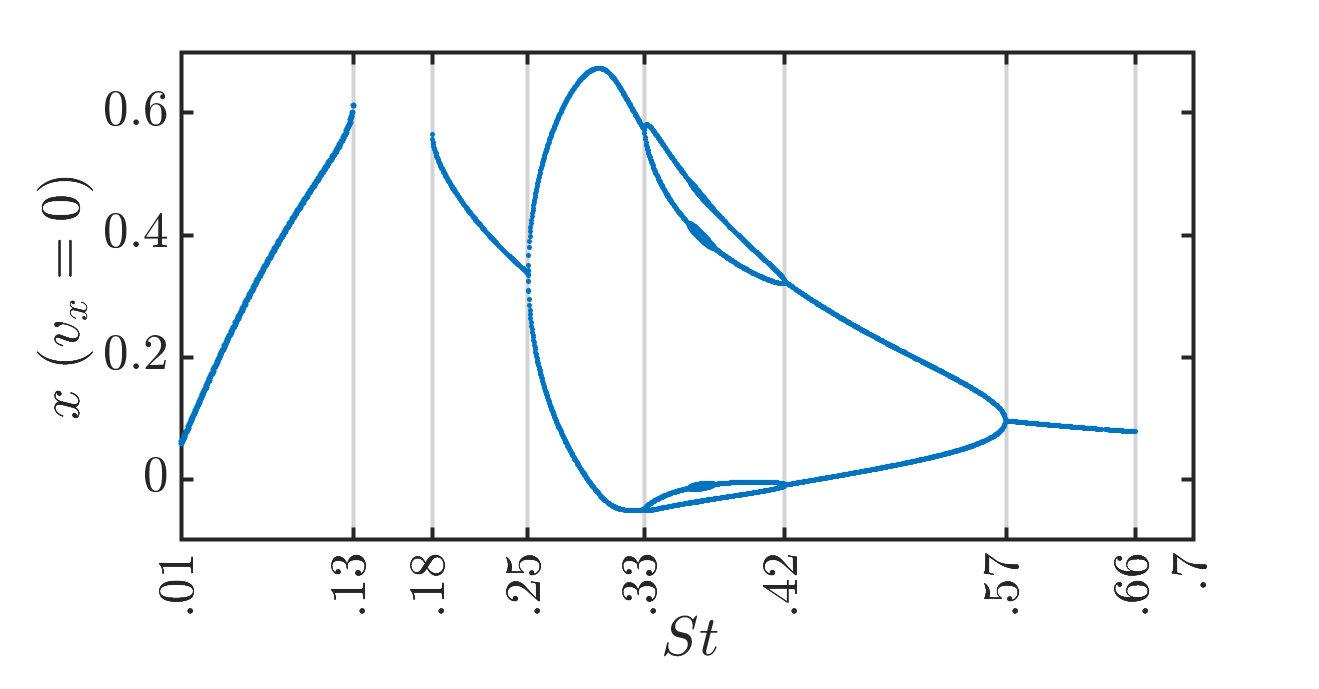}
    \subcaption{$R=0.48 \:\rev{(\rho_p/\rho_f \approx 2.6)}$ without BBH}   
    \end{minipage}
    \begin{minipage}[b]{\textwidth}
    \centering
    \includegraphics[width=.45\textwidth]{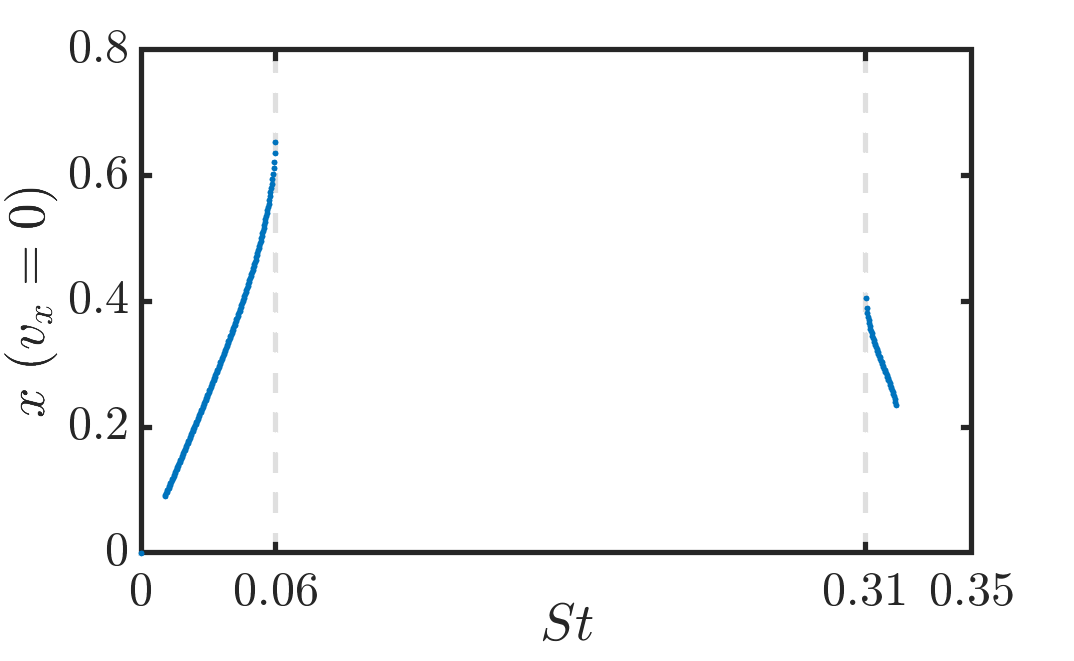} 
    \subcaption{\rev{$R=0.2 \:(\rho_p/\rho_f = 7)$ without BBH}}
    \end{minipage}
    \begin{minipage}[t]{.47\textwidth}
    \includegraphics[width=\textwidth]{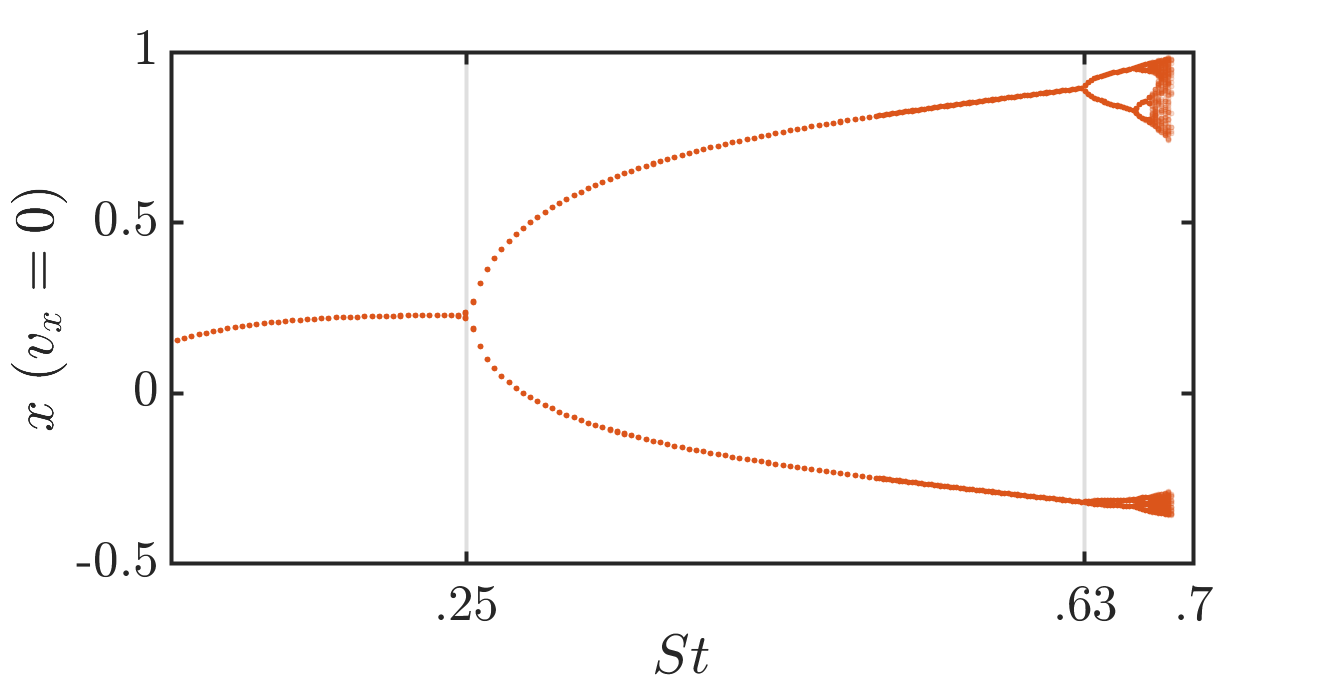}
    \subcaption{$R=0.5 \:\rev{(\rho_p/\rho_f = 2.5)}$ with BBH}
    \end{minipage}
    \begin{minipage}[t]{.47\textwidth}
    \includegraphics[width=\textwidth]{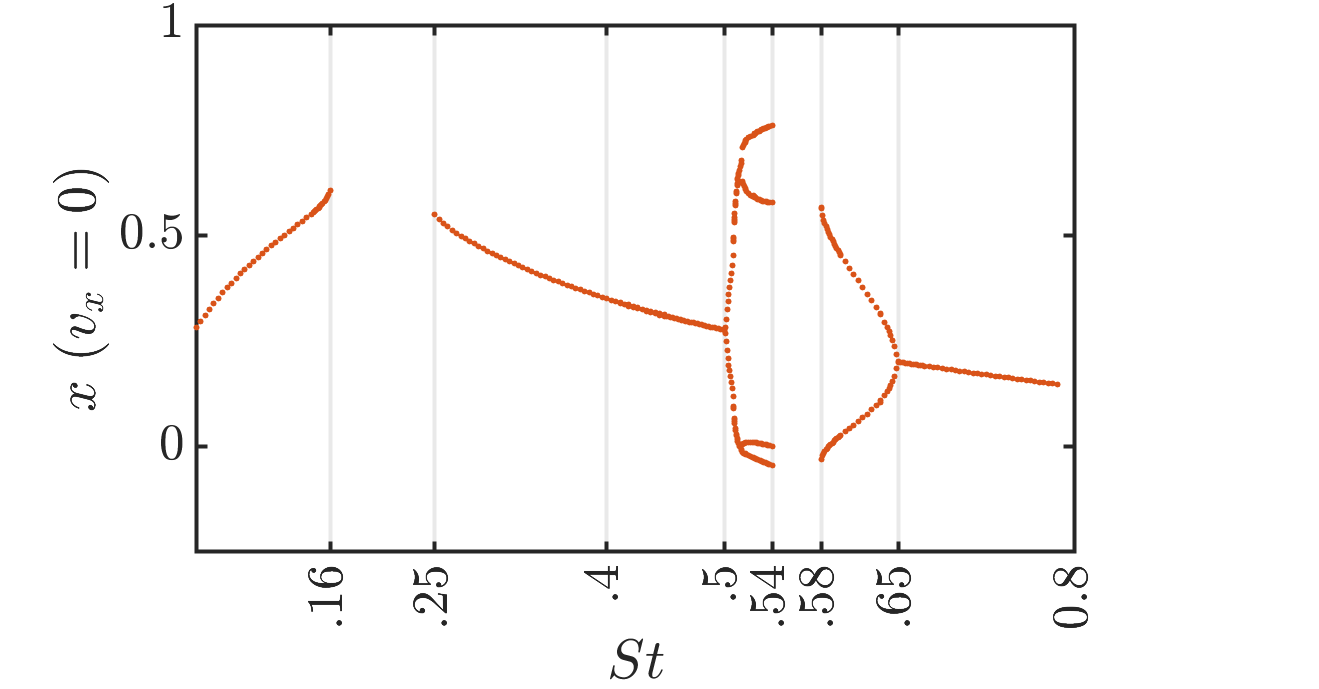}
    \subcaption{$R=0.2 \:\rev{(\rho_p/\rho_f = 7)}$ with BBH}
    \end{minipage} 
    \caption{Bifurcation diagrams for different \rev{representative} density ratios. In (a), (b) and (c) the BBH force is neglected, while in (d) and (e) it is included. On the ordinate are the extrema of the $x$-coordinate of the asymptotic trajectories. }
    \label{fig bifurcation diagrams}
\end{figure}

We move on to higher particle density, i.e., smaller R, with bifurcation plots shown in fig.~\ref{fig bifurcation diagrams}. Changing the density ratio introduces unexpected features in the dynamics. In figs.~\ref{fig bifurcation diagrams}(a) and (b) we see \rev{period-doubling} bifurcations followed by unusual period-halving bifurcations back to a fixed point at higher Stokes number. We also see that a small difference in density ratio changes the behaviour from chaotic to periodic. For the same density ratio as in fig.~\ref{fig bifurcation diagrams}(a), the inclusion of the BBH force converts the dynamics to that on a canonical period-doubling bifurcation route to chaos, as seen in fig.~\ref{fig bifurcation diagrams}(d). Interestingly, at this density ratio too, the BBH force does not significantly change the Stokes number at the first bifurcation occurs: going from fixed point to limit cycle. Through most of the range of $St$, a fixed point or a limit cycle persists, followed by a rapid breakdown into chaos within a short range of Stokes number. \rev{At the larger density ratio of $\sim 7$ (\cref{fig bifurcation diagrams}(c)), only two small regimes of particle trapping are seen. In contrast, with the inclusion of the BBH force, \rev{\cref{fig bifurcation diagrams}(d) and (e), trapping is more widespread across $St$.} A period-halving bifurcation occurs here too (fig.~\ref{fig bifurcation diagrams}(e)), but at a higher density ratio than without the BBH force.} Here too, at higher Stokes, we regain a fixed point as the sole attractor. Another non-standard feature seen in \rev{several of} these bifurcation diagrams is the existence of gaps. Within these windows, no particles are trapped in the vicinity of the vortices. In fact, in fig.~\ref{fig bifurcation diagrams}(e), two windows are visible, so the actual Stokes number of the period-halving bifurcation, from a $2$-cycle to a simple limit cycle, is not obtainable, though a period-$1$ limit cycle followed by a fixed point are evident at higher Stokes numbers. In particulate flows, we have come to expect a monotonic trend in \rev{complexity} as the Stokes number increases, so the transition, as we move up in $St$, from \rev{an attractor}, to no particles being trapped, and back to particle-trapping in \rev{an attractor}, is worthy of remark. \rev{Moreover during period-halving, an increase in $St$ simplifies the attractor}, and we hope that these findings alone will be intriguing enough to the reader to be motivated to explore particulate flows in this context. 

With these examples, we demonstrate the general trend in our system: the \rev{long-time} behaviour of inertial particles with the BBH force at a given density parameter $R$ is in broad qualitative agreement with the behaviour without BBH at a higher $R$. In other words, a denser particle, with the inclusion of the BBH force, behaves qualitatively like a lighter particle without the BBH force, \rev{asymptotically in time}.

\begin{figure}
    \centering
    \begin{minipage}[b]{.45\textwidth}
    \includegraphics[width=\textwidth]{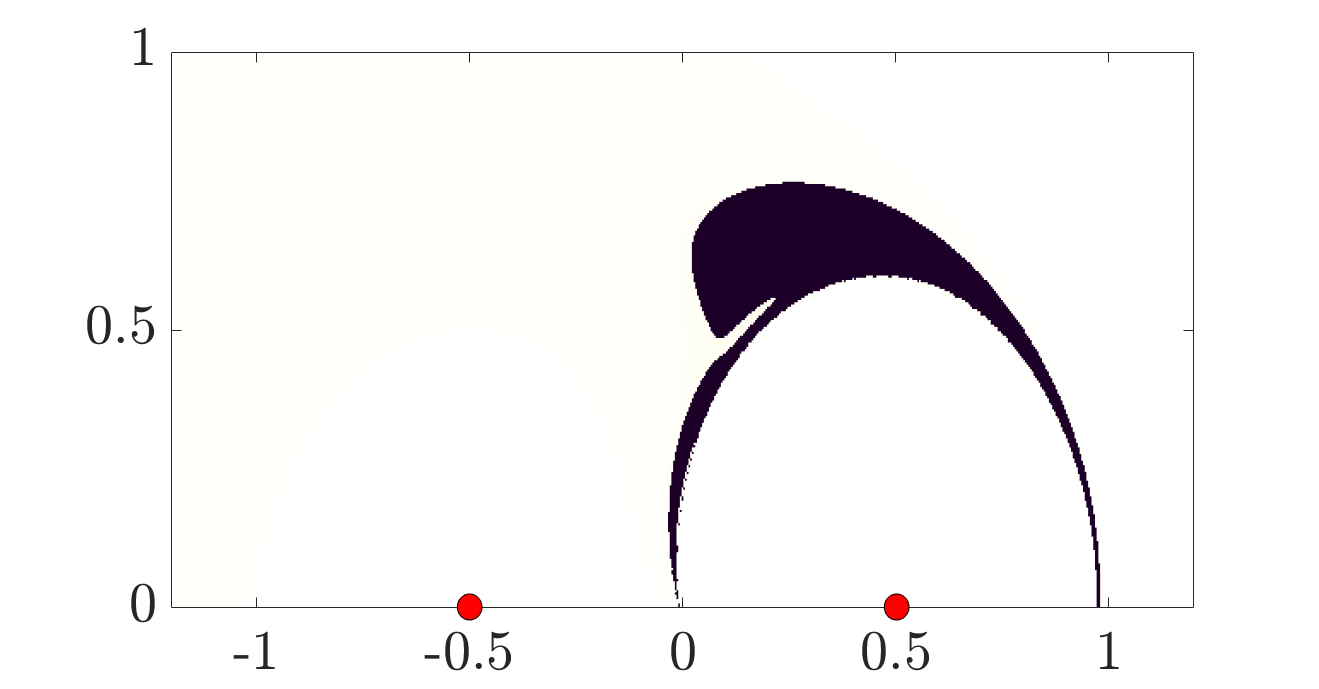}
    \subcaption{\rev{At period-doubling,} $St=0.320$}
    \end{minipage}
     \begin{minipage}[b]{.45\textwidth}
    \includegraphics[width=\textwidth]{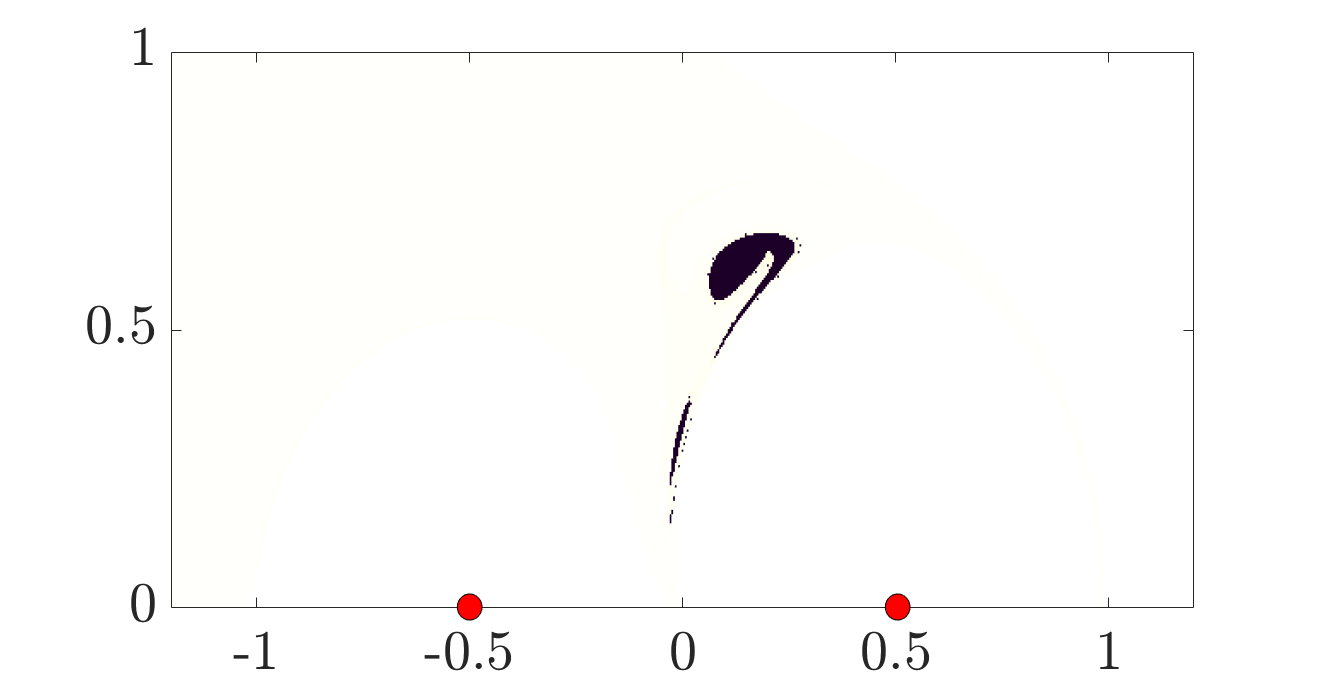}
    \subcaption{\rev{At period-halving,} $St=0.483$}
    \end{minipage}
     \begin{minipage}[b]{.45\textwidth}
\includegraphics[width=\textwidth]{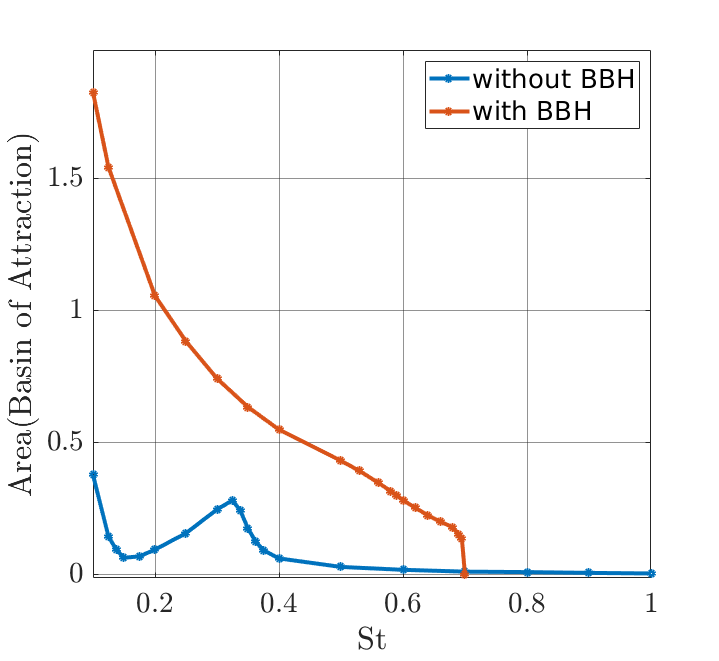}
\subcaption{}
    \end{minipage}
    \caption{(a-b) Basins of attraction for $R=0.5 \; \rev{(\rho_p/\rho_f = 2.5)}$ without the BBH force during period-doubling and period-halving respectively (see fig.~\ref{fig bifurcation diagrams}(a)). Notice the difference in sizes. (c) Variation of the size of the basin of attraction with Stokes number at $R=0.5$, with and without the BBH force.  }
    \label{fig boa size}
\end{figure}

\begin{figure}
    \centering
    \begin{minipage}[b]{.47\textwidth}   
    \includegraphics[width=\textwidth]{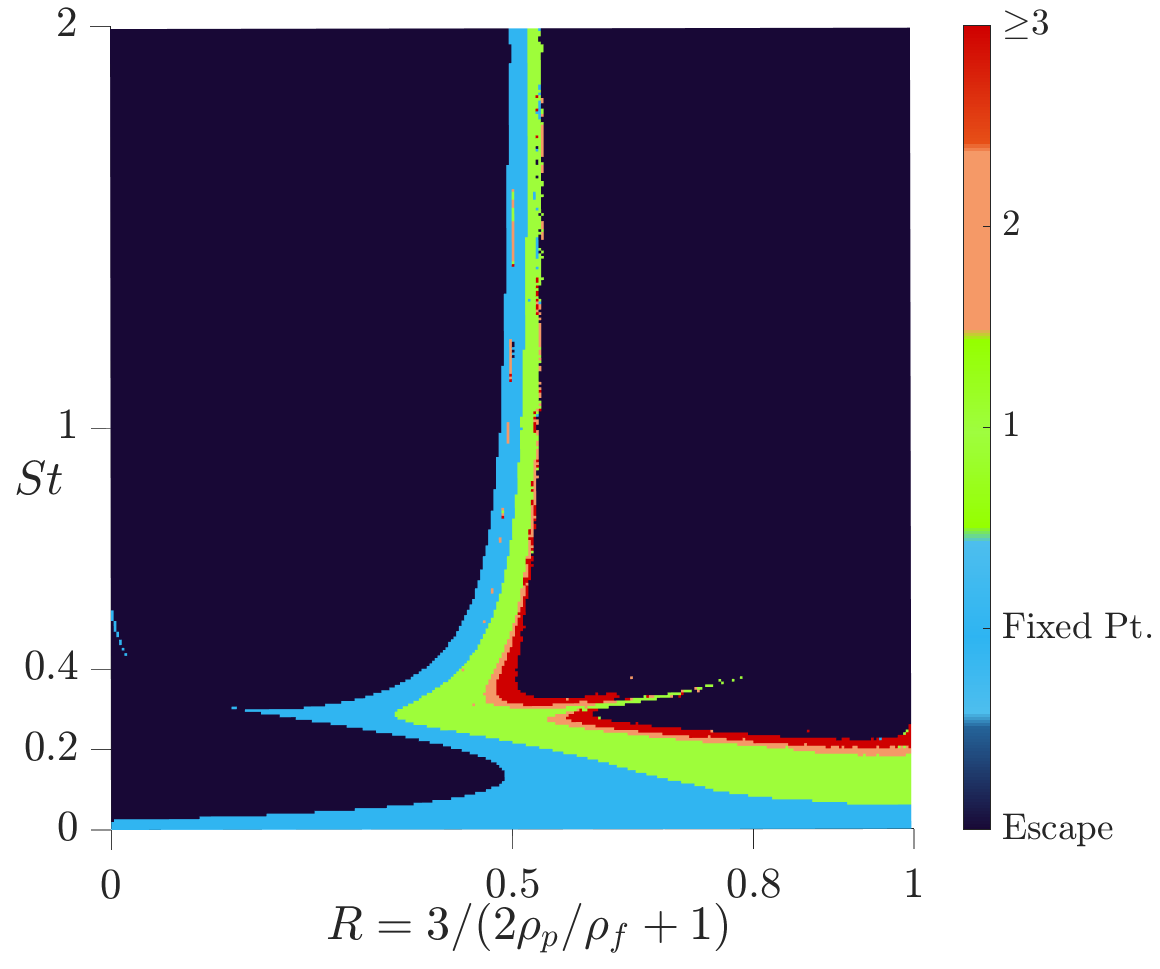}
    \subcaption{}
    \end{minipage}
     \begin{minipage}[b]{.48\textwidth}
    \includegraphics[width=\textwidth]{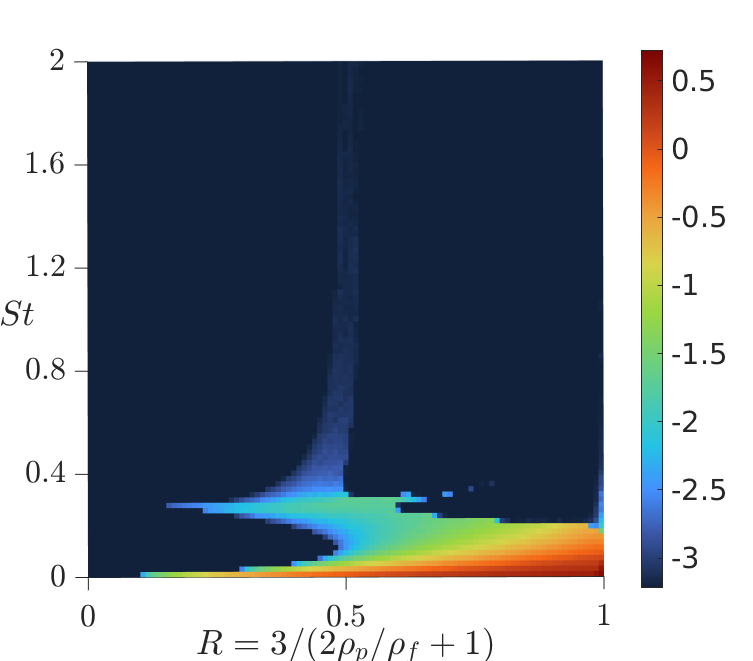}
    \subcaption{}
    \end{minipage}
    \caption{Phase plots for inertial particles, without the BBH force,  based on (a) the period of the attracting orbit. The regime occupied by limit cycles of period 2 and above is very narrow. (b) The logarithm of the size of the basin of attraction. Particles of near-neutral densities tend to stay longer. In both plots, there is anomalous behaviour at $R \sim 0.5$.}
    \label{fig_phase_plots}
\end{figure}
Two sample BoAs are shown without the BBH force in figs.~\ref{fig boa size}(a) and (b), to give a visual idea of how the BoA shrinks as we approach $St_{crit}$. The Stokes numbers chosen correspond to period-doubling and period-halving bifurcations, respectively. Over a range of $St$, the areas of the BoA are shown in fig.~\ref{fig boa size}(c),  with and without the BBH force, for a higher density ratio than in fig.~\ref{fig BoA_area_0.84}.  Again, with the BBH force, we see the trapping of particles of significantly higher $St$ than the dynamics we obtain by neglecting would suggest. Moreover, the BoA is significantly larger with the BBH force than without for the entire range of Stokes number. We may conclude that the neglect of the BBH force will seriously underestimate the number of particles trapped in the vicinity of vortices. Interestingly at this higher particle density, in the absence of the BBH force, the size of the BoA varies non-monotonically with Stokes number, and a small non-zero fraction of particles remains trapped even at high Stokes number. We do not have an explanation for this anomalous behaviour. 
The anomalous behaviour vanishes in this case upon the inclusion of the BBH force, showing a monotonic decrease \rev{of BoA size} with Stokes number and a rapid decrease to zero just before $St_{crit}$. We hasten to note that the dynamics including the BBH force too showed such anomalous behaviour \rev{elsewhere, as evidenced by the gap in fig.~\ref{fig bifurcation diagrams}(e) in the region $0.16<St<0.25$ where size of BoA drops to zero.} So this particular feature is not merely a consequence of neglecting the BBH force. \rev{Since particles of higher Stokes number get centrifuged out of the vicinity of vortices faster, we would have expected a shrinking BoA with increasing particle inertia. This canonical expectation} is belied over some ranges of density ratio.

It is relevant to mention that these broad findings on the effect of the BBH force are in contrast with those for the flow past a solid cylinder \citep{daitcheTel11,daitche2014}, where the inclusion of the BBH force reduces caustics as well as destroys attractors. Such reduction in clustering was also seen by \cite{guseva2013} in convective cell flow. Similarly, \citet{chong2013} studied finitely-dense inertial particles in a viscous streaming flow created by an oscillating cylinder, wherein they concluded that the BBH force resists particle trapping. Evidently, the physics of the BBH force cannot be oversimplified thus.

The case of the infinitely dense particle was studied by \rev{\cite{angilella2010physica}} upon neglecting the BBH force, where it was shown \rev{analytically} that there is a fixed point up to \rev{$St=(2-\sqrt{3})/2\pi$} and no attractor beyond. We repeated the calculations with the BBH force included for $\rho_p\gg\rho_f$, and found the critical Stokes number unchanged. \rev{Further, our computations for the location of the fixed point for all Stokes numbers below this are in excellent agreement with the analytical results of \cite{angilella2010physica}. We note the qualitative difference between infinitely dense particles and our largest density ratio of $R=0.2\ (\rho_p/\rho_f=7)$ (fig.~\ref{fig bifurcation diagrams}(e))}. We thus confirm that the BBH force has a noticeable effect on finitely-dense particles, especially when particle densities are of the same order of magnitude as that of the surrounding fluid. This indicates that the BBH force should be included as a significant force when studying solid-liquid systems such as microplastics in the ocean.  

To give an idea of the complexity in the solutions, we provide a phase plot in fig.~\ref{fig_phase_plots}, where the behaviour across density ratios and Stokes number without the BBH force is summarized. Fig.~\ref{fig_phase_plots}(a) shows the \rev{different} kinds of attracting orbits that \rev{one} obtains. At a given density ratio, as we move up in Stokes number, in some part of the regime, we go from attracting orbits to no attracting orbits, whereas in other portions we can go back to attracting fixed points or limit cycles over a range of Stokes numbers. \rev{We may identify the following three regimes}: for $1 > R \gtrsim 0.5$ we have a period-doubling route to chaos, for $0.5 > R > 0.35$ a period-doubling route, which may go all the way to chaos or may be limited to a few bifurcations, is followed by period halving, leading to a single fixed point, and for $0.35>R$ we have only attracting fixed point in the regime where we have trapped particles. With the BBH force, we have the three regimes, but the transitions all happen at lower values of $R$. \rev{In fig.\ref{fig_phase_plots}(a)} the density ratio of $R \sim 0.5$ is most interesting, where the existence of attracting orbits at large Stokes number is possible, and there is sensitive dependence on the density ratio. Chaotic attractors only exist at $R\gtrsim 0.5$, i.e., when the particle and fluid densities are comparable. And here too, the range of Stokes numbers at a given $R$ over which chaotic attractors are seen is very narrow. The corresponding areas of the BoA are shown as a phase plot in fig.~\ref{fig_phase_plots}(b). Broadly, at low Stokes numbers, as the particles become denser, the BoA shrinks. However, at intermediate Stokes number and density ratios, we see \rev{non-monotonic} behaviour. As $\;R \to 1 \rev{ (\text{i.e., } \rho_p \sim \rho_f)}$, the particles are near neutrally buoyant, and over a range of Stokes numbers,  the entire region II corresponds closely to the BoA. Invariably, in this limit, the attractor is a fixed point.  

\section{Particle Leakage}\label{sec:leak}
\begin{figure}
    \centering
    \includegraphics[width=0.4\textwidth]{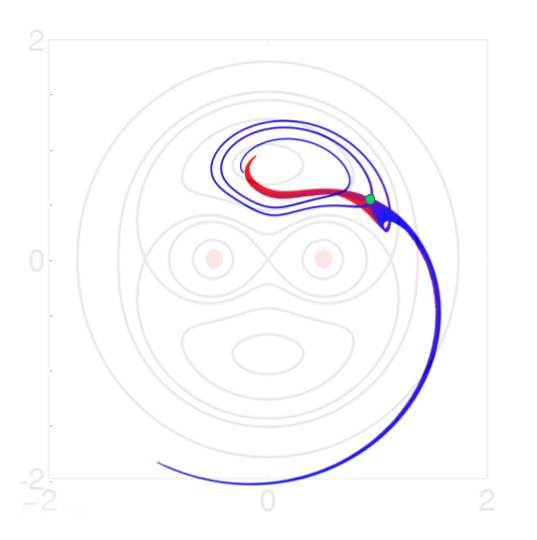}
    \caption{Leakage of particles past the trapping criteria without the BBH force, for the representative case of $R=0.84 \:\rev{(\rho_p/\rho_f \approx 1.3)}$ with $St_{crit}=0.24675$. The trajectories of a set of particles in the narrow range $St\in(0.2465,0.2470)$ about $St_{crit}$, with identical initial conditions, are shown. Trajectories for $St < St_{crit}$ are coloured red, where particles  are seen to remain trapped, while those for $St > St_{crit}$ are coloured blue, where particles escape. The green dot is the saddle fixed point at $St_{crit}$, whose location does not vary visibly for the narrow range of $St$ shown.}
    \label{intro_leak}
\end{figure}

We have seen that for every density ratio $R$, there is a critical Stokes number, $St_{crit}$, above which no particle remains indefinitely in the vicinity of the system. We now ask what happens beyond $St_{crit}$. Fig. \ref{intro_leak} shows two sets of particle trajectories, with the same initial conditions, but one with $St$ slightly less than $St_{crit}$ and the other with $St$ slightly greater than $St_{crit}$. The first set is trapped forever, whereas the second set escapes. \rev{At $St=St_{crit}$ one point  on the chaotic attractor in phase space coincides with the saddle point. This is termed a crisis}, \citep{grebogi1983}, where the chaotic attractor disappears, making way for a chaotic saddle. \rev{The dynamics near the chaotic saddle is `leaky', i.e., all particles near the chaotic saddle will leave the vicinity in finite time.} 

\begin{figure}
    \centering    
    \begin{minipage}[b]{.37\textwidth}
    \includegraphics[width=\textwidth]{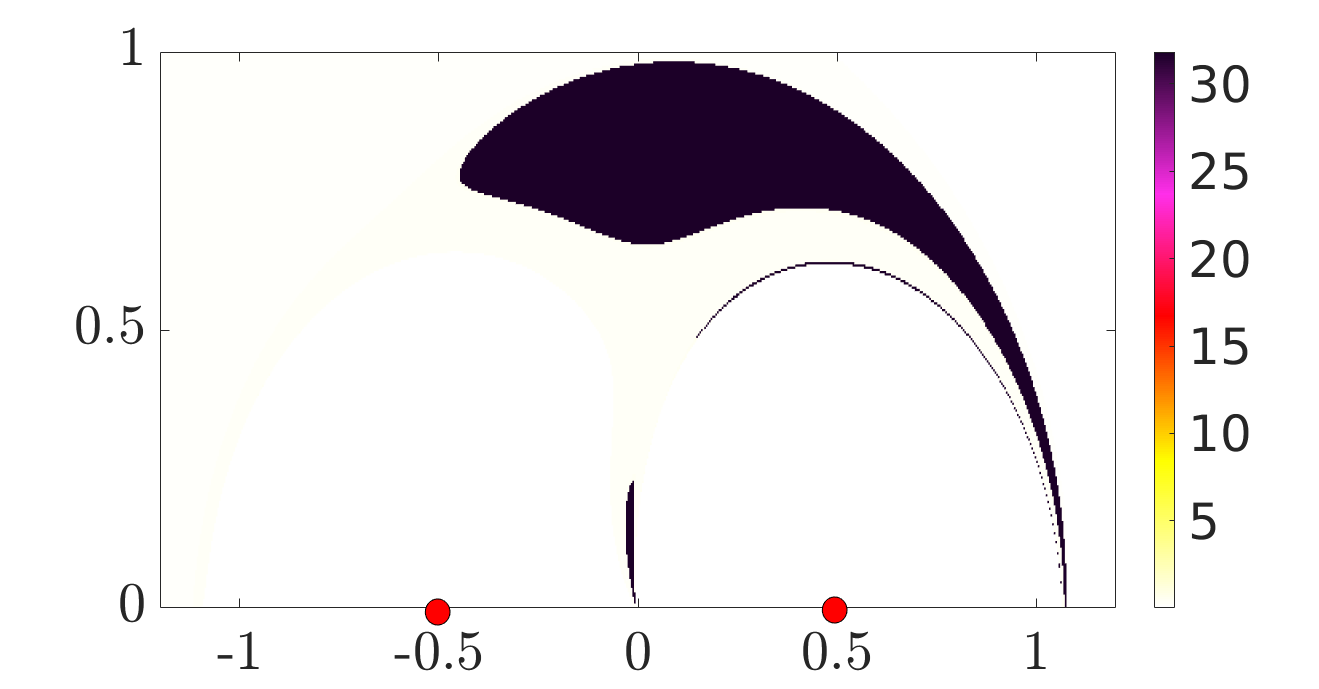}
    \subcaption{$St=0.24 \rev{< St_{crit}}$}
    \end{minipage}\\
    \begin{minipage}[b]{.45\textwidth}
    \includegraphics[width=\textwidth]{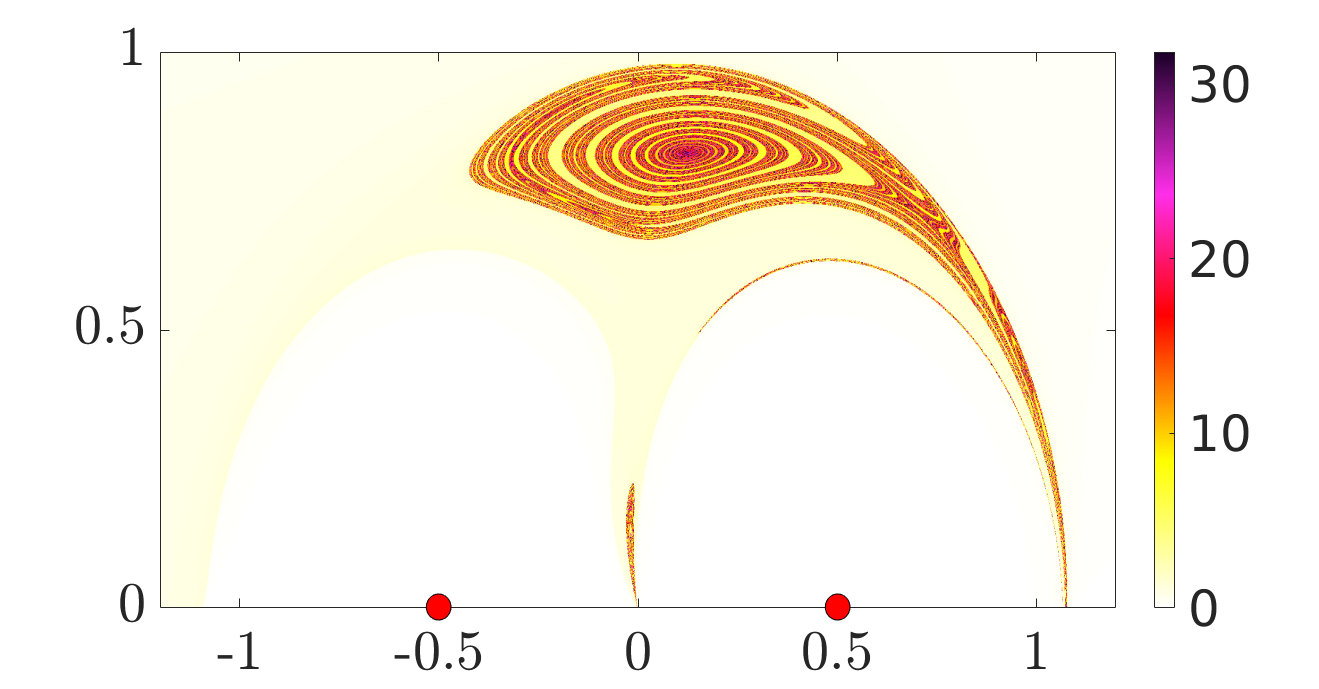}
    \subcaption{$St=0.25 \rev{> St_{crit}}$}
    \end{minipage}
    \begin{minipage}[b]{.45\textwidth}
    \includegraphics[width=\textwidth]{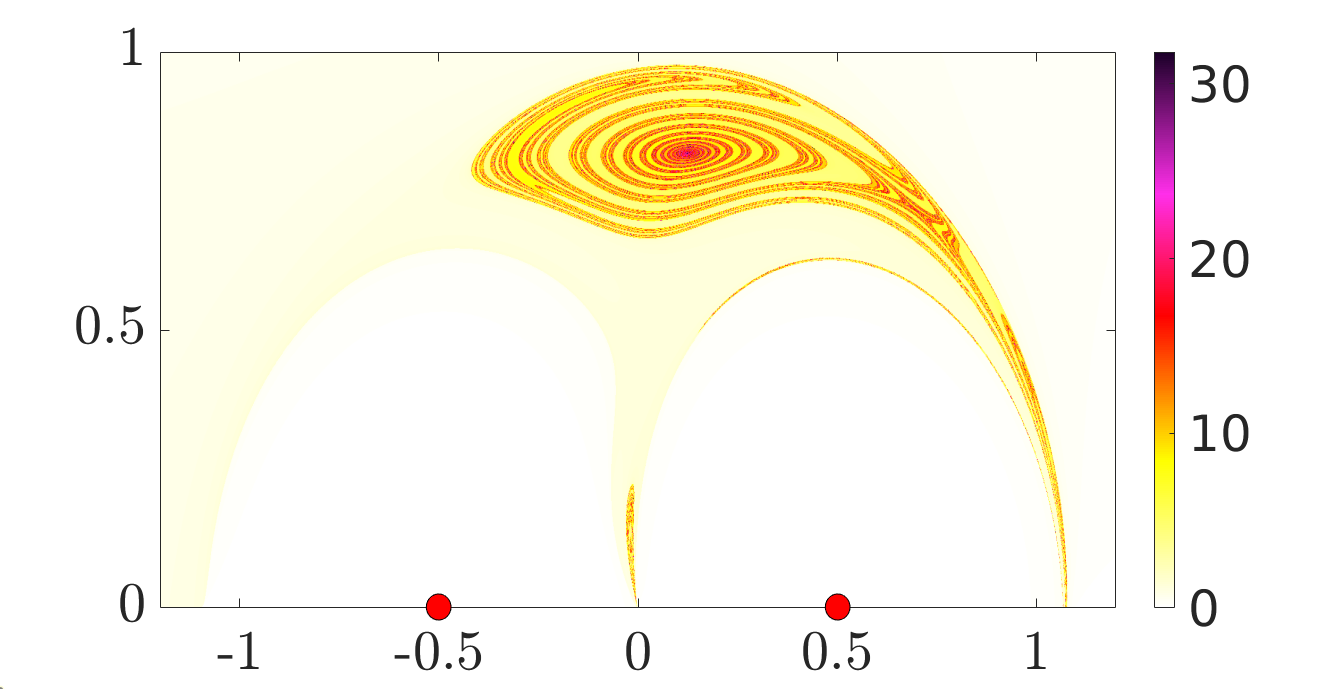}
    \subcaption{$St=0.255 \rev{> St_{crit}}$}
    \end{minipage}
    \caption{Residence times of inertial particles for the density ratio $R=0.84 \:\rev{(\rho_p/\rho_f \approx 1.3)}$, without the BBH force. The colour at a given position indicates the residence time of a particle starting at that location.}
    \label{fig residence time}
\end{figure}

A particle starting at a given location is traced until it leaves the system, and the time at which it leaves the system is noted down as its residence time within the region of interest. We define the `system' by a circle of radius $2$, centred at the origin. While the numbers for residence time depend weakly on this choice, a change in the definition will not change our conclusions. This residence time is plotted for all initial locations in fig.~\ref{fig residence time}, for three Stokes numbers. At $St=0.24$, which is less than $St_{crit}$ \rev{(fig.~\ref{fig residence time}(a))}, we have a patch of particles whose residence time is nominally equal to the simulation time $T_{max}$, and we have confirmed that this patch corresponds to the BoA whose particles are permanent residents. However, \rev{past the critical Stokes number,} at $St=0.25$, all particles escape at finite times \rev{(fig.~\ref{fig residence time}(b))}. There are sharp ridges in the figure, and
particles originating on these have a large residence time. These ridges are separated by valleys of very low particle residence times. Although the area of the BoA is zero, the fact that particles can remain in the vicinity and close to each other for tens of rotation time-scales signifies the enhanced opportunity for collisions even beyond $St_{crit}$. At \rev{the even higher} $St$ of $0.255$, in fig.~\ref{fig residence time}(c), the residence times have already dropped significantly.

\begin{figure}
    \centering
    \begin{minipage}[b]{.8\textwidth}
    \includegraphics[width=\textwidth]{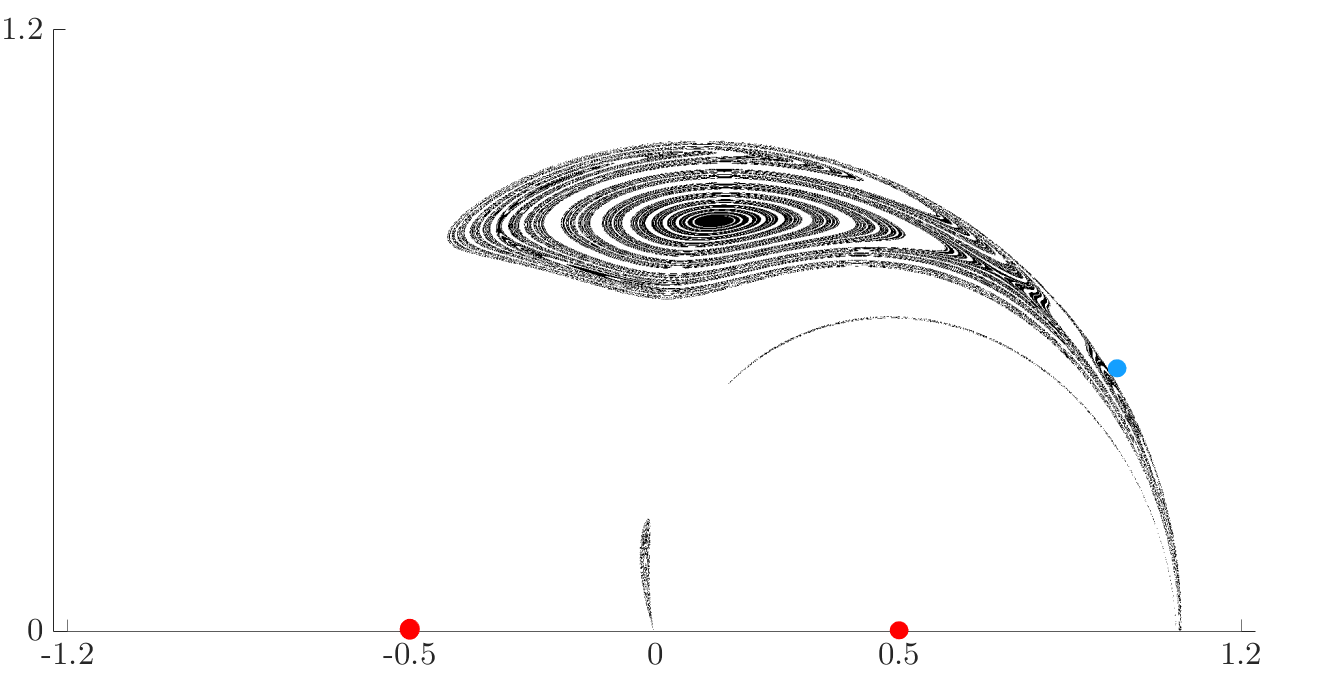}
    \subcaption{}
    \end{minipage}
    \begin{minipage}[b]{.45\textwidth}
    \includegraphics[width=\textwidth]{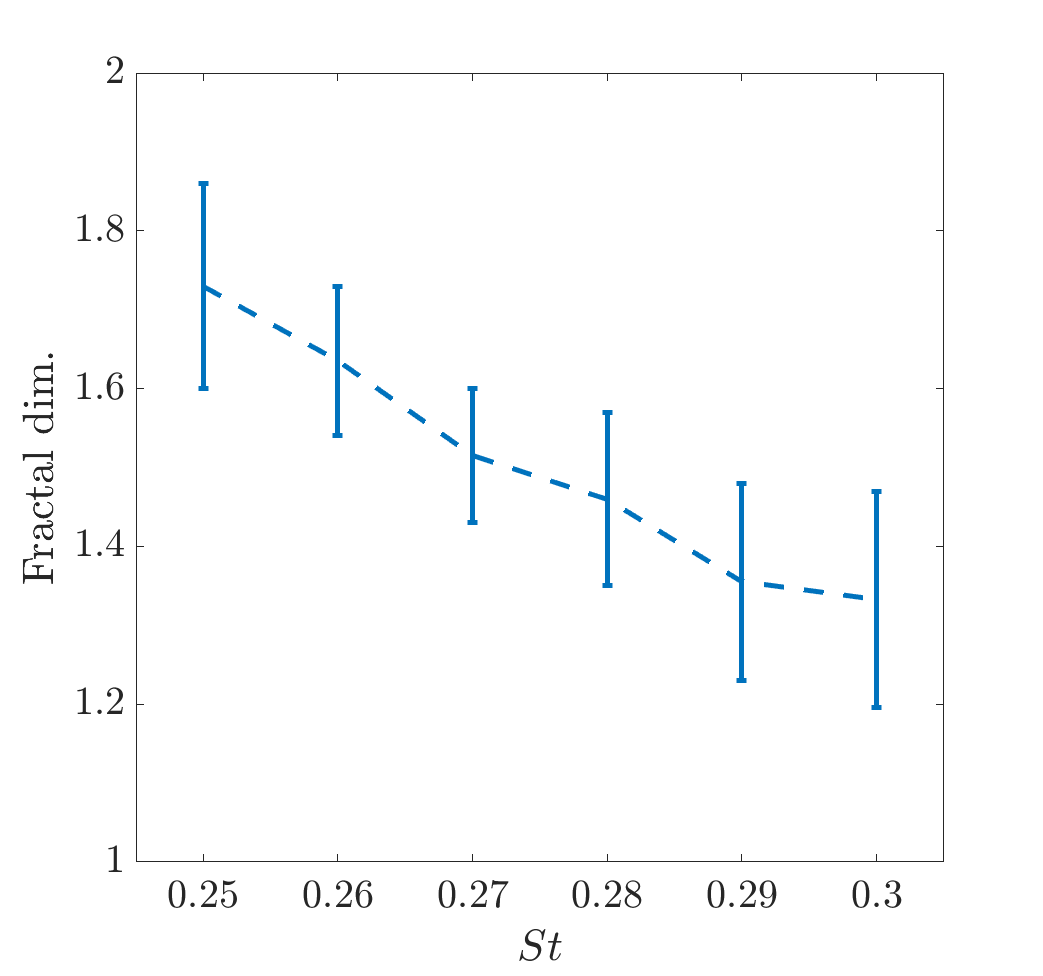}
    \subcaption{}
    \end{minipage}
    \caption{(a) Stable manifold of the saddle point shown by the blue dot (slightly shifted for clarity), for $R=0.84 \:(\rho_p/\rho_f \approx 1.3)$ and $St=0.25$, without the BBH force. The manifold shown is a fractal object with \rev{an approximate fractal dimension of $1.7$}. The vortex centres are indicated by red dots. \rev{(b) Fractal dimension, calculated using the box-counting method, for $St>St_{crit}$ for $R=0.84\ (\rho_p/\rho_f\approx 1.3)$. Error bars indicate the variation in the calculated dimension for different box sizes. The dotted line indicates the trend and is not an exact quantitative measure.} }
    \label{fig stable manifold}
\end{figure}

The stable manifold of the saddle point corresponding to \rev{the case in} fig.~\ref{fig residence time}(b), \rev{at $St=0.25$, is shown in \cref{fig stable manifold}(a) for $v_0=0$ particles}. It \rev{theoretically represents} the original locations \rev{(in phase space)} of particles which flow \rev{into} the saddle, approaching it as $t\to \infty$. \rev{It occupies zero area and has a fractal dimension of approximately $1.7$. A particle starting exactly on the stable manifold would have infinite residence time. Particles starting very close to the stable manifold will have have large but finite residence times. Therefore,} we see a close correspondence between the ridges of high residence times in fig.~\ref{fig residence time}(b) and the stable manifold \rev{in fig.~\ref{fig stable manifold}(a)}. The fractal nature of the stable manifold and the fractal distribution of ridges and valleys in the residence time plot are signatures of transient chaos \citep{tel2015}. \rev{In fig.~\ref{fig stable manifold}(b), we show that the fractal dimension of the stable manifold falls sharply as $St$ increases. A stable manifold of fractal dimension $1$ indicates no transient chaos, and the reduction in fractal dimension with increasing Stokes indicates a corresponding reduction in transient chaos.}

\begin{figure}
    \centering
    \begin{minipage}{0.45\textwidth}
    \includegraphics[width=\textwidth]{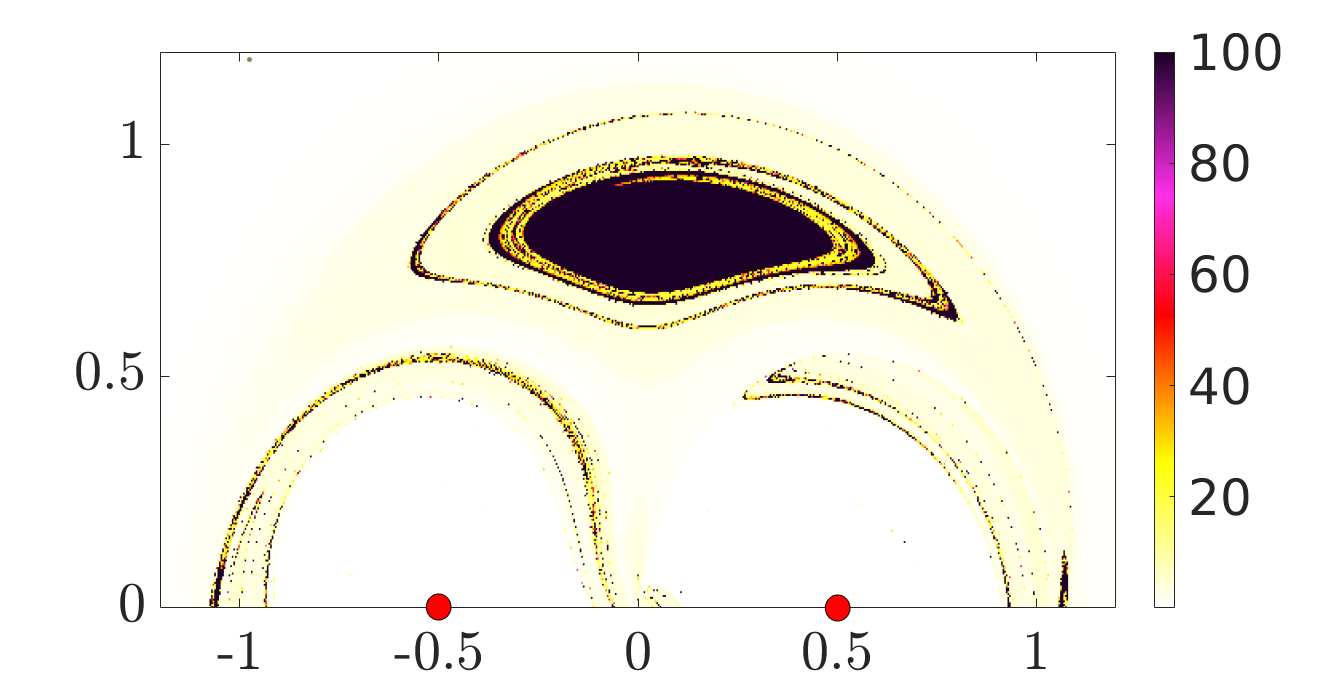}
    \subcaption{$R=0.84, ~St=0.495<St_{crit}$}
    \end{minipage}
    \begin{minipage}{0.45\textwidth}
    \includegraphics[width=\textwidth]{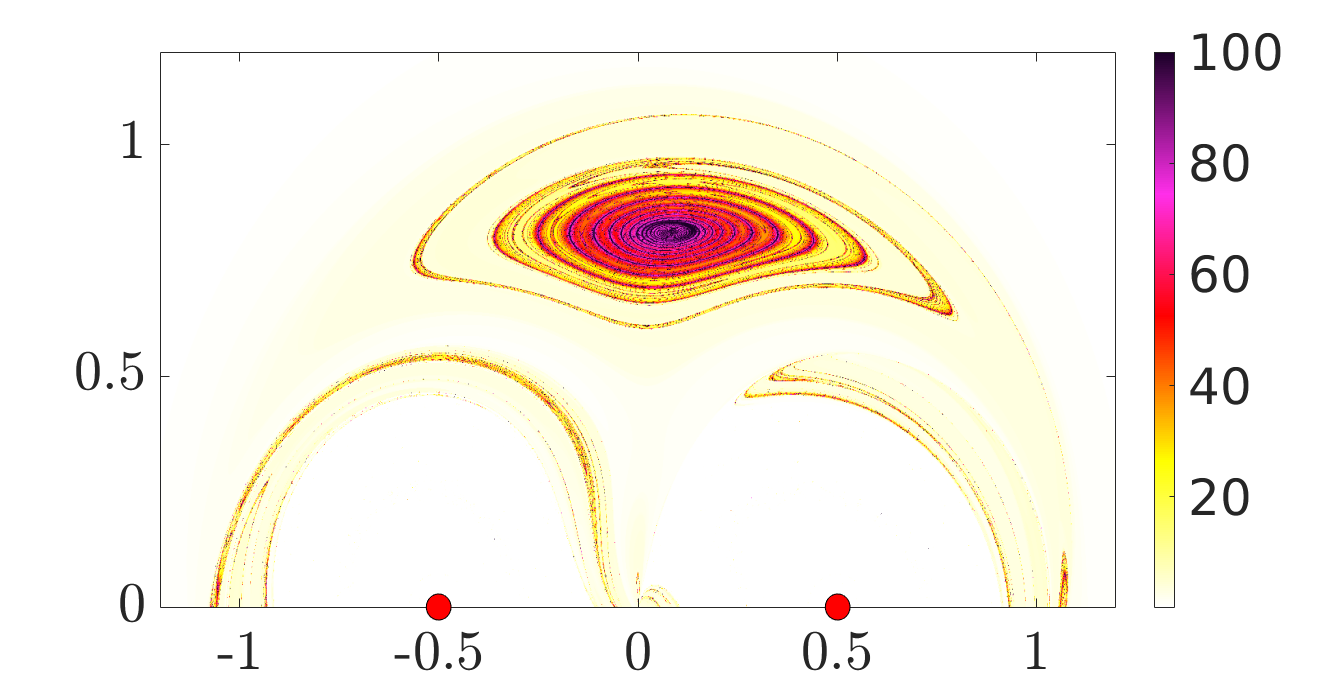}
    \subcaption{$R=0.84, ~St=0.508>St_{crit}$}
    \end{minipage}
    \begin{minipage}{0.45\textwidth}
    \includegraphics[width=\textwidth]{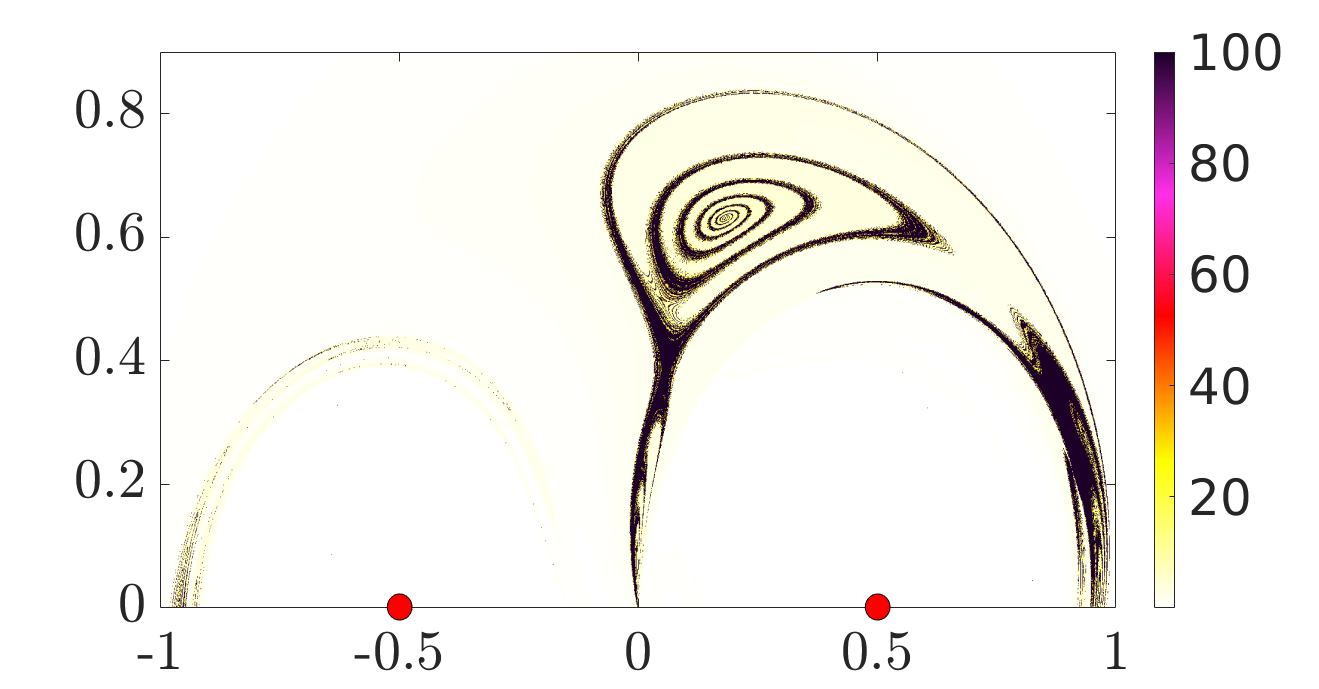}
    \subcaption{$R=0.50,~ St=0.685<St_{crit}$}
    \end{minipage}
    \begin{minipage}{0.45\textwidth}
    \includegraphics[width=\textwidth]{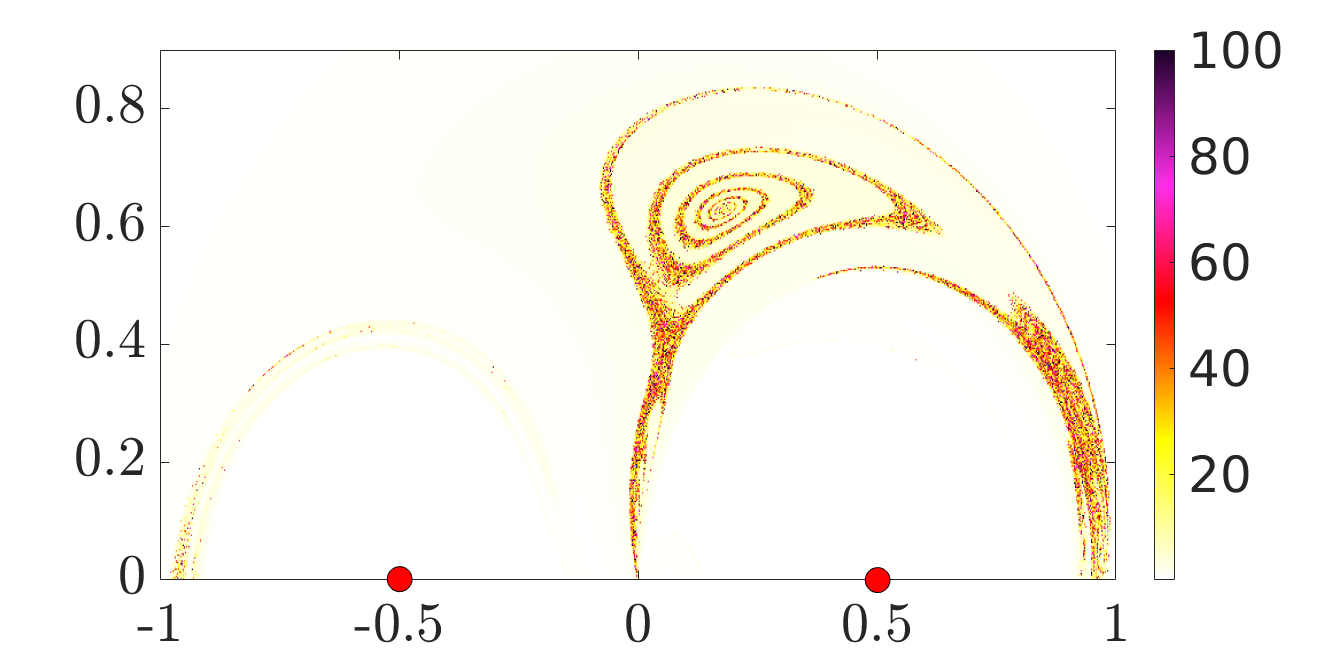}
    \subcaption{$R=0.50,~St=0.705>St_{crit}$}
    \end{minipage}
    \caption{Residence time plots with the inclusion of the BBH force for two density ratios around their corresponding critical Stokes numbers: $R=0.84\:\rev{(\rho_p/\rho_f \approx 1.3)}$ with $St_{crit}\approx0.5$ in the top panel (a-b) and  $R=0.50 \:\rev{(\rho_p/\rho_f = 2.5)}$ with $St_{crit}\approx 0.69$ in the bottom panel (c-d). For each $R$, a plot each is provided for a $St$ slightly below and slightly above the respective $St_{crit}$.}
    \label{fig bbh residence time}
\end{figure}

In figs.~\ref{fig bbh residence time}(a) and (b), we plot the residence times including the BBH force at $R=0.84 \:(\rho_p/\rho_f \approx 1.3)$, just below and above the $St_{crit}\rev{\approx 0.5}$. As before, when $St<St_{crit}$, the residence time for particles in the BoA, denoted by the dark patch \rev{in fig.~\ref{fig bbh residence time}(a)}, is infinite. \rev{We recall from fig.~\ref{bif_bbh_0.84} that the corresponding attractors are the co-existing simple period-1 limit cycle and a period-2 limit cycle, with the latter associated to a very small BoA.} In the case without the BBH force, we had seen that the residence times outside the BoA were very small, but here, we find long residence times in the valleys as well. These valleys spiral inward as we increase $St$  where the BoA shrinks rapidly, (as was seen in fig.~\ref{fig BoA_area_0.84}). Just beyond $St_{crit}$ the residence time plot is given by fig.~\ref{fig bbh residence time}(b). Again we find closely spaced ridges and valleys in the residence times. In contrast to the case without the BBH force, where we found a chaotic saddle with long-lasting transients close by, here we do not find any evidence of a chaotic saddle. Instead, a significant number of particles spend a long time near a structure which resembles period-1 orbit before eventually escaping. Although with the BBH force, we no longer have a standard dynamical system \rev{in the position-velocity state space}, this structure resembles a periodic-orbit saddle \rev{seen in standard dynamical systems}. 

In figs.~\ref{fig bbh residence time}(c) and (d), the residence time is provided for $R=0.5$, just below and just above $St_{crit}$ respectively. The BBH force is included. It will be recalled from Fig.~\ref{fig bifurcation diagrams}(d) that there is a chaotic attractor before the sudden  disappearance of the BoA. This is reminiscent of $R=0.84$ without the BBH force. The \rev{irregular} spatial distribution of residence times in fig.~\ref{fig bbh residence time}(d) is also similar to the plots in fig.~\ref{fig residence time}(b,c). All these are signatures of transient chaos.

For both $R=0.5$ and $R=0.84$, at $St$ just below $St_{crit}$ the basin boundary itself has an irregular fractal-like structure, which was absent without BBH. We note that fractal basin boundaries arise in the context of inertial particles in flow, such as, for heavy ($R=0$) inertial particles due to the interplay between transient chaos and fixed point attractors \citep{angilella2014}, and for finitely-dense inertial particles due to the addition of the BBH force \citep{guseva2013}.

\begin{figure}
    \centering
    \begin{minipage}[b]{.47\textwidth}
    \includegraphics[width=\textwidth]{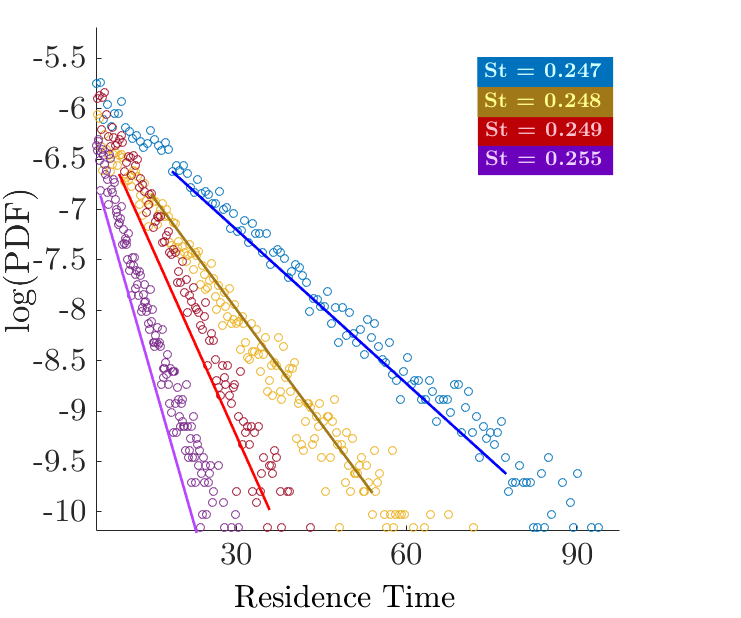}
    \end{minipage}
    \begin{minipage}[b]{.47\textwidth}
    \includegraphics[width=\textwidth]{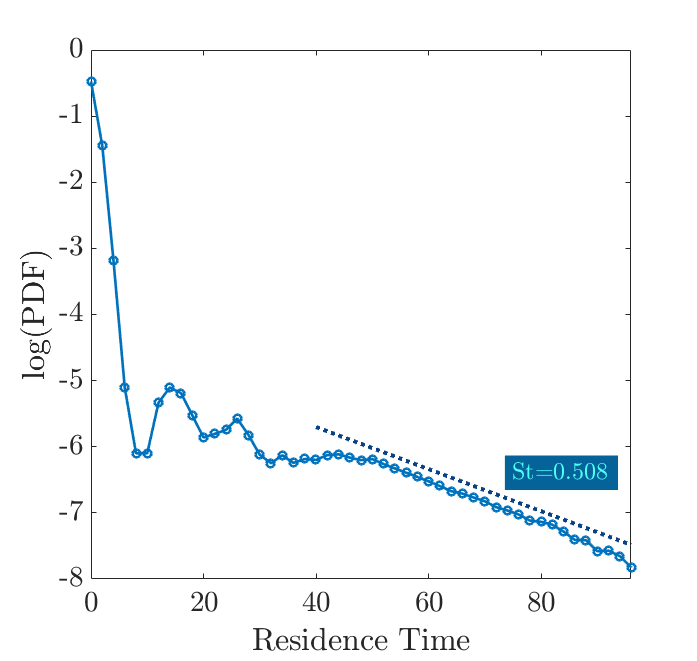}
    \end{minipage}
    \caption{Distributions of residence times of inertial particles of $R=0.84$, at  Stokes numbers just beyond the critical, as indicated in the legends. (a) In the absence of the BBH force, we see an exponential distribution of residence times.  The slope increases rapidly as $St$ moves further away from $St_{crit}(=0.24675)$. (b) In the presence of the BBH force. Particles which last a long time in the vicinity display an exponential distribution in residence times, whereas the probability of staying is non-monotonic in the residence time at smaller residence times. Here $St_{crit}\rev{\approx 0.5}$. } 
    \label{fig misc equal}
\end{figure}
\begin{figure}
    \centering
    \begin{minipage}[b]{0.47\textwidth}
    \includegraphics[trim={8cm 0 6cm 0}, width=\textwidth]{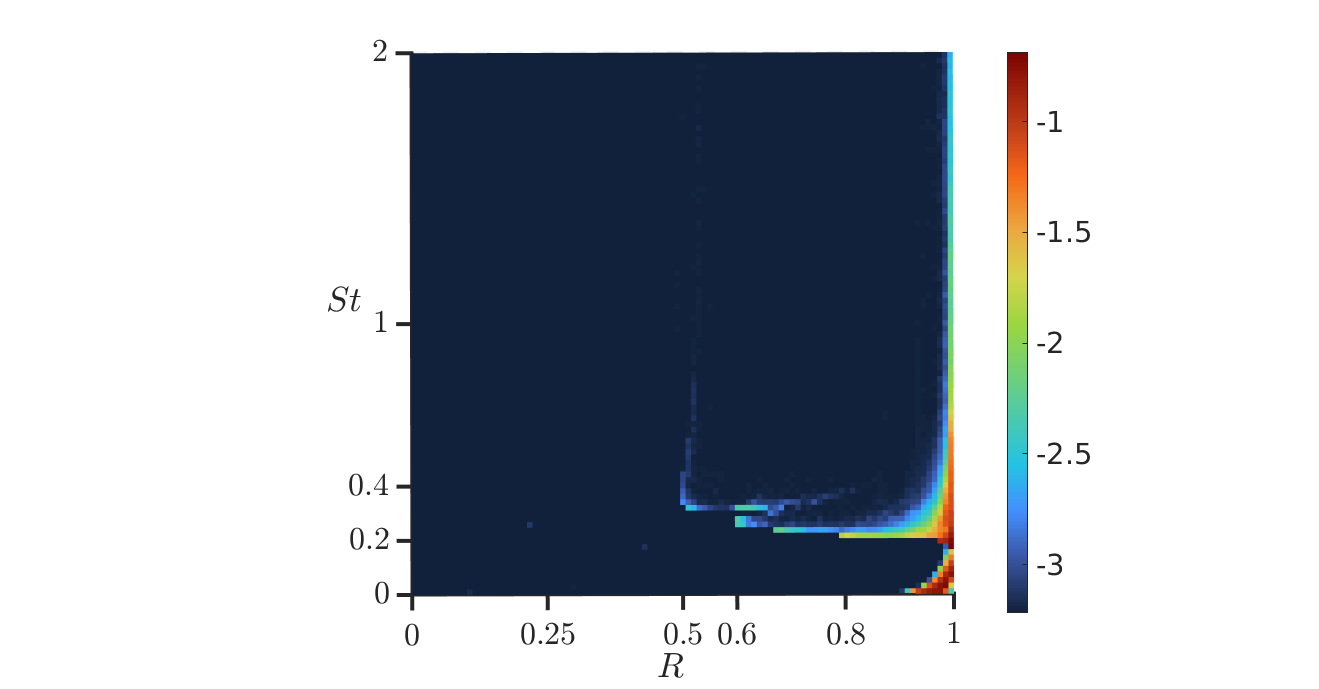}
    \end{minipage}
    \caption{Logarithm of the size of basins occupied by particles with long but finite residence time, i.e., $T_{esc}>10T$. Notice that as $R \to 1$, \rev{i.e., the particle becomes more neutrally buoyant,} a significant fraction of particles can have long residence times for a significant range of $St$.} 
    \label{fig:transients}
\end{figure}
We see that residence times can vary greatly, even for particles starting out at neighbouring locations. It is thus worthwhile to calculate the probability densities of residence times. Fig.~\ref{fig misc equal}(a) shows the distribution of  residence times, without the BBH force, for all particles initialized in the region of interest for four Stokes numbers above the critical. The distributions are all exponential, which is a signature of transient chaos, i.e., the existence of a chaotic saddle \citep{tel2015}. Note that even beyond $St_{crit}$ the residence times are still $O(10)$ time periods, which can be significant. However, note that residence time does decrease very rapidly as we move away from $St_{crit}$. Fig.~\ref{fig misc equal}(b) is a corresponding residence time distribution plot for the same density ratio, but with BBH, for particles whose Stokes number is just above the critical. While the long residence time particles display a short exponential-like feature, over moderate residence times, we see that the probability of a given residence time varies non-monotonically with the residence time. This is another feature we obtained only with the inclusion of the BBH force, which means further study is needed to understand the nature of the transients beyond $St_{crit}$ with the BBH force.

The regions in the $R-St$ parameter space where a significant number of particles have a large but finite residence time are shown in fig.~\ref{fig:transients}, without the BBH force. We note the non-monotonicity in Stokes number here too. Close to neutral density, long-lasting transient are seen to be extensive, and can be an important contributor to particle collisions. \rev{Away from neutral density ($R=1$), most of the particles which eventually escape do so within a few time periods ($O(T)$), so the phenomenon of trapping and clustering becomes evident early on. However, to visual accuracy, the convergence time of trapped particles to their respective attractors can be $O(10 T)$, or even higher with the BBH force.} 

\section{Summary and discussion}\label{sec:summary}

We examined the dynamics of inertial particles that are denser than the fluid in the simplest rotating and vortical system: that of two identical vortices \rev{in periodic circular motion}. We found significant levels of particle trapping in the vicinity, even for particles of significant Stokes number. The inclusion of the BBH force results in higher levels of trapping for a given Stokes number, i.e., much larger BoA. It also results in a wider range of Stokes number over which trapping occurs. These are in contrast to earlier findings on the effects of BBH in other flows. The trapped particles are attracted to fixed points or limit cycles of varying complexity, all the way to chaotic attractors. So, there is extreme clustering into lower-dimensional (zero area) manifolds, of particles initially in a finite BoA. Particles in a rotating system thus do not follow the expectation that they will constantly centrifuge out of vortical regions, and collect in regions of high strain. In fact, the attractors do not correspond to the highest strain regions. 

The trapped particles undergo a rich variety of dynamics in the $R-St$ parameter space. The BoA is in some part non-monotonic function of the Stokes number, and in fact we can have a range of Stokes numbers devoid of particle-trapping. A period-doubling route to chaos is observed in some parameter ranges, whereas an unusual period-halving back to a fixed point is seen in other parts of the parameter space. This regime is physically interesting because it belies the expectation gained so far that particles of higher inertia have a propensity to attain more complicated limiting trajectories. A given behaviour observed at a given particle density without the BBH force appears at a much higher particle density (lower $R$) with the inclusion of the BBH force. Moreover, a given bifurcation typically happens at a higher $St$ with the BBH force than without.

Close to a critical Stokes number $St_{crit}$, which depends on $R$, there is a sudden drop in the BoA. Beyond this, no particles are trapped forever, since a crisis occurs and the chaotic attractor becomes a chaotic saddle. Beyond this, the system displays alternating ridges of high but finite residence time, and valleys of low residence time. The range of Stokes number over which long-lasting transients are seen gets larger at particle densities close to neutral buoyancy. One remarkable qualitative difference with the addition of the BBH force is that the period doubling route to chaos can remain incomplete until the rapid disappearance of the attractor. Correspondingly, the transient dynamics beyond $St_{crit}$ resemble the dynamics near a non-chaotic saddle. Further, the distribution of residence times is exponential without the BBH force and also with the BBH force for higher values, but is non-monotonic with the BBH force just beyond $St_{crit}$. Just before the crisis, the basin boundaries with the BBH force appear to have a fractal nature, unlike without the BBH force. Thus, the case with the BBH force merits further inquiry in this context.

We now discuss the limitations of our model. We recall that the MRE is a model equation derived in the \rev{small particle Reynolds number limit,} $Re_p \rightarrow 0$. In practical scenarios where $Re_p$ is only finitely small, the effects due to flow inertia will become significant in long but finite time. Therefore, the validity of the MRE in various regards including the form of the history force \rev{and quasi-steady drag force} are questionable after long simulation-time. \rev{However, our results are valid and insightful in the cases where significant clustering is observed within short-times, subject to the condition $Re_p<1$. }

\rev{The MRE is widely used at Stokes numbers of $O(1)$ but the need to satisfy this requirement simultaneously with $Re_p \ll 1$, as well as keeping the particle size small, i.e., $a/d \ll 1$ imposes additional restrictions. By expressing $Re_p$ in terms of our control parameters $R$ and $St$, we obtain the scaling relation $Re_p (a/d) \sim R St |v_{rel}|$. When $St \sim O(1)$ we usually have $|v_{rel}| \sim O(1)$. In principle the requirements can be satisfied if $R \to 0$, but in an experiment, even in the extreme limit of solid particles in air, we typically have $R \sim 10^{-3}$ at the lowest. Demanding arbitrary smallness of the particle size $(a/d)$ is therefore penalized by large $Re_p$. For finite density ratios, as in our case, satisfying the requirements is even harder. }
From our computations of slip velocity, we find that the slip velocities with the BBH force are usually significantly lower than without it. At the highest Stokes numbers in our study, $Re_p \sim O(10)$ for $a/d \sim O(10^{-2})$ for some part of the dynamics. However, close to the fixed points, and often near the limit cycles, slip velocities are low and so is $Re_p$. The small $Re_p$ assumption is more reasonable for very dense and near-neutrally buoyant particles. Due to this, and the findings in \cite{maxeyWang1996} that show applicability of the MRE for $Re_p \lesssim 17$, we expect that our findings have qualitative significance. In fact, the MRE (with the BBH force) has been previously used to study chaotic dynamics for $Re_p\sim O(1)$ \citep{daitche2014,daitche2015} in a different flow. We note that estimates of $Re_p$ are rarely discussed in the literature, and need more attention.

\rev{Our model also assumes one-way coupling between the particles and the underlying fluid and therefore our results hold in the dilute limit of particle concentration. If the concentration of particles becomes high due to clustering itself, the feedback of particles on the fluid is no longer negligible. Feedback from particles can affect vortex merger \citep{Shuai_2024}, in which case a separate study on inertial particles in the ensuing flow dynamics under two-way coupling is required. We have ignored gravity from our model to isolate and study the effects of hydrodynamic forces on inertial particle dynamics. While co-rotating vortices do occur commonly in turbulence, the merger process will include effects from the rest of the flow, as well as viscous effects depending on the local flow Reynolds number. The relevance of our results for true turbulence will therefore be affected by these other factors.}

\rev{In spite of its limitations, our toy model study brings to attention an important aspect of extreme clustering in rotating flows. In fact if the rotation rate is externally applied rather than driven by a flow, we get an extra handle which can potentially mitigate the limitations.} The attracton of particles towards a manifold of dimension smaller than the spatial dimension, like a pair of fixed points or limit cycles, will result in extreme clustering. This can be a major contributor to particle collision and agglomeration, and in the case of tiny living organisms, to enabling communication and sexual reproduction. The relevance of these findings to particle aggregation in protoplanetary disks, in rotating systems in the ocean, and also in turbulent-flow neighbourhoods dominated by a few strong vortices are worth investigation. Future studies on simple model flows, consisting of characteristic combinations of vortices, with and without rotation of the entire system, in two and three dimensions, would be illuminating, and provide a platform to understand some features of particulate turbulent flows. We therefore hope that our work will motivate experiments and further theory on this important question. 

\section*{Acknowledgements}
The computations reported in the paper were partially performed on the Mario computing cluster at ICTS-TIFR. Research at ICTS-TIFR is supported by the Department of Atomic Energy, Government of India, under Project Identification No. RTI4001. We thank the anonymous referees for valuable suggestions which improved this paper. 

\section*{Declaration of interests}
The authors report no conflict of interest.

\appendix{}
\section{\rev{Dynamics in the vicinity of vortices of unequal strength}}\label{app:A}
\begin{figure}
    \centering
    \begin{minipage}[b]{0.32\textwidth}
    \includegraphics[width=\textwidth]{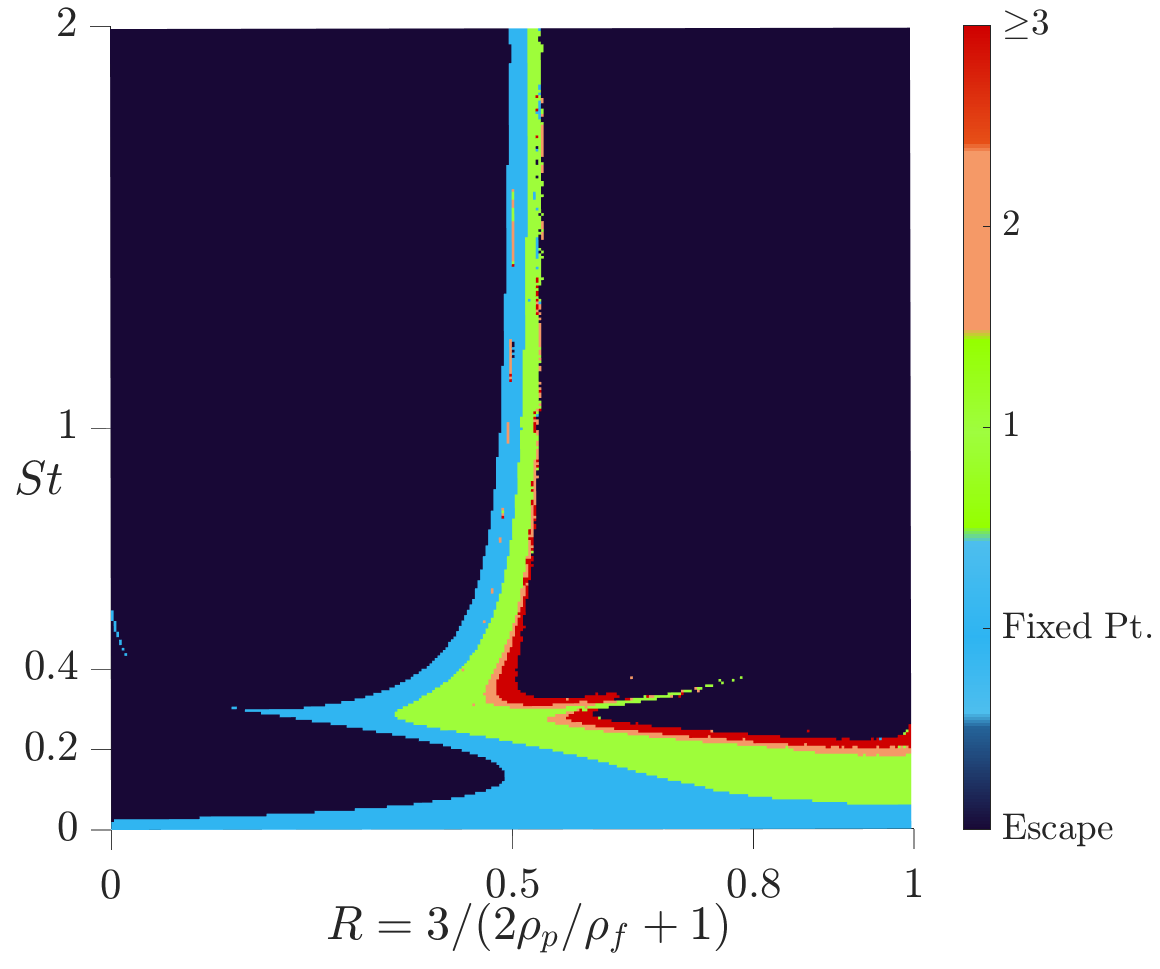}
    \subcaption{$\Gamma_1/\Gamma_2=1.0$}
    \end{minipage}
    \begin{minipage}[b]{0.32\textwidth}
    \includegraphics[width=\textwidth]{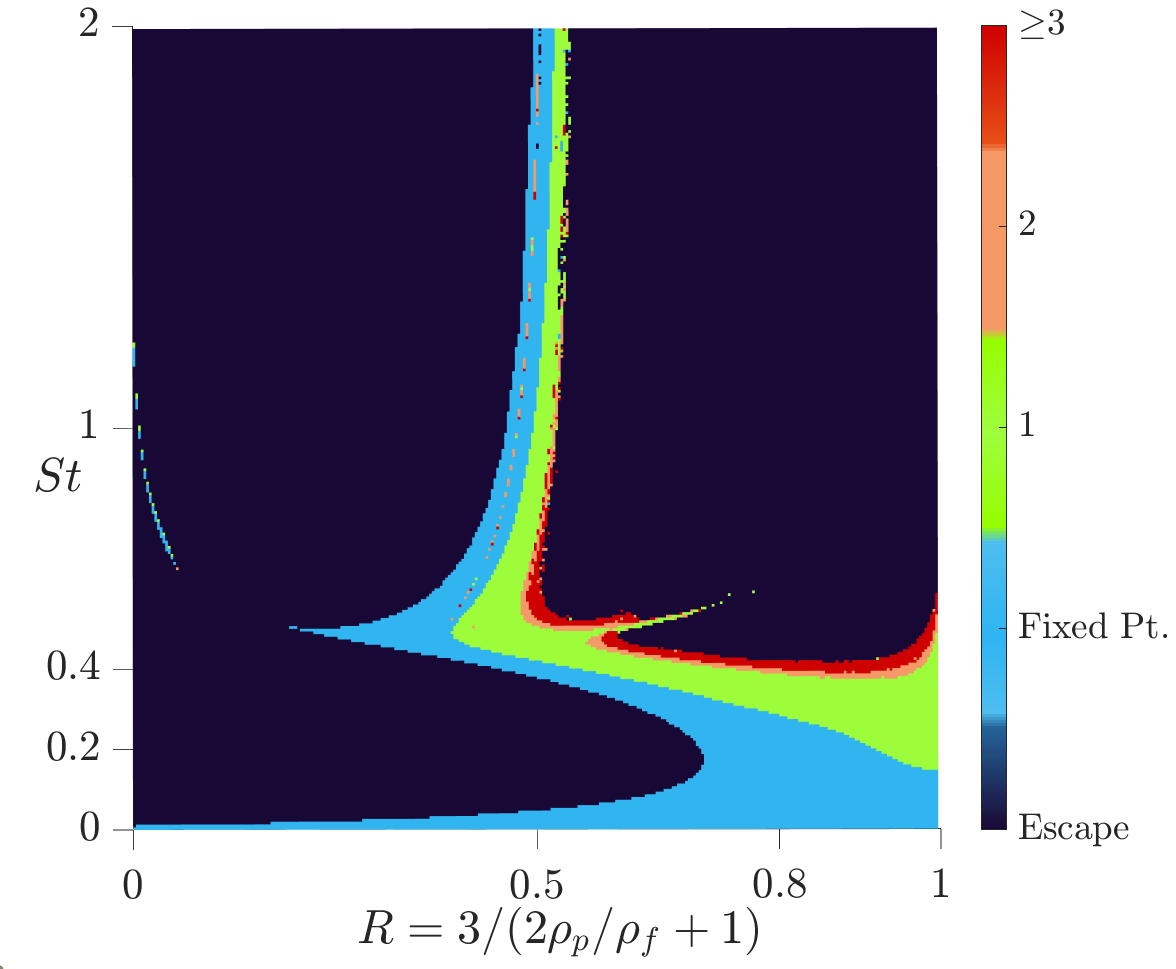}
    \subcaption{$\Gamma_1/\Gamma_2=0.5\ (top)$}
    \end{minipage}
    \begin{minipage}[b]{0.32\textwidth}
    \includegraphics[width=\textwidth]{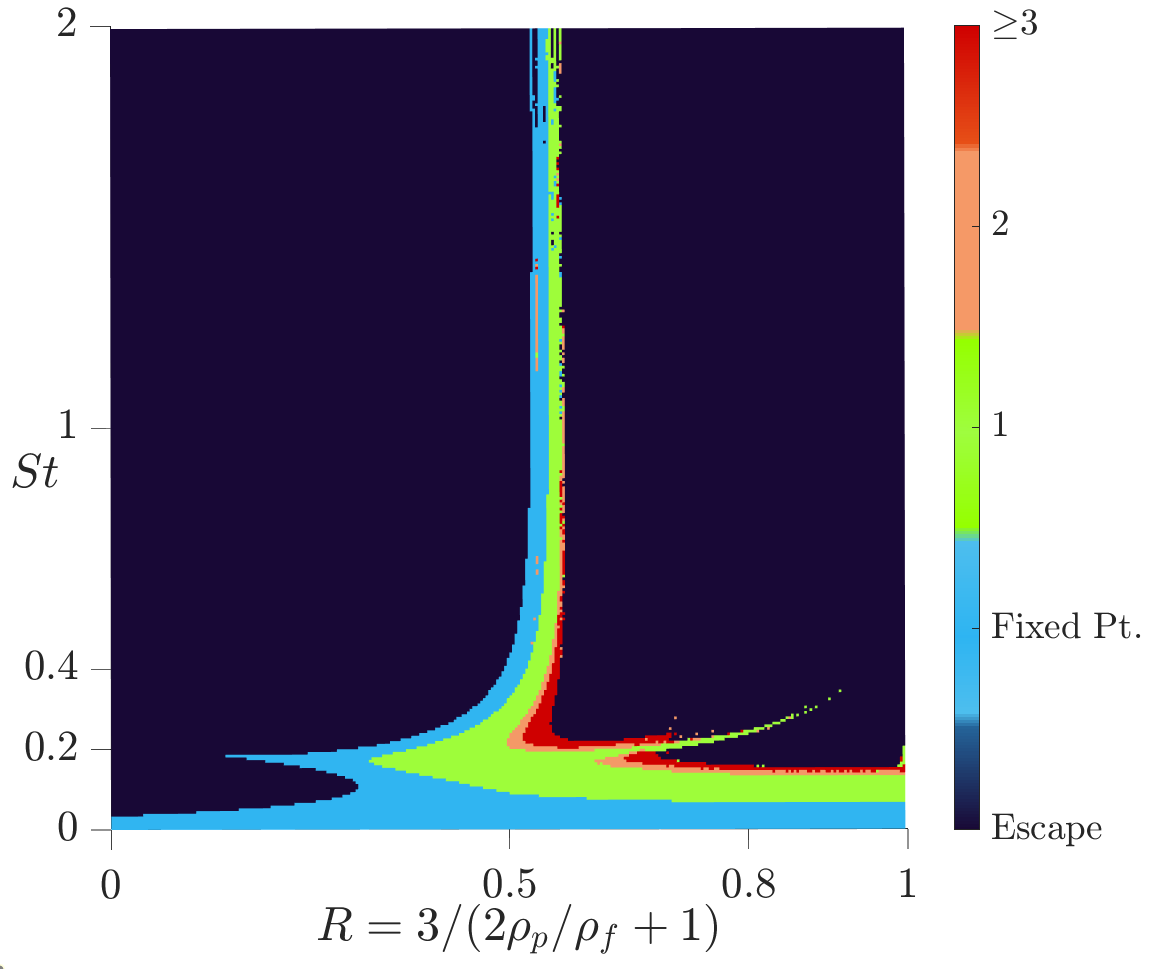}
    \subcaption{$\Gamma_1/\Gamma_2=0.5\ (bottom)$}
    \end{minipage}
    \caption{\rev{Trapping phase plots in the $R-St$ space based on the type of attractor for vortex strength ratios ($\Gamma_1/\Gamma_2$) of $1.0$ and $0.5$, without the BBH force. The former is presented here again to enable direct comparison. The latter case breaks the top-down symmetry, leading to different dynamics in the regions above and below the vortices shown in the different phase plots (b), (c).}}
    \label{fig:phaseplotunequal}
\end{figure}

\rev{In order to demonstrate the qualitative generalization of our findings to unequal vortices, we show the representative case of inertial particles near two co-rotating vortices with relative vortex strength ratio $\Gamma_1/\Gamma_2= 0.5$. The weaker vortex is placed on the left in the rotating frame, without loss of generality. It suffices to consider the case without the BBH force for qualitative comparisons. 

The unequal vortex strength breaks symmetry with respect to the line joining the centers of the vortices. Therefore, we have separate $R-St$ phase plots, namely \cref{fig:phaseplotunequal}(b) and \cref{fig:phaseplotunequal}(c), for the top and bottom half-planes respectively. For comparison, we have placed these plots alongside the case of identical vortices, $\Gamma_1/\Gamma_2=1$ (\cref{fig:phaseplotunequal}(a)). Despite significant quantitative differences, the overall similarity in the structure of these phase plots is unmistakable. Whether for an equal-strength or unequal-strength pair, inertial particles near co-rotating vortex pair of any vortex strength ratio can get trapped into attracting orbits of differing complexities depending on the Stokes number $St$ and density ratio parameter $R$, and the non-monotonic variation of trapping behaviour with increasing Stokes number can be noted for a range of $R$. Further, we note that our phase plots at $R\to 0$ recover the trapping states found in \cite{angilella2010} for the vortex strength ratio $0.5$, namely there are two attracting fixed points at low $St$ whereas there is only one at a higher $St$. }

\begin{figure}
    \centering
    \begin{minipage}[b]{0.4\textwidth}
    \includegraphics[width=\textwidth]{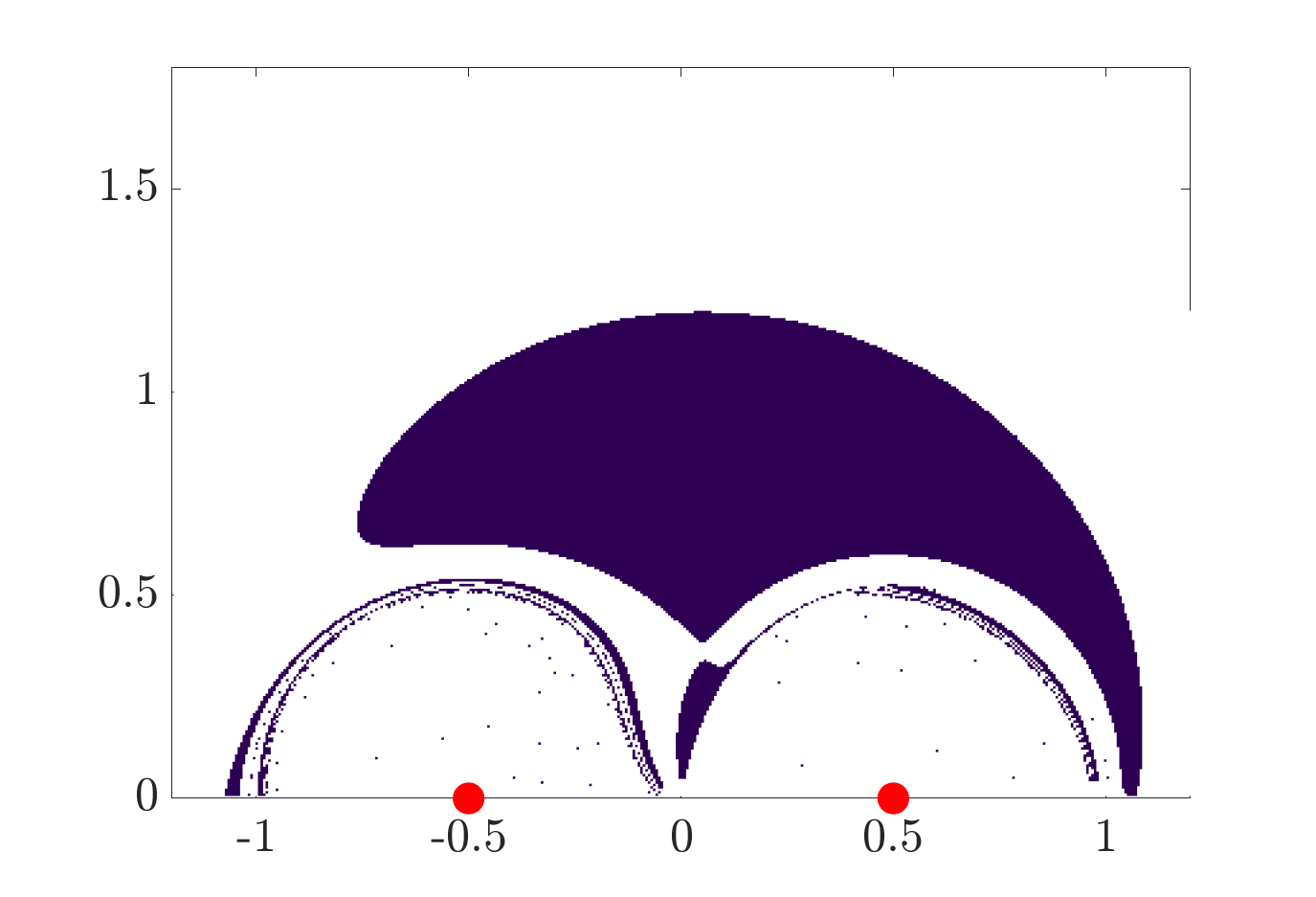}
    \subcaption{$St=0.09\ \ v_0=0_{rot}$}
    \end{minipage}
    \begin{minipage}[b]{0.4\textwidth}
    \includegraphics[width=\textwidth]{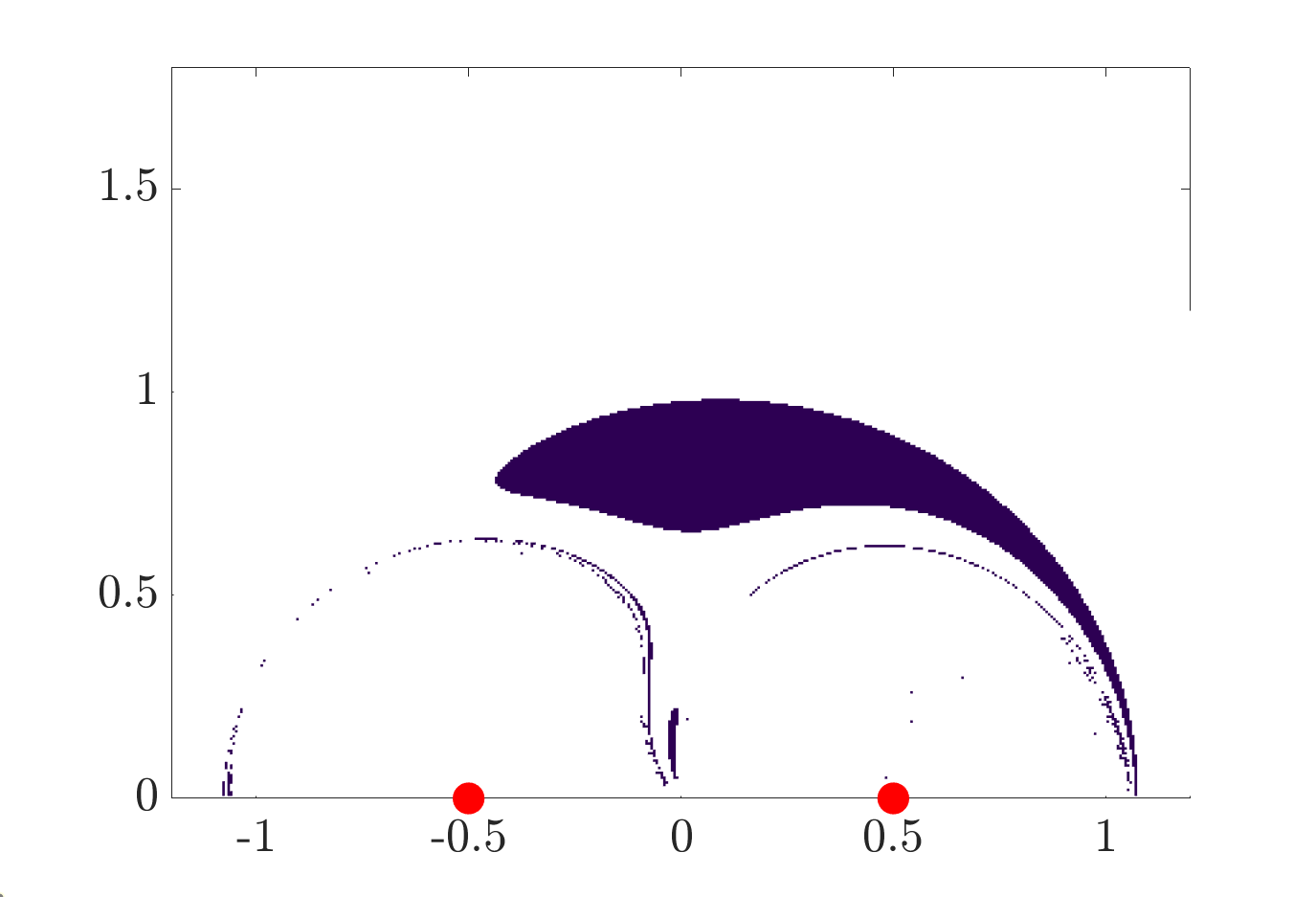}
    \subcaption{$St=0.24\ \ v_0=0_{rot}$}
    \end{minipage}
    \begin{minipage}[b]{0.4\textwidth}
    \includegraphics[width=\textwidth]{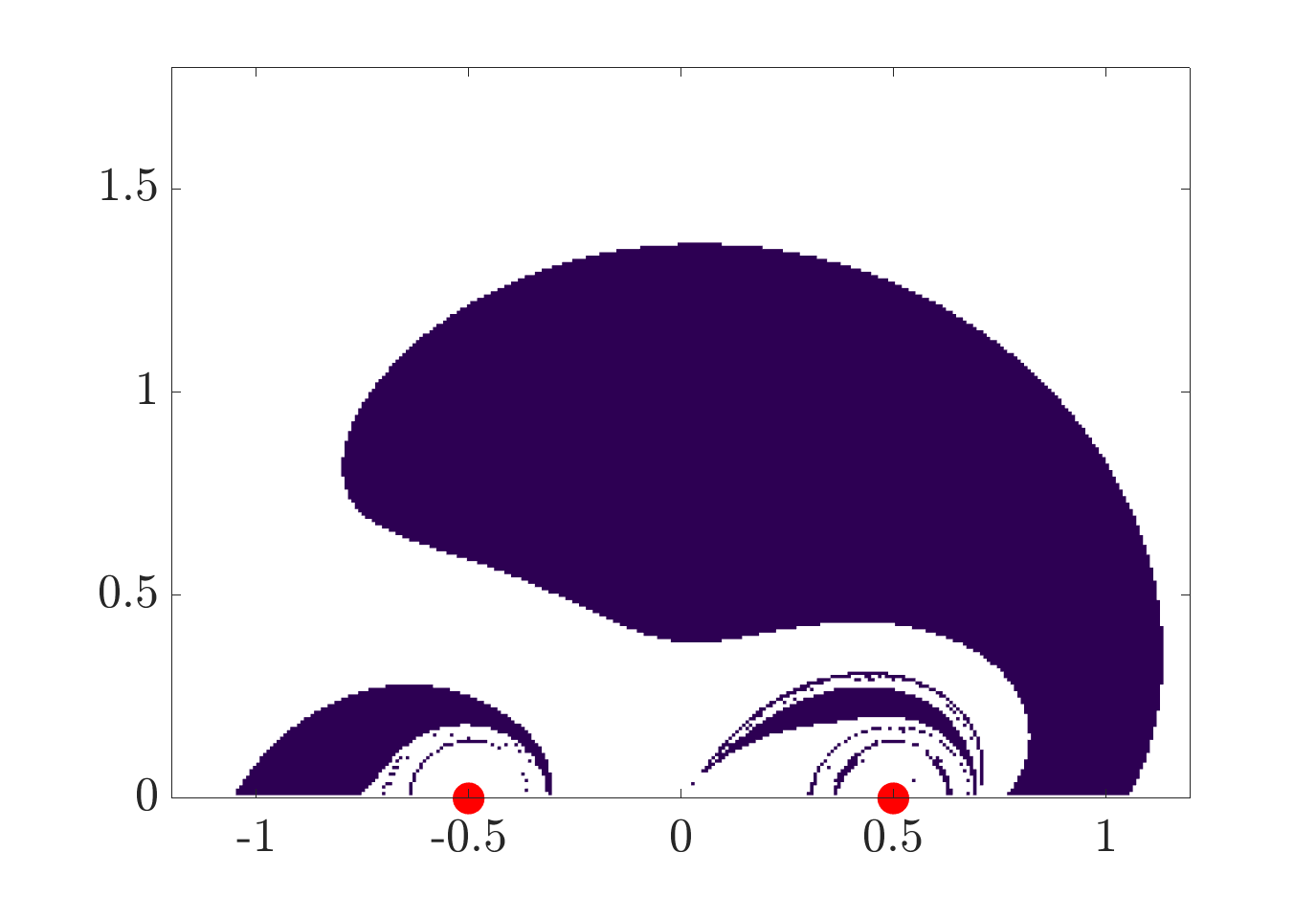}
    \subcaption{$St=0.09\ \ v_0=u$}
    \end{minipage}
    \begin{minipage}[b]{0.4\textwidth}
    \includegraphics[width=\textwidth]{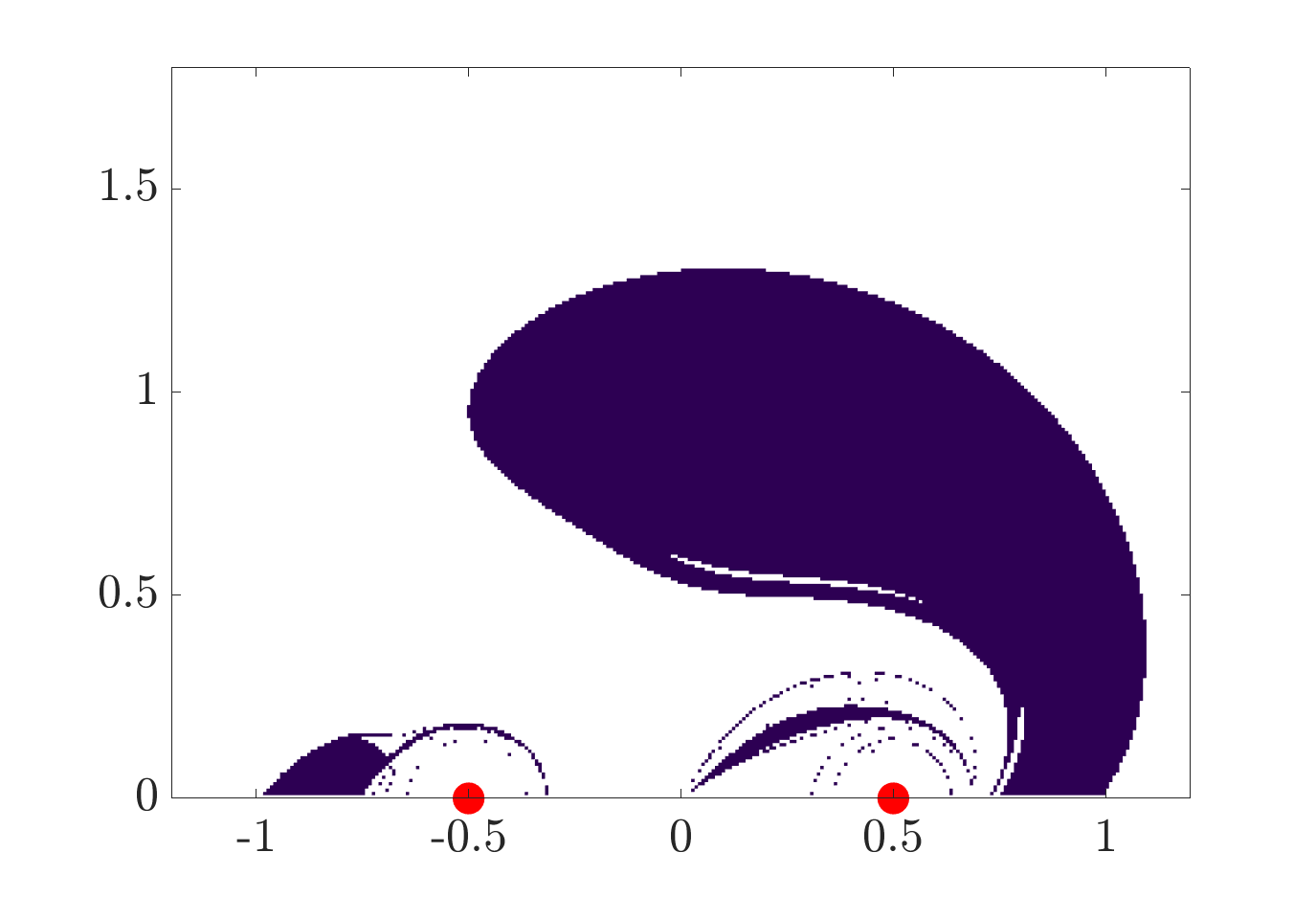}
    \subcaption{$St=0.24\ \ v_0=u$}
    \end{minipage}
    \caption{\rev{Basins of attraction without the BBH force at $R=0.84\ (\rho_p/\rho_f\approx 1.3)$ for $St=0.09$ and $St=0.24$, with the initial conditions: (i) $v_0=0_{rot}$, (ii) $v_0=u$. For this density ratio, we have $BoA_{(ii)}>BoA_{(i)}$. But the BoA sizes are comparable for smaller Stokes number.}}
    \label{fig:inicondwithoutbbhR=0.84}
\end{figure}
\begin{figure}
    \centering
    \begin{minipage}[b]{0.4\textwidth}
    \includegraphics[width=\textwidth]{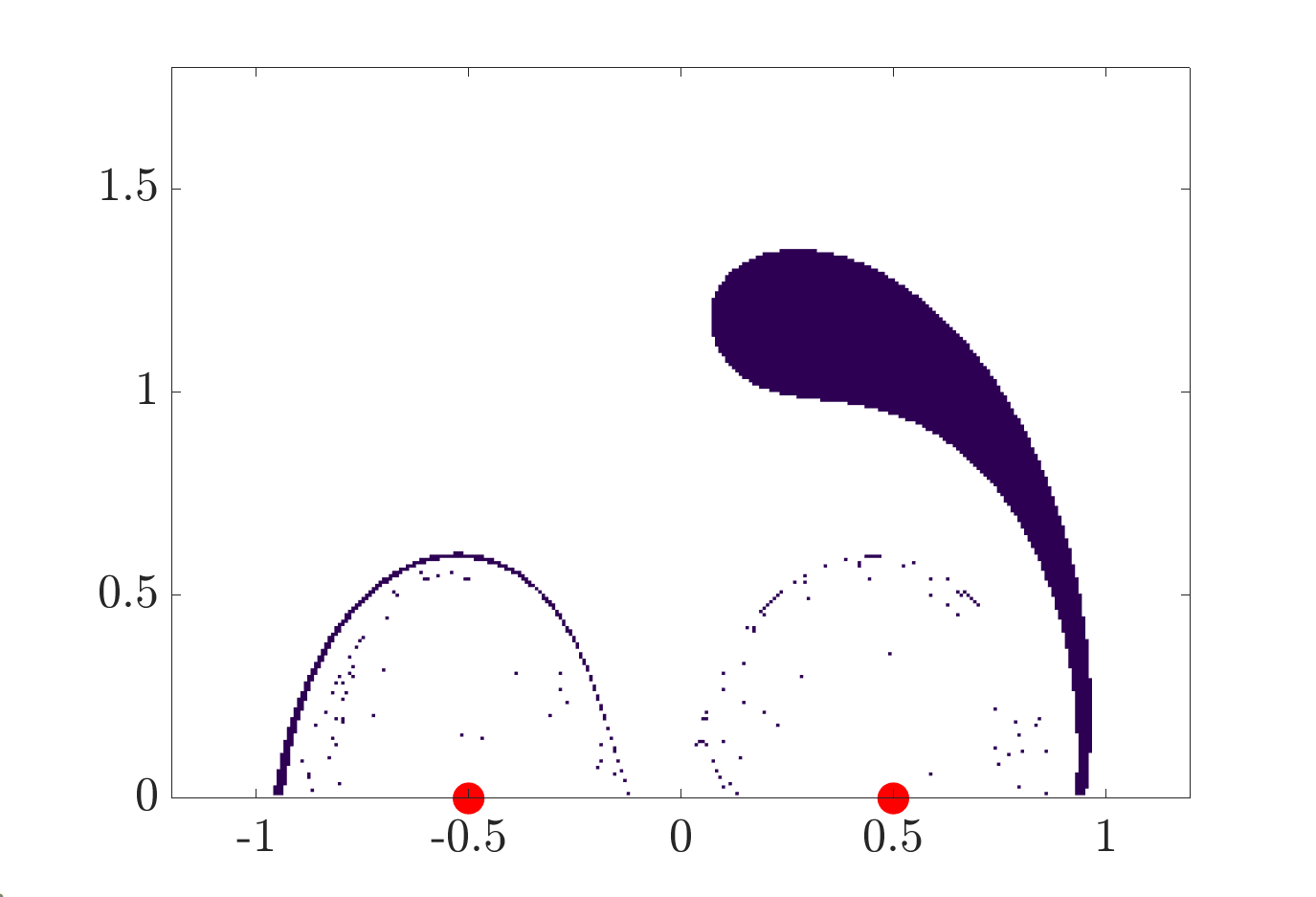}
    \subcaption{$St=0.10\ \ v_0=0_{rot}$}
    \end{minipage}
    \begin{minipage}[b]{0.4\textwidth}
    \includegraphics[width=\textwidth]{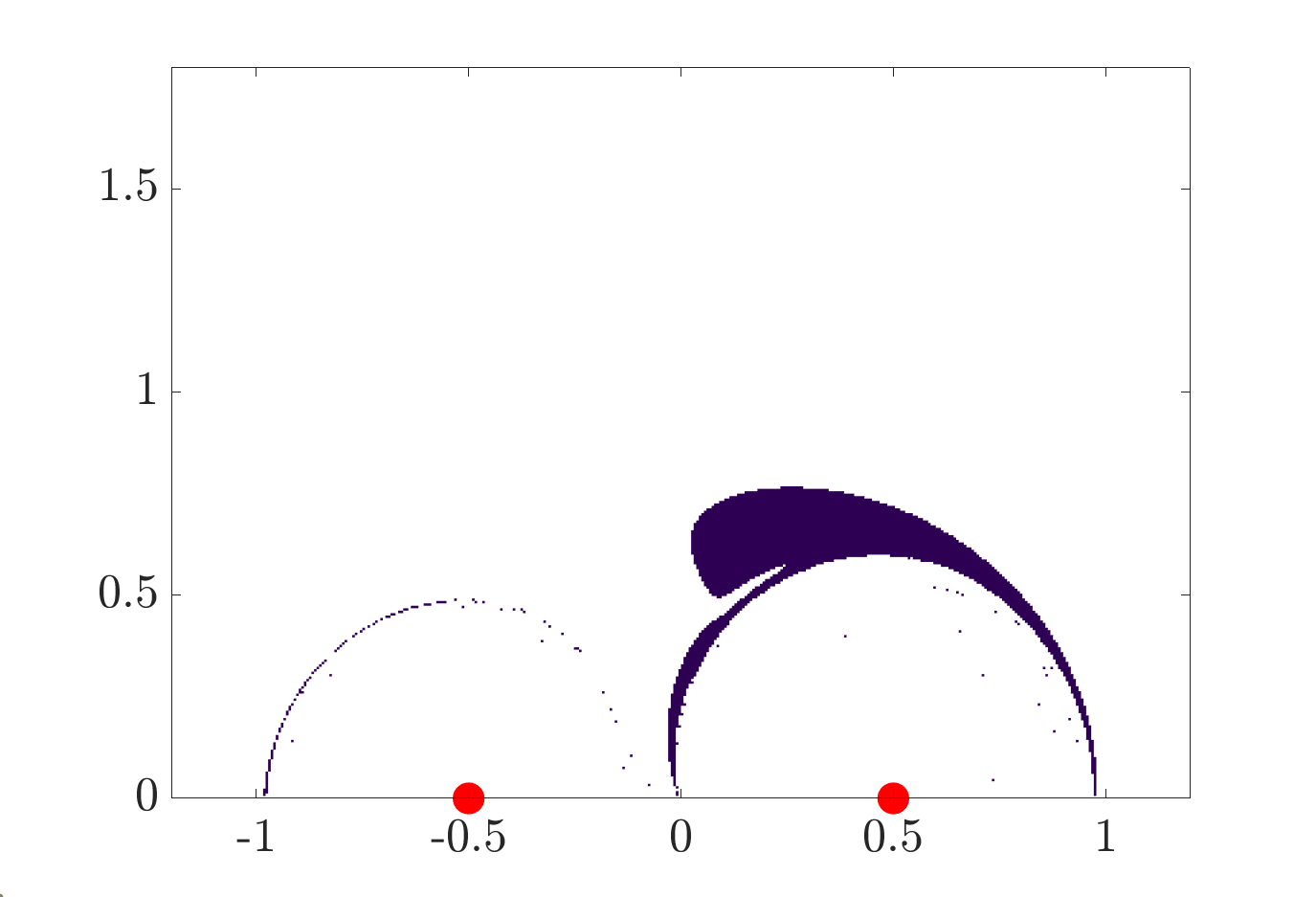}
    \subcaption{$St=0.32\ \ v_0=0_{rot}$}
    \end{minipage}
    \begin{minipage}[b]{0.4\textwidth}
    \includegraphics[width=\textwidth]{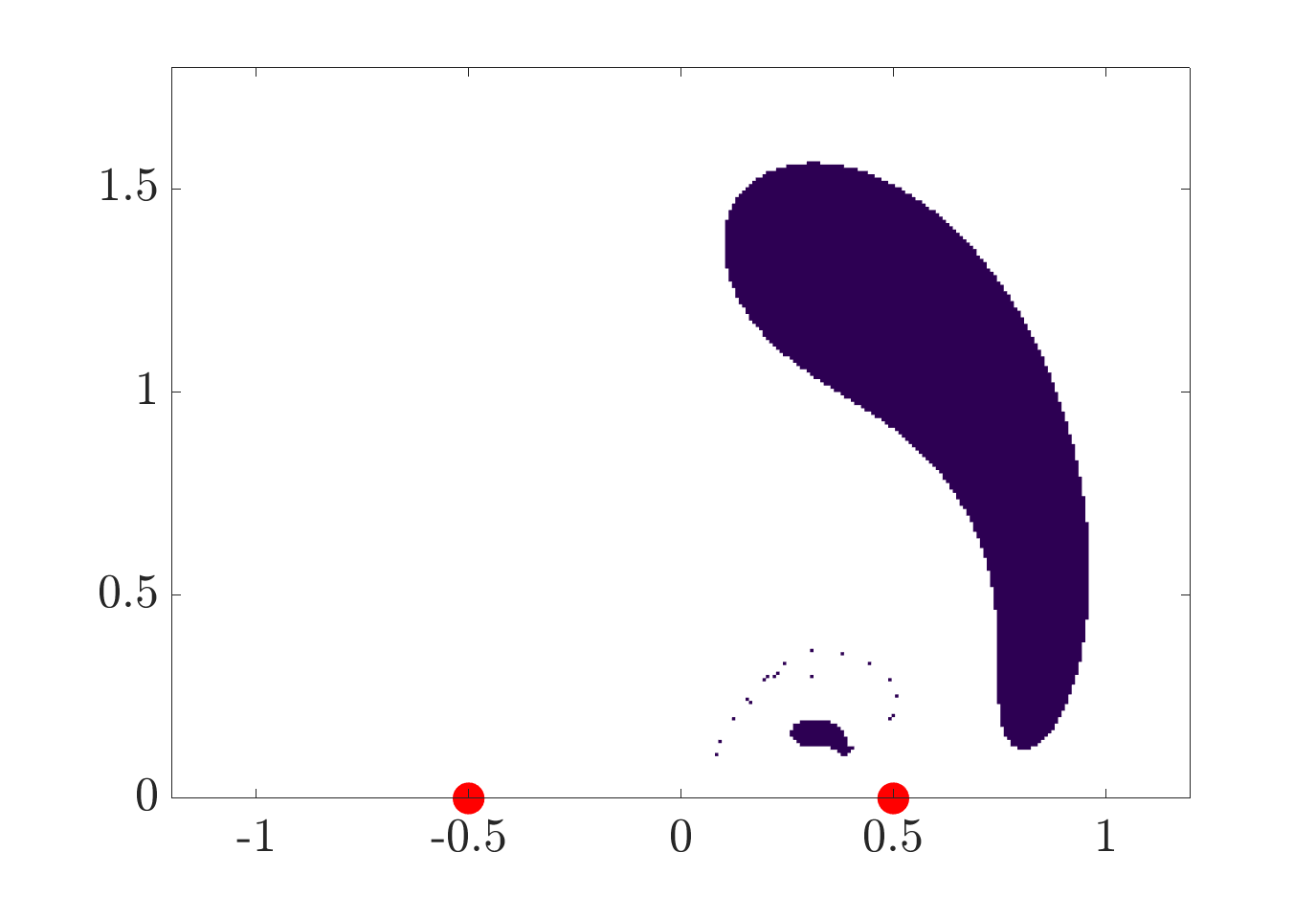}
    \subcaption{$St=0.10\ \ v_0=u$}
    \end{minipage}
    \begin{minipage}[b]{0.4\textwidth}
    \includegraphics[width=\textwidth]{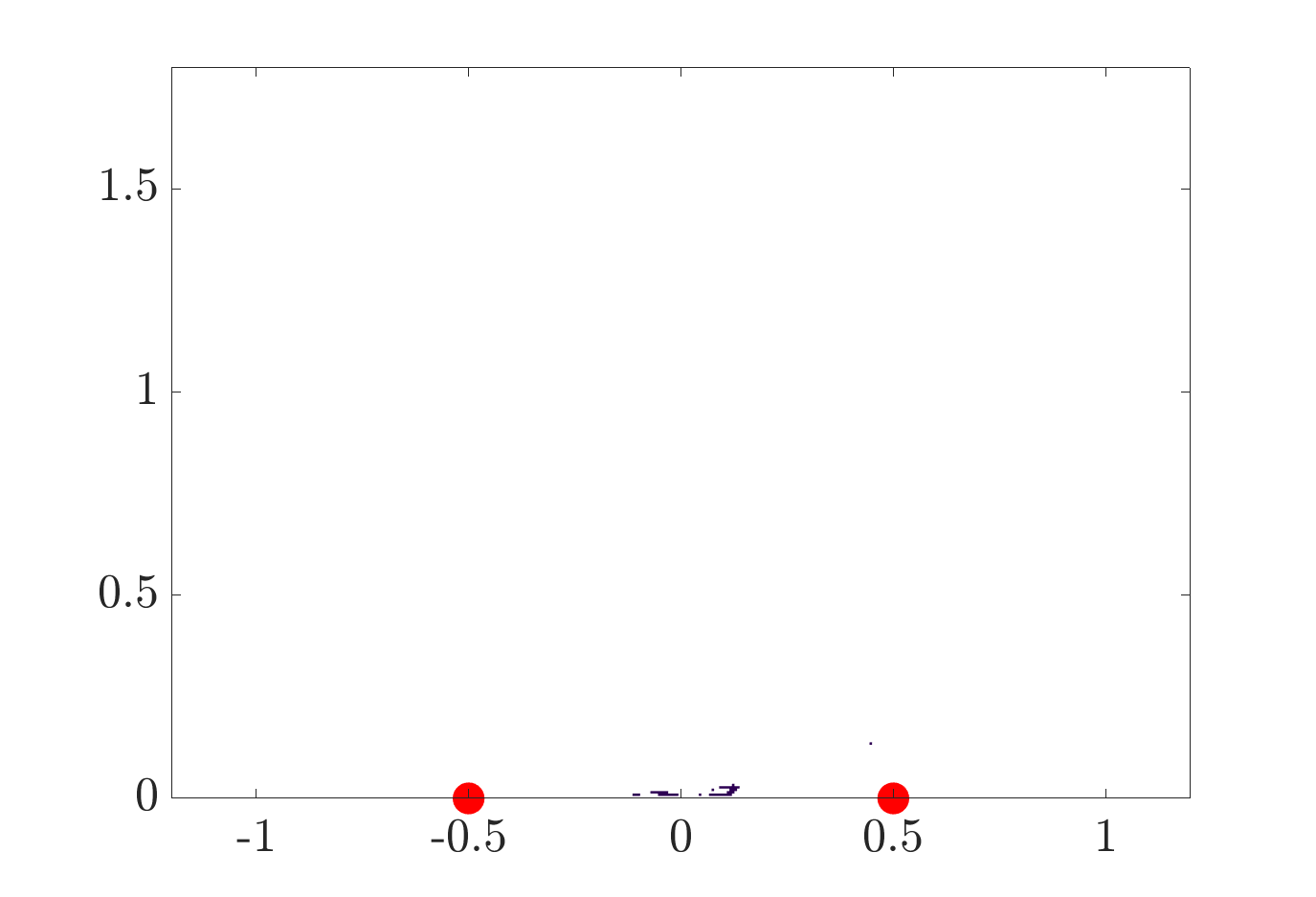}
    \subcaption{$St=0.32\ \ v_0=u$}
    \end{minipage}
    \caption{\rev{Basins of attraction without the BBH force at $R=0.50\ (\rho_p/\rho_f = 2.5)$ for $St=0.10$ and $St=0.32$, with the initial conditions: (i) $v_0=0_{rot}$, (ii) $v_0=u$. Already at $St=0.1$ there is a visible quantitative difference, but the BoA sizes are comparable. Notably, at $St=0.32$, BoAs are insignificant for the initial condition (ii). In contrast to the case of $R=0.84$, we have $BoA_{(i)}\gg BoA_{(ii)}$ for the higher Stokes number.}}
    \label{fig:inicondwithoutbbhR=0.50}
\end{figure}
\begin{figure}
    \centering
    \begin{minipage}[b]{0.3\textwidth}
    \includegraphics[width=\textwidth]{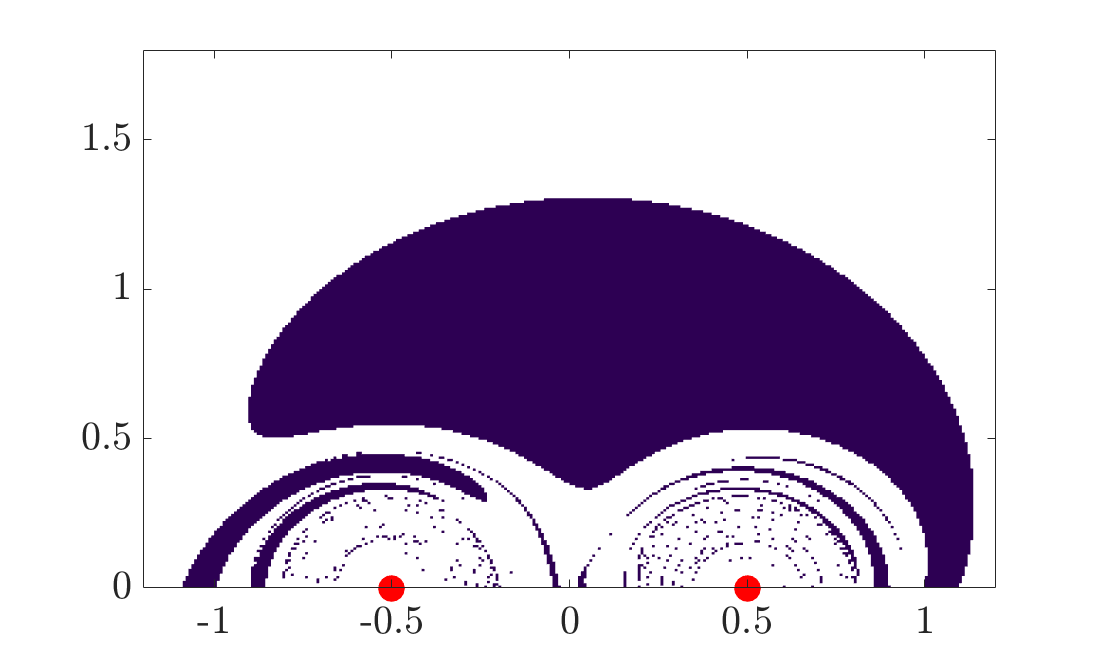}
    \subcaption{$St=0.09\ \ v_0=0_{rot}$}
    \end{minipage}
    \begin{minipage}[b]{0.3\textwidth}
    \includegraphics[width=\textwidth]{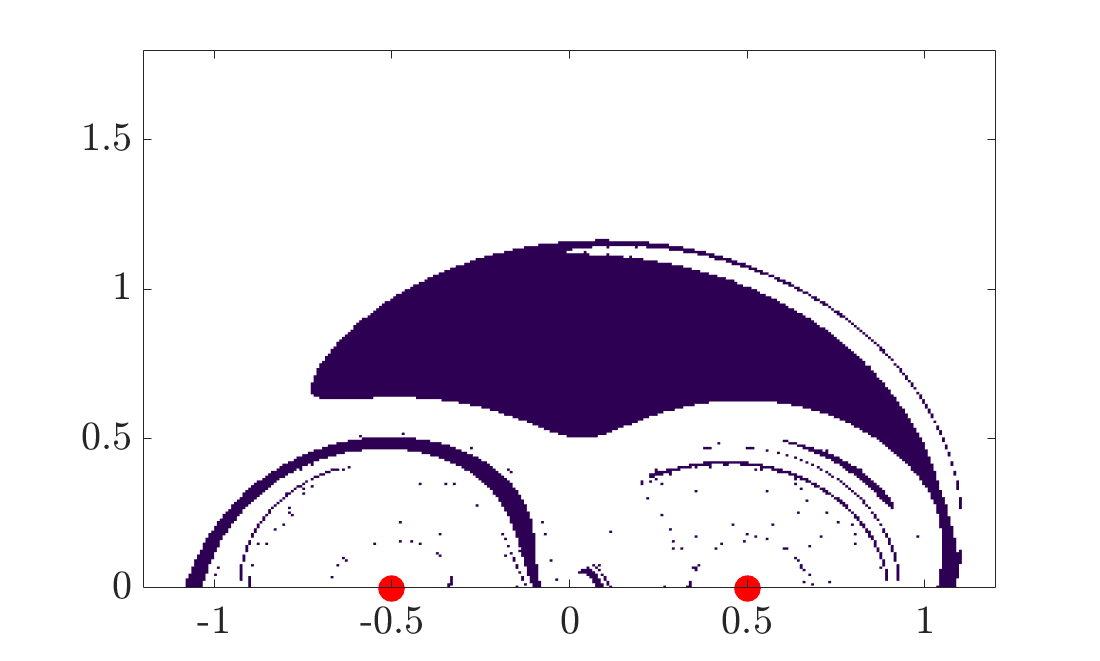}
    \subcaption{$St=0.24\ \ v_0=0_{rot}$}
    \end{minipage}
    \begin{minipage}[b]{0.3\textwidth}
    \includegraphics[width=\textwidth]{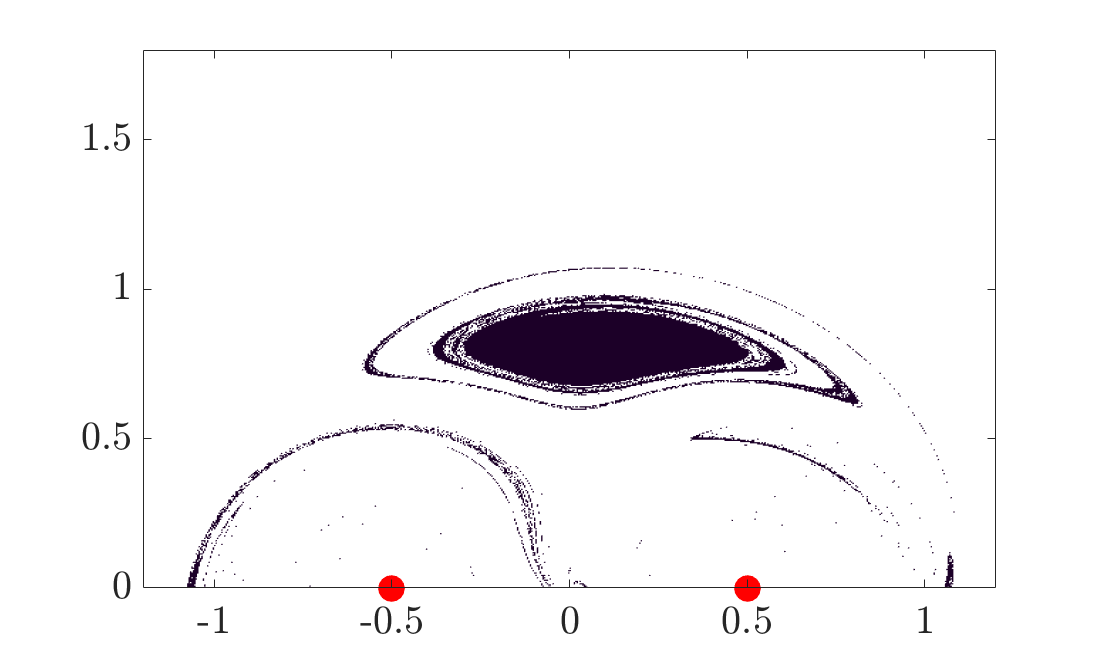}
    \subcaption{$St=0.495\ \ v_0=0_{rot}$}
    \end{minipage}
    \begin{minipage}[b]{0.3\textwidth}
    \includegraphics[width=\textwidth]{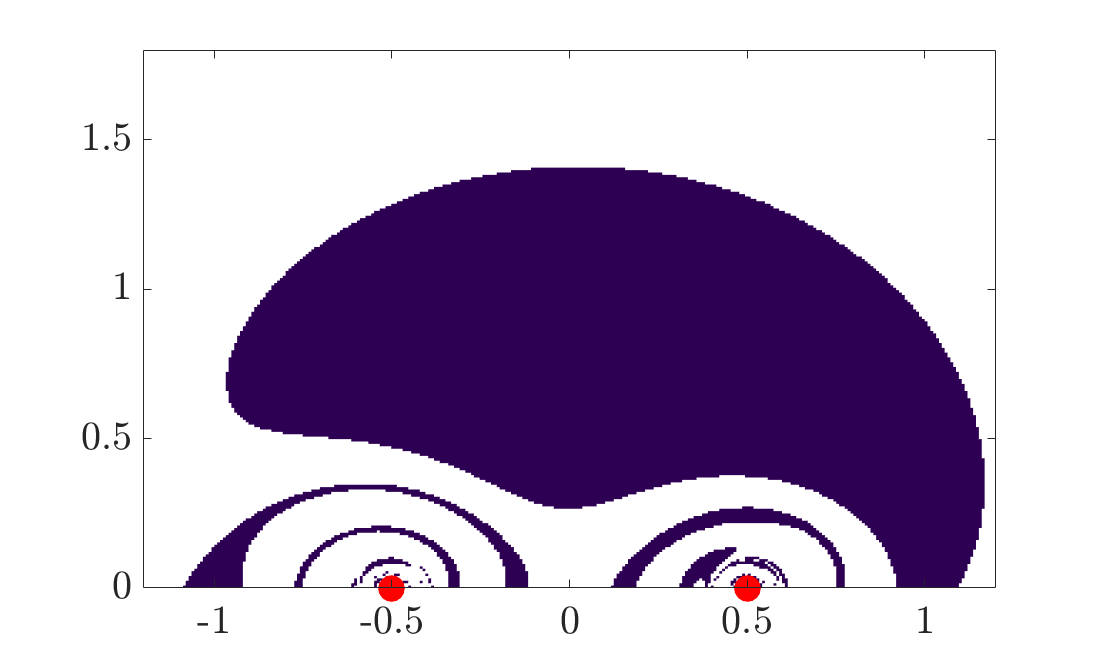}
    \subcaption{$St=0.09\ \ v_0=u$}
    \end{minipage}
    \begin{minipage}[b]{0.3\textwidth}
    \includegraphics[width=\textwidth]{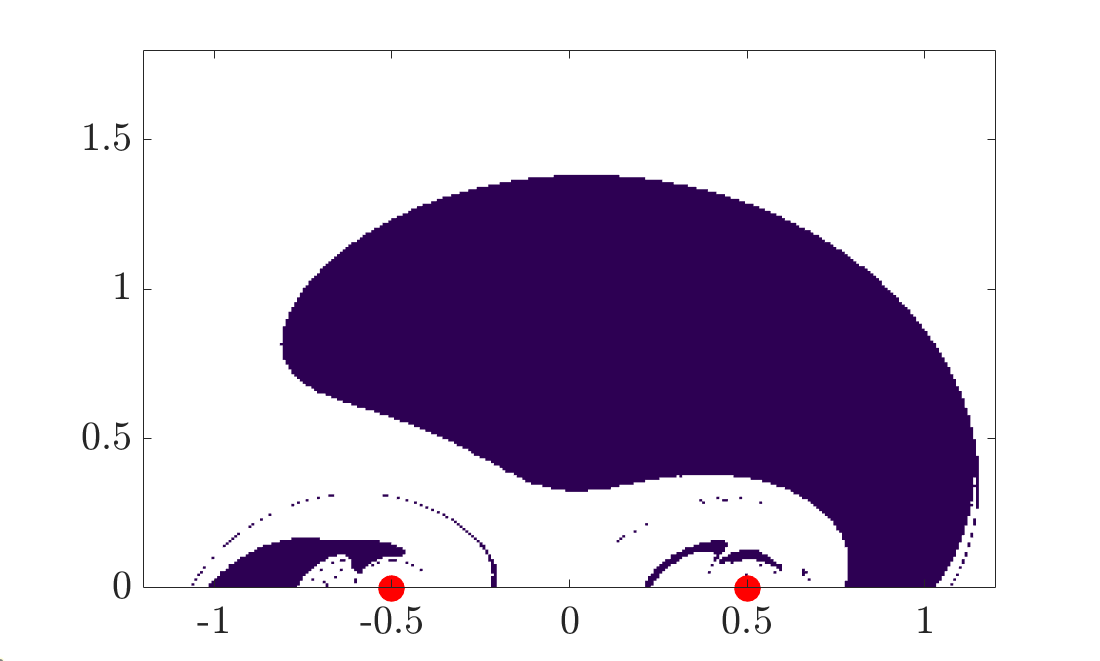}
    \subcaption{$St=0.24\ \ v_0=u$}
    \end{minipage}
    \begin{minipage}[b]{0.3\textwidth}
    \includegraphics[width=\textwidth]{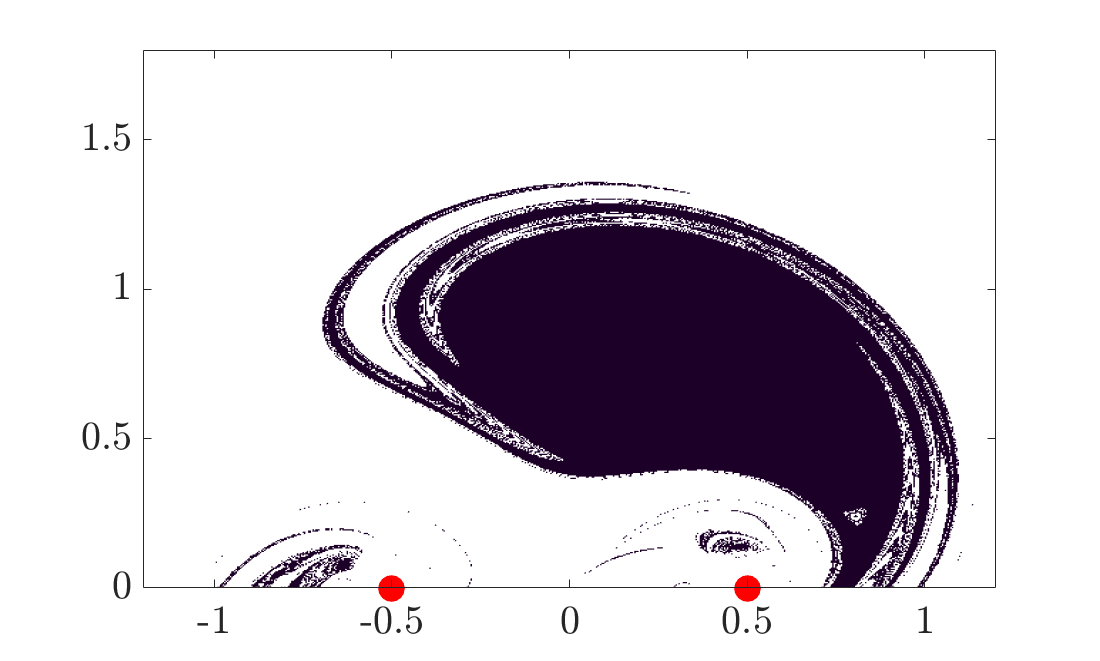}
    \subcaption{$St=0.495\ \ v_0=u$}
    \end{minipage}
    \caption{\rev{Basins of attraction with the BBH force at $R=0.84\ (\rho_p/\rho_f\approx 1.3)$ for $St=0.09$, $St=0.24$ and $St=0.495$, with the initial conditions, (i) $v_0=0$ and (ii) $v_0=u$. For $St=0.09$ as well as for $St=0.24$, the BoAs are comparable in size for the different initial conditions. But there is a significant difference at the larger $St=0.495$.}}
    \label{fig:inicondwithbbhR=0.84}
\end{figure}
\begin{figure}
    \centering
    \begin{minipage}[b]{0.4\textwidth}
    \includegraphics[width=\textwidth]{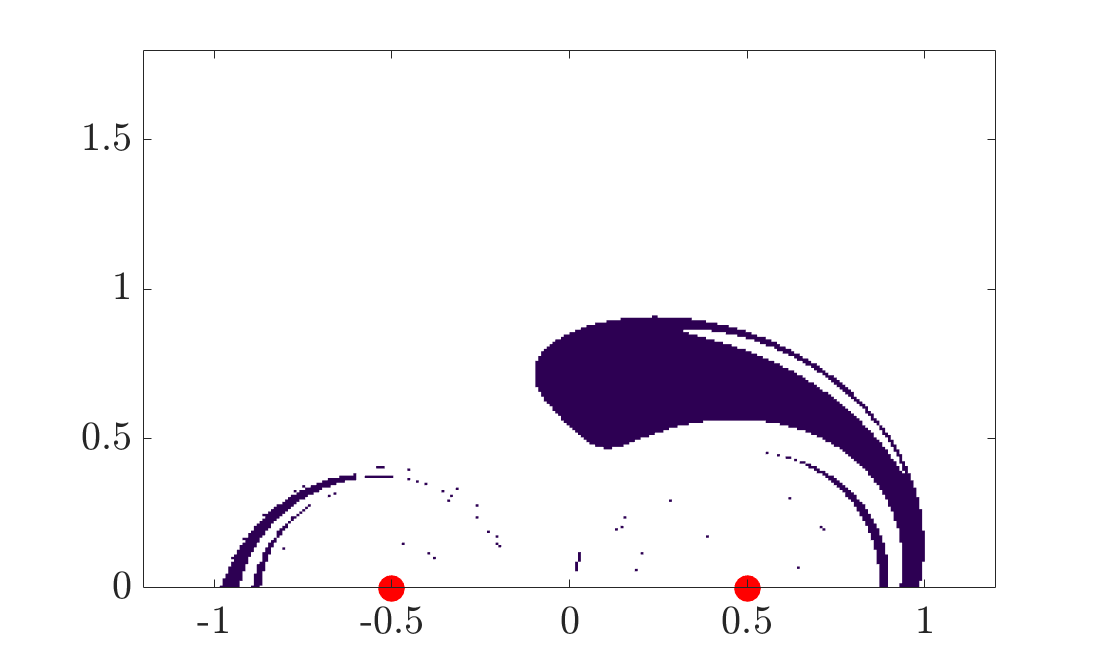}
    \subcaption{$St=0.32\ \ v_0=0_{rot}$}
    \end{minipage}
    \begin{minipage}[b]{0.4\textwidth}
    \includegraphics[width=\textwidth]{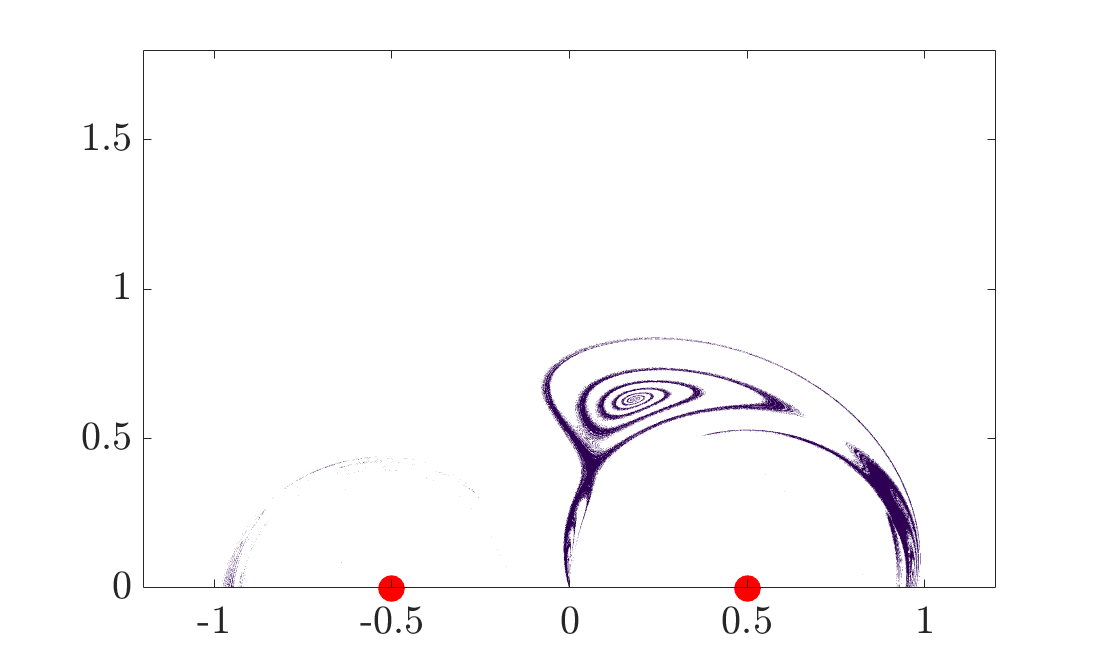}
    \subcaption{$St=0.685\ \ v_0=0_{rot}$}
    \end{minipage}
    \begin{minipage}[b]{0.4\textwidth}
    \includegraphics[width=\textwidth]{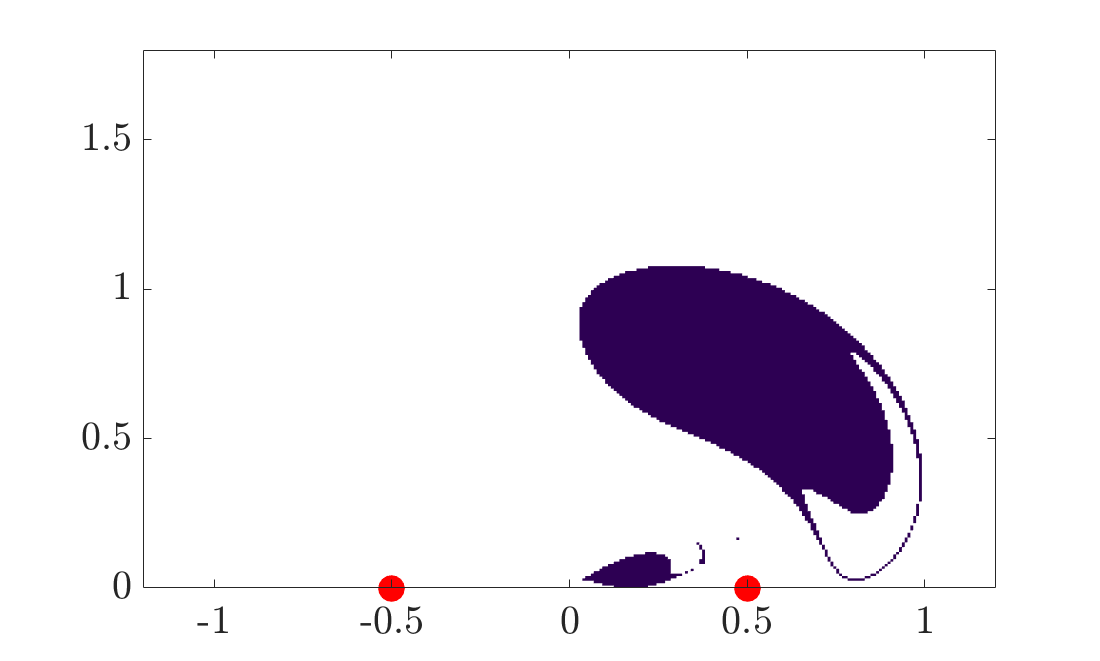}
    \subcaption{$St=0.32\ \ v_0=u$}
    \end{minipage}
    \begin{minipage}[b]{0.4\textwidth}
    \includegraphics[width=\textwidth]{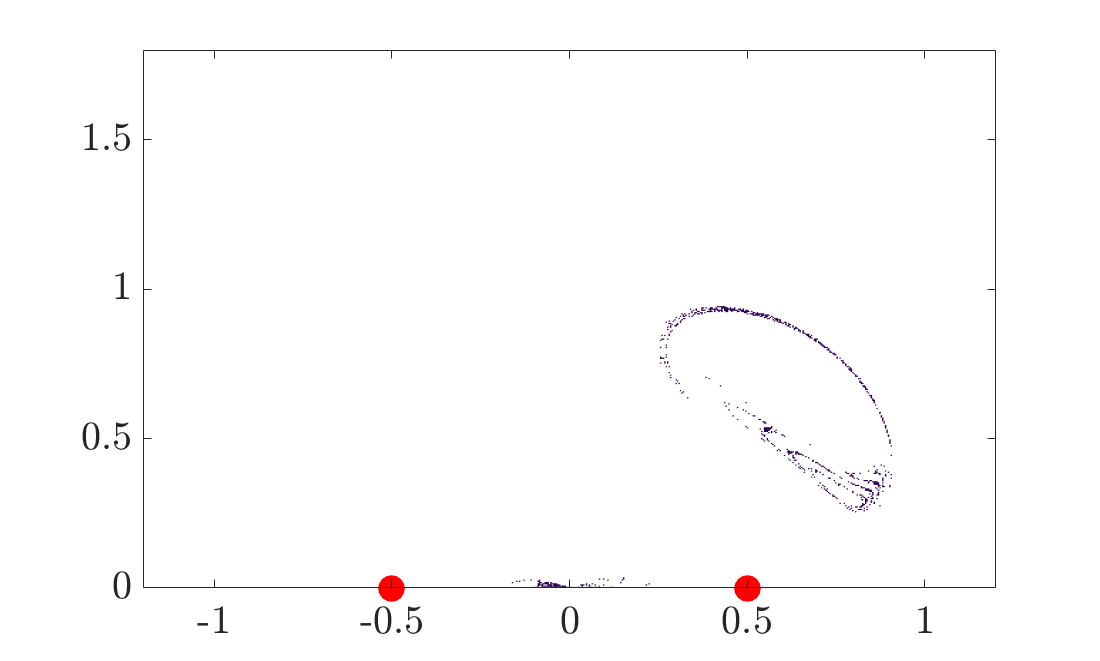}
    \subcaption{$St=0.685\ \ v_0=u$}
    \end{minipage}
    \caption{\rev{Basins of attraction with the BBH force at $R=0.50\ (\rho_p/\rho_f = 2.5)$ for $St=0.32$ and $St=0.685$, with the initial conditions (i) $v_0=0$ and (ii) $v_0=u$. For $St=0.32$, BoA sizes are comparable for the two initial conditions. But for the higher $St=0.685$ there is a large difference.}}
    \label{fig:inicondwithbbhR=0.50}
\end{figure}

\section{\rev{Effects of particle initial velocity on the basins of attraction}}\label{app:B}
\rev{In real-life scenarios, it is impractical to control the initial conditions, and each particle's initial velocity could be different from that of its neighbors. Therefore, it serves us well to study two typical initial velocities, namely (i) $\bf{v}_0=\bf{0}$ in the rotating frame, and (ii) $\bf{v}_0=\bf{u}(\bf{y}_0)$ (zero relative velocity). We denote their respective BoAs by the appropriate subscript. \rev{But we must note that the attractors themselves, and quantities like $St_{crit}$, are properties of the flow geometry and particle $St$ and $R$, and not of the initial conditions. Initial condition (i) is chosen for its simplicity for the majority of this study, while (ii) describes particles which, just before the computations begin, have zero Stokes number, but at $t=0$ go through a sudden growth to become more inertial, e.g., due to agglomeration, or, in the case of cloud water droplets, due to condensation.}

\rev{Without the BBH force, BoAs for two density ratios and two Stokes numbers each are shown in 
figs.~\ref{fig:inicondwithoutbbhR=0.84} and~\ref{fig:inicondwithoutbbhR=0.50} for the two initial conditions. We find that sizes of BoAs are relatively insensitive to initial conditions for small Stokes numbers, $St\lesssim 0.1$, whereas at moderate Stokes numbers there is a significant effect. With the BBH force, the effect of initial conditions is weak up to a larger Stokes number, as evident from figs.~\ref{fig:inicondwithbbhR=0.84} and ~\ref{fig:inicondwithbbhR=0.50}. }

At higher Stokes, the effect of changing initial condition from (i) to (ii) on the BoA depends on the density ratio itself. For closer to neutrally buoyant particles, BoA$_{(ii)}>$BoA$_{(i)}$ (see \cref{fig:inicondwithoutbbhR=0.84}(b,d) for without the BBH force and \cref{fig:inicondwithbbhR=0.84}(c,f) for with the BBH force), while for denser particles BoA$_{(i)}>$BoA$_{(ii)}$. In fact, at higher density ratios and higher $St$, we find BoA$_{(ii)}\approx 0$, making condition (i) ideal to detect attractors. This happens because an inertial particle on an attracting fixed point always has $v=0$ in the rotating frame. Hence the corresponding BoA is guaranteed to have a non zero area. A similar trend for higher $St$ is observed with the BBH force. For  $R=0.84\ (\rho_p/\rho_f\approx 1.3)$, BoA$_{(ii)}>$BoA$_{(i)}$ (see \cref{fig:inicondwithbbhR=0.84}(c,f)). For $R=0.5\ (\rho_p/\rho_f = 2.5)$, BoA$_{(i)}>$BoA$_{(ii)}$ (see \cref{fig:inicondwithbbhR=0.50}(b,d)), where, again, BoA$_{(ii)}$ is much smaller and harder to detect. }

\rev{Further the trend of the basins of attraction with the BBH force being larger than those without persists with initial condition (ii) as well (compare fig.\ref{fig:inicondwithoutbbhR=0.84}(c,d) fig.\ref{fig:inicondwithbbhR=0.84}(d,e)), however to a differing quantitative extent.}

\rev{To summarize, the sizes of the BoAs are only weakly affected by initial conditions at small Stokes numbers but are quite sensitive at high $St$. The sensitivity increases as the particles become heavier, and is lower with the BBH force than with it. However, the trend of BoA size being larger with BBH than without it persists.}

\bibliographystyle{jfm} 
\bibliography{jfm-refs}

\begin{thebibliography}{46}
\expandafter\ifx\csname natexlab\endcsname\relax\def\natexlab#1{#1}\fi
\def\au#1{#1} \def\ed#1{#1} \def\yr#1{#1}\def\at#1{#1}\def\jt#1{\textit{#1}} \def\bt#1{#1}\def\bvol#1{\textbf{#1}} \def\vol#1{#1} \def\pg#1{#1} \def\publ#1{#1}\def\arxiv#1{#1}\def\org#1{#1}\def\st#1{\textit{#1}}

\bibitem[Aliseda {\em et~al.\/}(2002)Aliseda, Cartellier, Hainaux \& Lasheras]{aliseda02}
{\sc \au{Aliseda, A.}, \au{Cartellier, A.}, \au{Hainaux, F.} \& \au{Lasheras, J.~C.}} \yr{2002}  \at{Effect of preferential concentration on the settling velocity of heavy particles in homogeneous isotropic turbulence}.  \jt{Journal of Fluid Mechanics}  \bvol{468},  \pg{77–105}.

\bibitem[Angilella(2010)]{angilella2010physica}
{\sc \au{Angilella, Jean-Régis}} \yr{2010}  \at{Dust trapping in vortex pairs}.  \jt{Physica D: Nonlinear Phenomena}  \bvol{239}~(18),  \pg{1789--1797}.

\bibitem[Angilella {\em et~al.\/}(2014)Angilella, Vilela \& Motter]{angilella2014}
{\sc \au{Angilella, J.~R.}, \au{Vilela, R.~D.} \& \au{Motter, A.~E.}} \yr{2014}  \at{Inertial particle trapping in an open vortical flow}.  \jt{Journal of Fluid Mechanics}  \bvol{744},  \pg{183–216}.

\bibitem[Bec {\em et~al.\/}(2007)Bec, Biferale, Cencini, Lanotte, Musacchio \& Toschi]{bec07}
{\sc \au{Bec, J.}, \au{Biferale, L.}, \au{Cencini, M.}, \au{Lanotte, A.}, \au{Musacchio, S.} \& \au{Toschi, F.}} \yr{2007}  \at{Heavy particle concentration in turbulence at dissipative and inertial scales}.  \jt{Phys. Rev. Lett.}  \bvol{98},  \pg{084502}.

\bibitem[Bracco {\em et~al.\/}(1999)Bracco, Chavanis, Provenzale \& Spiegel]{Bracco1999}
{\sc \au{Bracco, A.}, \au{Chavanis, P.~H.}, \au{Provenzale, A.} \& \au{Spiegel, E.~A.}} \yr{1999}  \at{{Particle aggregation in a turbulent Keplerian flow}}.  \jt{Physics of Fluids}  \bvol{11}~(8),  \pg{2280--2287}.

\bibitem[Cerretelli \& Williamson(2003)]{cerretelliWilliamson2003}
{\sc \au{Cerretelli, C.} \& \au{Williamson, C. H.~K.}} \yr{2003}  \at{The physical mechanism for vortex merging}.  \jt{Journal of Fluid Mechanics}  \bvol{475},  \pg{41–77}.

\bibitem[Chong {\em et~al.\/}(2013)Chong, Kelly, Smith \& Eldredge]{chong2013}
{\sc \au{Chong, K.}, \au{Kelly, S.~D.}, \au{Smith, S.} \& \au{Eldredge, J.~D.}} \yr{2013}  \at{{Inertial particle trapping in viscous streaming}}.  \jt{Physics of Fluids}  \bvol{25}~(3),  \pg{033602}.

\bibitem[Daitche(2015)]{daitche2015}
{\sc \au{Daitche, A.}} \yr{2015}  \at{On the role of the history force for inertial particles in turbulence}.  \jt{Journal of Fluid Mechanics}  \bvol{782},  \pg{567–593}.

\bibitem[Daitche \& T\'el(2011)]{daitcheTel11}
{\sc \au{Daitche, A.} \& \au{T\'el, T.}} \yr{2011}  \at{Memory effects are relevant for chaotic advection of inertial particles}.  \jt{Physical Review Letters}  \bvol{107},  \pg{244501}.

\bibitem[Daitche \& T\'el(2014)]{daitche2014}
{\sc \au{Daitche, A.} \& \au{T\'el, T.}} \yr{2014}  \at{Memory effects in chaotic advection of inertial particles}.  \jt{New Journal of Physics}  \bvol{16}~(7),  \pg{073008}.

\bibitem[Druzhinin(1995)]{fieldapprox_Druzhinin95}
{\sc \au{Druzhinin, O.~A.}} \yr{1995}  \at{On the two-way interaction in two-dimensional particle-laden flows: the accumulation of particles and flow modification}.  \jt{Journal of Fluid Mechanics}  \bvol{297},  \pg{49–76}.

\bibitem[Falkovich {\em et~al.\/}(2002)Falkovich, Fouxon \& Stepanov]{Falkovich2002}
{\sc \au{Falkovich, G}, \au{Fouxon, A} \& \au{Stepanov, M~G}} \yr{2002}  \at{Acceleration of rain initiation by cloud turbulence}.  \jt{Nature}  \bvol{419}~(6903),  \pg{151--154}.

\bibitem[Ferry \& Balachandar(2001)]{fieldapprox_Ferry01}
{\sc \au{Ferry, Jim} \& \au{Balachandar, S.}} \yr{2001}  \at{A fast eulerian method for disperse two-phase flow}.  \jt{International Journal of Multiphase Flow}  \bvol{27}~(7),  \pg{1199--1226}.

\bibitem[Gallay \& Wayne(2002)]{gallayWayne02}
{\sc \au{Gallay, T.} \& \au{Wayne, C.~E.}} \yr{2002}  \at{Invariant manifolds and the long-time asymptotics of the {N}avier-{S}tokes and vorticity equations on {R}2}.  \jt{Archive for Rational Mechanics and Analysis}  \bvol{163},  \pg{209–258}.

\bibitem[Gatignol(1983)]{gatignol1983}
{\sc \au{Gatignol, R}} \yr{1983}  \at{The {F}a\'xen formulae for a rigid particle in an unsteady non-uniform {S}tokes flow}.  \jt{J. Mec. Theor. Appl.}  \bvol{2},  \pg{143--160}.

\bibitem[Gerosa {\em et~al.\/}(2023)Gerosa, Méheut \& Bec]{bec2023}
{\sc \au{Gerosa, F.~A.}, \au{Méheut, H.} \& \au{Bec, J.}} \yr{2023}  \at{Clusters of heavy particles in two-dimensional {K}eplerian turbulence}.  \jt{European Physical Journal Plus}  \bvol{138}~(9).

\bibitem[Grebogi {\em et~al.\/}(1983)Grebogi, Ott \& Yorke]{grebogi1983}
{\sc \au{Grebogi, C.}, \au{Ott, E.} \& \au{Yorke, J.~A.}} \yr{1983}  \at{Crises, sudden changes in chaotic attractors, and transient chaos}.  \jt{Physica D: Nonlinear Phenomena}  \bvol{7}~(1),  \pg{181--200}.

\bibitem[Guasto {\em et~al.\/}(2012)Guasto, Rusconi \& Stocker]{guastoReview12}
{\sc \au{Guasto, Jeffrey~S.}, \au{Rusconi, Roberto} \& \au{Stocker, Roman}} \yr{2012}  \at{Fluid mechanics of planktonic microorganisms}.  \jt{Annual Review of Fluid Mechanics}  \bvol{44}~(Volume 44, 2012),  \pg{373--400}.

\bibitem[Guseva {\em et~al.\/}(2013)Guseva, Feudel \& T\'el]{guseva2013}
{\sc \au{Guseva, K.}, \au{Feudel, U.} \& \au{T\'el, T.}} \yr{2013}  \at{Influence of the history force on inertial particle advection: Gravitational effects and horizontal diffusion}.  \jt{Physical Review E}  \bvol{88},  \pg{042909}.

\bibitem[Haller \& Sapsis(2008)]{haller2008}
{\sc \au{Haller, G.} \& \au{Sapsis, T.}} \yr{2008}  \at{Where do inertial particles go in fluid flows?}  \jt{Physica D: Nonlinear Phenomena}  \bvol{237}~(5),  \pg{573--583}.

\bibitem[Jaganathan {\em et~al.\/}(2023)Jaganathan, Govindarajan \& Vasan]{jaganathan2023}
{\sc \au{Jaganathan, D.}, \au{Govindarajan, R.} \& \au{Vasan, V.}} \yr{2023}  \at{Explicit {R}unge-{K}utta algorithm to solve non-local equations with memory effects: case of the {M}axey-{R}iley-{G}atignol equation} ,  \arxiv{arXiv: 2308.09714}.

\bibitem[Lasheras \& Tio(1994)]{kiong94}
{\sc \au{Lasheras, Juan~C.} \& \au{Tio, Kek-Kiong}} \yr{1994}  \at{{Dynamics of a Small Spherical Particle in Steady Two-Dimensional Vortex Flows}}.  \jt{Applied Mechanics Reviews}  \bvol{47}~(6S),  \pg{S61--S69}.

\bibitem[Marcu {\em et~al.\/}(1995)Marcu, Meiburg \& Newton]{MMN1995}
{\sc \au{Marcu, B.}, \au{Meiburg, E.} \& \au{Newton, P.~K.}} \yr{1995}  \at{{Dynamics of heavy particles in a Burgers vortex}}.  \jt{Physics of Fluids}  \bvol{7}~(2),  \pg{400--410}.

\bibitem[Marshall(1998)]{Marshall1998}
{\sc \au{Marshall, J.~S.}} \yr{1998}  \at{{A model of heavy particle dispersion by organized vortex structures wrapped around a columnar vortex core}}.  \jt{Physics of Fluids}  \bvol{10}~(12),  \pg{3236--3238}.

\bibitem[Marshall(2005)]{Marshall2005}
{\sc \au{Marshall, J.~S.}} \yr{2005}  \at{{Particle dispersion in a turbulent vortex core}}.  \jt{Physics of Fluids}  \bvol{17}~(2),  \pg{025104}.

\bibitem[Maxey {\em et~al.\/}(1996)Maxey, Chang \& Wang]{maxeyWang1996}
{\sc \au{Maxey, M.R.}, \au{Chang, E.J.} \& \au{Wang, L-P.}} \yr{1996}  \at{Interactions of particles and microbubbles with turbulence}.  \jt{Experimental Thermal and Fluid Science}  \bvol{12}~(4),  \pg{417--425}.

\bibitem[Maxey(1987)]{fieldapprox_Maxey87}
{\sc \au{Maxey, M.~R.}} \yr{1987}  \at{The gravitational settling of aerosol particles in homogeneous turbulence and random flow fields}.  \jt{Journal of Fluid Mechanics}  \bvol{174},  \pg{441–465}.

\bibitem[Maxey \& Riley(1983)]{maxey1983}
{\sc \au{Maxey, M.~R.} \& \au{Riley, J.~J.}} \yr{1983}  \at{Equation of motion for a small rigid sphere in a nonuniform flow}.  \jt{Physics of Fluids}  \bvol{26}~(4),  \pg{883--889}.

\bibitem[Moffatt {\em et~al.\/}(1994)Moffatt, Kida \& Ohkitani]{moffatt1994}
{\sc \au{Moffatt, H.~K.}, \au{Kida, S.} \& \au{Ohkitani, K.}} \yr{1994}  \at{Stretched vortices – the sinews of turbulence; large-{R}eynolds-number asymptotics}.  \jt{Journal of Fluid Mechanics}  \bvol{259},  \pg{241–264}.

\bibitem[Nath \& Roy(2024)]{nath2024clustering}
{\sc \au{Nath, Anu V.~S.} \& \au{Roy, Anubhab}} \yr{2024} Clustering and chaotic motion of heavy inertial particles in an isolated non-axisymmetric vortex,  \arxiv{arXiv: 2403.10011}.

\bibitem[Nizkaya {\em et~al.\/}(2010)Nizkaya, Angilella \& Buès]{angilella2010}
{\sc \au{Nizkaya, T.}, \au{Angilella, J.~R.} \& \au{Buès, M.}} \yr{2010}  \at{{Note on dust trapping in point vortex pairs with unequal strengths}}.  \jt{Physics of Fluids}  \bvol{22}~(11),  \pg{113301}.

\bibitem[Prasath {\em et~al.\/}(2019)Prasath, Vasan \& Govindarajan]{prasath2019}
{\sc \au{Prasath, S.~G.}, \au{Vasan, V.} \& \au{Govindarajan, R.}} \yr{2019}  \at{Accurate solution method for the {M}axey–{R}iley equation, and the effects of {B}asset history}.  \jt{Journal of Fluid Mechanics}  \bvol{868},  \pg{428–460}.

\bibitem[Raju \& Meiburg(1997)]{rajuMeiburg97}
{\sc \au{Raju, N.} \& \au{Meiburg, E.}} \yr{1997}  \at{{Dynamics of small, spherical particles in vortical and stagnation point flow fields}}.  \jt{Physics of Fluids}  \bvol{9}~(2),  \pg{299--314},  \arxiv{arXiv: https://pubs.aip.org/aip/pof/article-pdf/9/2/299/19133848/299\_1\_online.pdf}.

\bibitem[Ramadugu {\em et~al.\/}(2022)Ramadugu, Perlekar \& Govindarajan]{ramadugu2022}
{\sc \au{Ramadugu, R.}, \au{Perlekar, P.} \& \au{Govindarajan, R.}} \yr{2022}  \at{Surface tension as the destabiliser of a vortical interface}.  \jt{Journal of Fluid Mechanics}  \bvol{936},  \pg{A45}.

\bibitem[Ravichandran \& Govindarajan(2015)]{croorSingleVortex2015}
{\sc \au{Ravichandran, S.} \& \au{Govindarajan, R.}} \yr{2015}  \at{{Caustics and clustering in the vicinity of a vortex}}.  \jt{Physics of Fluids}  \bvol{27}~(3),  \pg{033305}.

\bibitem[Ravichandran {\em et~al.\/}(2014)Ravichandran, Perlekar \& Govindarajan]{croorVortexPair2014}
{\sc \au{Ravichandran, S.}, \au{Perlekar, P.} \& \au{Govindarajan, R.}} \yr{2014}  \at{{Attracting fixed points for heavy particles in the vicinity of a vortex pair}}.  \jt{Physics of Fluids}  \bvol{26},  \pg{013303}.

\bibitem[Reade \& Collins(2000)]{readeCollins00}
{\sc \au{Reade, W.~C.} \& \au{Collins, L.~R.}} \yr{2000}  \at{{Effect of preferential concentration on turbulent collision rates}}.  \jt{Physics of Fluids}  \bvol{12}~(10),  \pg{2530--2540}.

\bibitem[Sapsis \& Haller(2010)]{haller2010}
{\sc \au{Sapsis, T.} \& \au{Haller, G.}} \yr{2010}  \at{{Clustering criterion for inertial particles in two-dimensional time-periodic and three-dimensional steady flows}}.  \jt{Chaos: An Interdisciplinary Journal of Nonlinear Science}  \bvol{20}~(1),  \pg{017515}.

\bibitem[Shuai {\em et~al.\/}(2024)Shuai, Roy \& Kasbaoui]{Shuai_2024}
{\sc \au{Shuai, Shuai}, \au{Roy, Anubhab} \& \au{Kasbaoui, M.~Houssem}} \yr{2024}  \at{The merger of co-rotating vortices in dusty flows}.  \jt{Journal of Fluid Mechanics}  \bvol{981}.

\bibitem[Squires \& Eaton(1991)]{squiresEaton91}
{\sc \au{Squires, K.~D.} \& \au{Eaton, J.~K.}} \yr{1991}  \at{{Preferential concentration of particles by turbulence}}.  \jt{Physics of Fluids A: Fluid Dynamics}  \bvol{3}~(5),  \pg{1169--1178}.

\bibitem[Tanga {\em et~al.\/}(1996)Tanga, Babiano, Dubrulle \& Provenzale]{tanga96}
{\sc \au{Tanga, P.}, \au{Babiano, A.}, \au{Dubrulle, B.} \& \au{Provenzale, A.}} \yr{1996}  \at{Forming planetesimals in vortices}.  \jt{Icarus}  \bvol{121}~(1),  \pg{158--170}.

\bibitem[Tio {\em et~al.\/}(1993)Tio, Liñán, Lasheras \& Gañán-Calvo]{calvo1993}
{\sc \au{Tio, Kek-Kiong}, \au{Liñán, Amable}, \au{Lasheras, Juan~C.} \& \au{Gañán-Calvo, Alfonso~M.}} \yr{1993}  \at{On the dynamics of buoyant and heavy particles in a periodic stuart vortex flow}.  \jt{Journal of Fluid Mechanics}  \bvol{254},  \pg{671–699}.

\bibitem[Tél(2015)]{tel2015}
{\sc \au{Tél, T.}} \yr{2015}  \at{{The joy of transient chaos}}.  \jt{Chaos: An Interdisciplinary Journal of Nonlinear Science}  \bvol{25}~(9),  \pg{097619}.

\bibitem[Varaksin \& Ryzhkov(2022)]{alekseyReview22}
{\sc \au{Varaksin, A.~Y.} \& \au{Ryzhkov, S.~V.}} \yr{2022}  \at{Vortex flows with particles and droplets (a review)}.  \jt{Symmetry}  \bvol{14}~(10).

\bibitem[Wang \& Maxey(1993)]{wangMaxey93}
{\sc \au{Wang, L.~P.} \& \au{Maxey, M.~R.}} \yr{1993}  \at{Settling velocity and concentration distribution of heavy particles in homogeneous isotropic turbulence}.  \jt{Journal of Fluid Mechanics}  \bvol{256},  \pg{27–68}.

\bibitem[Wilkinson {\em et~al.\/}(2006)Wilkinson, Mehlig \& Bezuglyy]{Wilkinson2006}
{\sc \au{Wilkinson, Michael}, \au{Mehlig, Bernhard} \& \au{Bezuglyy, Vlad}} \yr{2006}  \at{Caustic activation of rain showers}.  \jt{Phys. Rev. Lett.}  \bvol{97},  \pg{048501}.

\end{thebibliography}

\end{document}